\documentclass{emulateapj}
\usepackage{natbib}
\slugcomment{To appear in the Astrophysical Journal}
\shortauthors{Fitzpatrick}
\shorttitle{UV Spectrum of Vega}

\begin{document}

\newcommand{\atlas}{{ATLAS9}}
\newcommand{\tlusty}{{TLUSTY}}
\newcommand{\synspec}{{SYNSPEC}}
\newcommand{\spectrum}{{SPECTRUM}}

\newcommand{\iras}{{\it IRAS}}
\newcommand{\hip}{{\it Hipparcos}}
\newcommand{\cop}{{\it Copernicus}}
\newcommand{\ans}{{\it ANS}}
\newcommand{\hst}{{\it HST}}
\newcommand{\iue}{{\it IUE}}
\newcommand{\oao}{{\it OAO-2}}
\newcommand{\tmass}{{2MASS}} 

\newcommand{\jhk}{{\it JHK}}  
\newcommand{\ubv}{{\it UBV}}                                                                                                                                                                                                                            
\newcommand{\ebv}{\mbox{$E(B\!-\!V)$}}
\newcommand{\klam}{\mbox{$k(\lambda \!-\!V)$}}

\newcommand{\mast}{{MAST}}
\newcommand{\irsa}{{IRSA}}
\newcommand{\simbad}{{SIMBAD}}

\newcommand{\teff}{\mbox{$T_{\rm eff}$}}
\newcommand{\logg}{{$\log g$}}
\newcommand{\abund}{[m/H]}
\newcommand{\vturb}{$v_{turb}$}
\newcommand{\vrot}{$v \sin i$}
\newcommand{\vrad}{$v_{rad}$}

\newcommand{\msun}{${\rm M}_\sun$}
\newcommand{\rsun}{${\rm R}_\sun$}

\newcommand{\mic}{\mbox{$\mu{\rm m}$}}
\newcommand{\invmic}{\mbox{$\mu{\rm m}^{-1}$}}
\newcommand{\kms}{km\,s$^{-1}$}

\title{UV Spectral Synthesis of Vega}

\author{E.L.~Fitzpatrick}
\affil{Department of Astronomy \& Astrophysics, Villanova
University, 800 Lancaster Avenue, Villanova, PA 19085, USA; 
edward.fitzpatrick@villanova.edu}

\begin{abstract}
We show that the UV spectrum (1280--3200 \AA) of the ``superficially normal'' A-star Vega, as observed by the \iue\ satellite at a resolution comparable to the star's rotational broadening width, can be fit remarkably well by a single-temperature synthetic spectrum based on LTE atmosphere models and a newly constructed UV line list.  If Vega were a normal, equator-on, slow-rotating star, then its spectrum and our analysis would indicate a temperature of \teff\ $\simeq 9550$ K, surface gravity of \logg\ $\simeq 3.7$, general surface metallicity of [m/H] $\simeq -0.5$, and a microturbulence velocity of \vturb\ $\simeq 2.0$ \kms.  Given its rapid rotation and nearly pole-on orientation, however, these parameters must be regarded as representing averages across the observed hemisphere.  Modeling the complex UV line spectrum has allowed us to determine the specific surface abundances for 17 different chemical elements, including CNO, the light metals, and the iron group elements.  The resultant abundance pattern agrees in general with previous results, although there is considerable scatter in the literature. Despite its peculiarities, Vega has turned out to provide a powerful test of the extent of our abilities to model the atmospheric properties of the early A-stars, particularly the detailed UV line spectrum.  The value of the measurements from this pilot study will increase as this analysis is extended to more objects in the rich high-dispersion \iue\ data archive, including both normal and peculiar objects.
\end{abstract}


\section{INTRODUCTION} \label{sec_INTRO}

The main sequence A-stars (including the range from late-B through early-F) include a fascinating array of distinct, chemically peculiar objects. Among the specific types are the Am stars, HgMn stars, He-weak stars, and the $\lambda$ Bootis stars \citep[see, e.g., the discussion by][]{Wolff1983}. The lack of strong stellar winds in this spectral range, combined with the absence of a vigorous convective zone, results in a quiet outer atmosphere in which relatively slow diffusive processes can operate, potentially producing both vertical and horizontal chemical inhomogeneities in the line-forming region of the stars' photospheres. A number of specific processes have been suggested as contributing to the various forms of peculiarities, including magnetic fields, rotation, gravitational settling, accretion from interstellar space, and accretion from circumstellar disks. Indeed, a recent IAU symposium was entirely devoted to study of the ``A-Star Puzzle'' \citep{Zverko2004}, and \citet{Landstreet2004} notes how studies of A-stars can lead to insights into the physical processes producing the observed phenomena.  As noted by \citet[][also \citealt{Cowley1982}]{Cowley1980}, the general diffusion model for the chemically peculiar stars holds that they evolve from stars with normal chemical composition.  Likewise for the case of accretion.  The importance of ``normal'' A-stars is thus that they set the compositional benchmarks against which peculiar abundance patterns are measured.  Even the normal A-stars, however, pose challenges --- particularly in identifying exactly what normal is for an A-star.  Over 40 years ago \citet[][and see references within]{CSII} detected compositional anomalies in A-stars classified as normal. These stars have since become known as the ``superficially normal'' stars, i.e., stars whose peculiarities are not recognized until quantitative spectral analyses are performed.  The poster child for such stars is Vega (HD~172167), the photometric and spectral classification standard which, nevertheless, is now well known to be a pole-on, rapidly rotating star with extreme gravity darkening, decidedly non-solar surface abundances, an infrared excess, and a dusty disk \citep[e.g.,][]{Sadakane1981,Aumann1984,Gray1985,Gulliver1994,Wilner2002,Peterson2006}. Such results lead to the question of whether truly ``normal'' A-stars even exist \citep{Hill1991}! 

This paper represents a pilot study to determine the feasibility of modeling the full UV spectrum of the A-stars (both normal and peculiar) in order to extract information on the basic stellar properties (e.g., \teff\ and \logg) and the atmospheric surface abundances for a wide variety of chemical elements.  The ultraviolet, shortward of $\sim$3200 \AA, is an extremely rich spectral region for such a study, as it contains thousands of individual absorption lines from many chemical elements in multiple stages of ionization, generally including the dominant stages.  Of course, numerous studies of A-star abundances have been performed in the past including, for example, the work by Adelman and collaborators \citep[][and earlier papers]{Adelman2006}, \citet[][and earlier papers]{Hill1995}, and \citet{Varenne1999}.  These studies generally involve short segments of high-resolution ground-based spectra, one to $\sim$20 elements, and dozens to (occasionally) a few hundred individual spectral features. Only a very few quantitative studies have utilized UV spectra (e.g., \citealt{Smith1997} and earlier papers in the series; and \citealt{Adelman2004}) and these have concentrated on small segments of data containing spectral features for a small number of elements.  The new aspect of this program is that we plan to utilize essentially all the spectral information present in the wavelength range $\sim$1200--3200 \AA.  This study will complement the previous work, adding a valuable check on results derived from the optical spectral range, adding access to additional elements, and potentially revealing heretofore undetected behaviors in the surface composition of the A-stars.

The chief data set for this study is to be the archive of high-dispersion observations made by the {\it International Ultraviolet Explorer} (\iue) satellite \citep{Boggess1978a,Boggess1978b}.  These data cover the wavelength range 1150--3200 \AA\ with a spectral resolution of $\sim$20 \kms, utilizing observations with a short wavelength camera (1150--2000 \AA) and a long-wavelength camera (1900--3200 \AA).  A search of the \iue\ database, managed by the Multimission Archive at STScI (MAST), reveals $\sim$300 stars in object classes 22 (B6-B9.5 V-IV) and 30 (A0-A3 V-IV) for which data from both the short- and long-wavelength \iue\ cameras are available.  Of these, $\sim$100 are nominally normal stars, and the rest exhibit some compositional anomaly.  The great majority of these spectra have never been fully analyzed and thus this archive constitutes a treasure trove of virtually untapped information on the surface properties of the A stars.

The specific goals of this planned study are:

\begin{enumerate}
\item Determine the degree to which the full UV spectrum can be reproduced using the standard tools of LTE atmospheric models and LTE spectral synthesis;
\item Search for systematic discrepancies which may signal shortcomings in the input physics or assumptions, or both, of the standard tools or, alternatively, specific peculiarities with the target stars; 
\item Examine the completeness and accuracy of the line lists used to generate the synthetic spectra;
\item Develop better estimates of physical properties of the target stars and, particularly, confirm, refine, and expand surface abundance determinations using a different technique and a generally different set of spectral features than most previous analyses;
\item Examine the abundance patterns revealed from the study of a large group of normal and peculiar A-stars to gain insight into the processes driving the compositional anomalies.
\end{enumerate}

The target for this pilot study is the aforementioned Vega.  Given its special nature, its use for the pilot study might be questioned.  Echoing the thoughts of \citet{GarciaGil2005}, however, we note that the quality and quantity of data available for Vega are unique.  We will be able to compare our results to a large number of previous studies --- based on similar physical assumptions --- to help determine the reliability of the current analysis, while perhaps adding additional insights into this interesting star.  This paper is not expected nor intended to supersede other studies which more fully take into account Vega's particular physical peculiarities, notably, the high rotation rate which is partially masked by a near pole-on orientation \citep[e.g.,][]{Hill2010}.  In concert with such studies, however, this analysis will help in determining the limitations in accuracy which may be expected when simplified, though still computationally-challenging, models are used in the analysis of complex objects.  

The outline of this paper is straightforward.  In \S \ref{sec_DATA} we describe the data sets assembled for this study, including the \iue\ data noted above, as well as a limited number of UV spectra from other instruments.  The modeling process, through which we attempt to reproduce the full UV spectrum, is presented \S \ref{sec_ANALYSIS}.  The results of the modeling process are described in \S \ref{sec_BESTFITMODEL}  and \S \ref{sec_SED}, giving both a closeup of the detailed spectrum and a broad view of the spectral energy distribution, respectively.  The surface composition of Vega's atmosphere implied by our analysis is discussed in \S \ref{sec_COMPOSITION}, along with a comparison with previous results.  Some final comments are given in \S \ref{sec_SUMMARY}.

\section{THE DATA} \label{sec_DATA}

The primary data set used here consists of \iue\ spectrophotometry obtained using the large-aperture, high-resolution echelle observing mode with both the short-wavelength SWP camera (1150--2000 \AA) and the long-wavelength LWP camera (1900--3200 \AA). The spectral resolutions of the SWP and LWP data, $\sim$23 \kms\/ and $\sim$16 \kms, respectively, are comparable to the rotational broadening observed for Vega (see \S \ref{sec_ANALYSIS}).  The specific \iue\/ spectra used are listed in Table \ref{tab_DATA}.  The six SWP observations were all obtained during the time period 1988--1992 and have similar exposure times and data quality.  Two other large-aperture, high-resolution SWP spectra exist, from earlier time periods and with different exposure times, but were excluded in the interests of uniform data quality.  The five LWP observations listed in Table \ref{tab_DATA} were obtained between 1986 and 1992 and all have identical exposure times.  One other LWP spectrum, from an earlier time period and with a shorter exposure time, was excluded for the sake of data uniformity.  For the same reason, three high-resolution spectra from the LWR camera, which has a different photometric response and a different spectral resolution from the LWP camera, were not included.

The 11 \iue\/ spectra in Table \ref{tab_DATA} were all acquired from the \iue\/ data archive as hosted by MAST\footnote{The MAST Archive was accessed at http://archive.stsci.edu/hst/}.  These spectra were processed with the New Spectral Image Processing System \citep[NEWSIPS;][]{Nichols1996}.  We produced a single mean SWP spectrum by averaging the six individual SWP spectra and, likewise, produced a single mean LWP spectrum.  The averaging was done using routines in the \iue\/ Data Analysis Center (IUEDAC) software suite, written in the {\it Interactive Data Language} (IDL) and downloaded from MAST.  In particular, we used the routine CRSCOR to measure any small wavelength shifts that might exist within the sets of SWP and LWP spectra and then used COADD to correct the shifts and average together the individual spectra.  The averaging was done on an echelle order-by-order basis, with the mean SWP spectrum consisting of 51 separate orders (covering the wavelength range 1151--1978 \AA) and the LWP spectrum consisting of 47 orders (covering 1899--3097 \AA).  The final step in the data processing was to trim the spectra to eliminate low-quality data and overlap between orders.  Specifically, we: 1) removed the first and last 10\% of the data points from each order, 2) removed points which overlapped with the next order (in the sense of increasing wavelength), 3) eliminated all SWP data shortward of 1282 \AA\ (where the noise level is high), and 4) eliminated all LWP data shortward of 1979 \AA\ (which overlaps with the higher quality SWP data).  The result of this procedure is nearly complete coverage of the UV between 1282 and 3097 \AA, with a spectral resolution comparable to the expected intrinsic width of the stellar features and consisting of data from 80 different \iue\/ echelle orders (38 from the SWP and 42 from the LWP).  This spectrum is shown in the top panel in Figure \ref{fig_IUEDATA}.  The apparent noisiness of the spectrum is mainly the presence of many strong absorption features, although the short wavelength end of the LWP data (1980-2300 \AA) does have a significantly lower signal-to-noise ratio than other regions of the spectrum.  Small gaps at the long wavelength ends of the SWP and LWP data arise from incomplete overlap between adjacent echelle orders.  Some medium frequency structure in the data, most noticeable around 2200 \AA, arises from incomplete removal of the echelle ``ripple,'' a long-standing problem with high-resolution \iue\/ data.  The analysis described in \S \ref{sec_ANALYSIS} below compensates for this effect.  The model spectrum in the bottom panel of Figure \ref{fig_IUEDATA} will be discussed in \S \ref{sec_BESTFITMODEL}.

Our analysis also utilizes a small number of high quality, medium- and high-resolution UV spectra of Vega obtained by the Goddard High Resolution Spectrograph (GHRS) aboard the Hubble Space Telescope (\hst).   Likewise, we include a far-UV spectrum obtained by the Berkeley Extreme and Far-Ultraviolet Spectrometer (BEFS).  The BEFS instrument was flown aboard the ORPHEUS telescope on the {\it ORPHEUS-SPAS} space shuttle missions in 1993 and 1996 and is described fully by \citet{Dixon2002}.  The relevant GHRS and BEFS observations are listed in Table \ref{tab_DATA}.  All these data were obtained from MAST and no additional processing was performed.  

Finally, low-resolution UV spectra from the \iue\/ satellite provide a reference for the absolute flux levels of Vega, which will be examined in \S \ref{sec_SED}.  To facilitate this discussion, we obtained from MAST a set of 7 low-resolution SWP spectra, consisting of images SWP27024, SWP29864, SWP29866, SWP29867, SWP30548, SWP30549, SWP32906, and SWP32907.  These data were all acquired by \iue\ using the ``fast-trail'' technique developed for bright sources and the list includes all the SWP spectra for Vega for which no detector saturation occurred.  We corrected the data --- which were processed using the NEWSIPS system --- for residual systematic calibration errors and placed them onto the HST/FOS flux scale of \citet{Bohlin1996} using the corrections and algorithms described by \citet[][hereafter MF00]{massa2000}.  The individual spectra were then averaged to form a single far-UV spectrum covering the wavelength range $\sim$1250--2000 \AA\ and with a spectral resolution of approximately 6 \AA.

\section{THE ANALYSIS} \label{sec_ANALYSIS}

Our analysis consists of computing a rotationally- and instrumentally-broadened LTE synthetic UV spectrum from a pre-existing stellar atmosphere model and comparing it with the observed high-resolution UV spectrum of Vega.  Adjustments are then made iteratively in the various stellar properties which determine the synthetic spectrum until the best possible agreement is obtained with the observations.  For the atmosphere models, we use the ATLAS9 grid from \citet{kurucz1991}, computed with a microturbulence velocity of 2 \kms\/ and a wide range of \teff, \logg, and [m/H] values, spanning the region of interest.  To compute the synthetic spectrum, we use the program SPECTRUM (Version 2.76e) developed by Richard Gray \citep[see][]{Gray1994}\footnote{SPECTRUM was obtained from http://www1.appstate.edu/dept/physics/spectrum/spectrum.html}.

The main analysis was performed on the high-resolution \iue\/ data described in the previous section, which consist of 80 separate echelle orders covering the range 1282--3097 \AA, with no overlap between orders and all normalized by the mean flux as computed over the entire region.  The analysis includes a set of global stellar parameters and a set of order-specific parameters.  The former include the effective temperature \teff, surface gravity \logg, and overall abundance [m/H] --- all of which specify the \atlas\ atmospheric structure model --- as well as a set of individual elemental abundances X/H, the microturbulence velocity \vturb, and the rotational broadening velocity \vrot\ --- all of which are used in the running of SPECTRUM and the construction of the synthetic spectrum. The order-specific parameters (i.e., those determined individually for each echelle order) include the radial velocity \vrad\ and a 5-point ``continuum'' used to match the low-frequency shape of the individual echelle order data and the synthetic spectrum.  $V_{rad}$ is taken as an order-dependent quantity to help compensate for order-specific wavelength calibration uncertainties.  The ``continuum'' is computed using a cubic spline interpolation between 5 equally-spaced anchor points (for each order) whose values are determined in the analysis.  The use of this continuum removes the influence of low- and medium-frequency phenomena on our analysis, including interstellar extinction, shortcomings in the absolute calibration of the data, and shortcomings in the removal of the echelle ``ripple'' signature from the \iue\/ data.   Our results therefore are dependent only on the high-frequency information in the data, i.e., the spectral lines, and insensitive to the general level or shape of the spectral energy distribution (SED).  In the analysis, predetermined values can be adopted for any or all of the parameters described above, or best-fit values can be determined iteratively.

Beginning with an initial set of estimates for the full set of free parameters, the analysis proceeds as follows:
\begin{enumerate}
\item Construct a stellar atmosphere model for the desired set of \teff, \logg, and [m/H] values.  We produce this model by logarithmically interpolating the structure variables within the ATLAS9 2 \kms\/ grid using the set of models on the vertices of the grid which surround the desired parameter set.  The result is then outputted to a file in the format of the ATLAS9 models and serves as input for the second step.

\item Compute the full resolution synthetic UV spectrum covering the full wavelength range of the data using the program SPECTRUM, whose inputs include the atmosphere model, a set of elemental abundances, a value of \vturb, and a line list containing the atomic data for the relevant set of absorption features. 

\item Rotationally broaden the full spectrum, using the desired desired value of \vrot.

\item Slice the rotationally broadened spectrum into 80 segments covering the wavelength ranges of the individual \iue\/ orders.  These segments range from 12 to 34 \AA\ wide.  Broaden each segment with the appropriate instrumental resolution, velocity-shift them according to the order-dependent value of \vrad, and the scale them by the interpolated 5-point ``continuum.'' 

\item Compare this set of theoretical spectra with the observed \iue\/ data and make adjustments to the free parameters to improve the agreement.  

\item Go back to step 1 and repeat the process until no further improvement in agreement between theory and observation can be obtained. The final set of free parameters constitutes our ``best-fit'' model.
\end{enumerate}

The iterative comparison between theoretical and observed spectra is handled by the non-linear least squares procedure MPFIT written in the {\it IDL} by \citet{Markwardt2009}.  Weighting factors for the analysis were determined from measurements of the signal-to-noise ratio across each echelle order.  Weights were set to zero for cases of detector blemishes and in the cores of strong resonance lines which might be expected to have some interstellar contamination.  Also, the weights were set to zero for regions including some \ion{C}{1} and \ion{Si}{1} lines which are found to have suspect atomic parameters (see \S \ref{sec_COMPOSITION}).

Rotational and instrumental broadening is performed with the programs SMOOTH2 and AVSINI which come with Gray's SPECTRUM package.  Note that the rotational broadening step is only an approximation, since it assumes that Vega is an equator-on star.  However, departures of the line shapes from the classical rotational profile seem important only for weak lines of non-dominant species \citep[e.g,][]{Gulliver1991}, which play only a small role in our analysis.  The appropriate FWHM values of the instrumental broadening for high-resolution \iue\/ LWP and SWP data were taken from Figures 2.21 and 2.23, respectively, of the \iue\/ NEWSIPS Information Manual, as obtained from the MAST Archives.  For the SWP, a value of 3 pixels was adopted for all orders, corresponding to 0.10 \AA\ ($\sim$23 \kms) at the shortest wavelengths and 0.15 \AA\ ($\sim$23 \kms) at the longest wavelengths.  For the LWP, the resolution is somewhat wavelength-dependent, varying between 2.2 and 3.0 pixels.  This produces a range of 0.12 to 0.19 \AA\/ in wavelength space and 16 to 20 \kms\ in velocity space.  For each of the 80 \iue\/ spectral orders, a single value of the spectral resolution, specified in \AA\ and computed for the mean wavelength of the order, was used in SMOOTH2.  The broadening profile is assumed to be Gaussian.  Vacuum wavelengths were used for all UV calculations.      

Reproducing a high-resolution spectrum requires a reliable list of absorption lines in the spectral range of interest and an accurate set of atomic parameters for those lines.  For this study, and for a parallel one of the bright A-star Sirius A (Fitzpatrick \& McClain, in preparation; hereafter Paper II) we constructed a new list, containing the most recent updates to the atomic transition knowledge base.  We formed this list by compiling the data for all atomic transitions producing spectral lines between 1000 \AA\ and 3400 \AA\ and with absolute line strength information (i.e., $\log gf$), utilizing the Vienna Atomic Line Database \citep[VALD;][]{Kupka2000}\footnote{The VALD was accessed at http://vald.astro.univie.ac.at/~vald/php/vald.php} and the National Institute of Standards and Technology Atomic Spectra Database \citep[NIST/ASD;][]{NIST}\footnote{The NIST Atomic Spectra Database was accessed at http://physics.nist.gov/asd3.}.  The NIST/ASD values of $\log gf$, which represent experimental determinations, were adopted in cases where they and VALD disagreed.  This process yielded a list of over 253,000 features from elements H through U, many with radiative, Stark, and van der Waals broadening half-widths available.  The H lines (Lyman-$\alpha$ and Lyman-$\beta$) were removed from the list since they are treated separately by \spectrum.  Examination of synthetic spectra produced by this initial list revealed a number of clear shortcomings, generally in the form of predicted features where no (or only very weak) observed features were present, suggesting severe errors in the $\log gf$ values.  We treated these by either removing the predicted lines from the list or by empirically measuring $\log gf$ using the observed strengths of the lines.  The details of these and other modifications made to the list are given in Paper II, where we also provide a table of the empirical $\log gf$ values, which may be of interest to other investigators.  Likewise, the final line list will be made available from the author.  As will be noted further in the detailed discussion of the derived abundances (\S \ref{sec_COMPOSITION}), lines with empirically-derived $\log gf$ values were masked in the main analysis and do not contribute to the results.  

\section{The Best-Fit Model} \label{sec_BESTFITMODEL}

After considerable experimentation with the analysis technique, we determined that we could solve for robust values of \teff, \logg, \vturb, \vrad, and the abundances of 17 chemical elements, while adopting fixed, predetermined values for \vrot\ and the metallicity of the \atlas\ structure  models.  The results of this analysis are presented in Table \ref{tab_RESULTS}.  

The stellar properties listed in column 1 are mostly self-explanatory although a few require some elaboration.  These are:  (1) the quantity \abund, which refers to the general metallicity of the ATLAS9 models used to provide the atmospheric structure; (2) the radial velocities, \vrad(SWP) and \vrad(LWP), which are mean values for the SWP and LWP cameras, respectively; and (3) the individual elemental abundances, e.g., C/H, which are the values determined by the analysis and are given in the form ${\rm X/H} \equiv \log{\rm (N_X/N_H)} + 12.0$.  The parameters of our ``best-fit'' model for the UV absorption line spectrum of Vega, as represented by the high-resolution \iue\/ SWP and LWP observations, are listed in column 2 of the Table (``FIT A'').  For this fit, we adopted the ATLAS9 model grid with roughly 1/3 solar abundances (i.e., ${\rm [m/H]} = -0.5$) and assumed a value of \vrot\ = 21.8 \kms\/ from \citet{Gulliver1994}.  (The sensitivity of the results to both assumptions will be examined below.)  All other parameters in the analysis, including the low frequency mapping of the surface fluxes to the observed fluxes, were free parameters determined by the $\chi^2$ minimization procedure.  We fixed \vrot\ due to concerns that the \iue\ spectral resolution was not high-enough nor well-enough determined to allow us to defend a new measurement of the rotational broadening.  In view of the insensitivity of the our results to \vrot, the adoption of a fixed value for the main analysis is reasonable.  For the same reason,  we also made no attempt to model the effects of macroturbulence, whose presence has been suggested by the work of \citet{Yoon2008} and \citet{Hill2010}.  As experience with other stars accumulates, we will attempt to refine the measurements of the \iue\/'s spectral resolution, possibly allowing more information to be extracted from the observed line profiles.  The final column of Table \ref{tab_RESULTS} lists the approximate number of spectral features available for the individual abundance determinations.  Numerous weak features have not been included in this tally and, in some cases, the abundances are dominated by a smaller number of prominent features (see \S \ref{sec_COMPOSITION}). 

The temperature we find from the UV line spectrum, $9547\pm17$ K, is generally consistent with results already in the literature, mostly based on the level or shape of the SED in the optical-through-UV spectral region.  See Table 3 of \citet{GarciaGil2005} for a summary of such results, which generally lie in the range 9600--9700 K.  A different approach, that of modeling the rotationally broadened \ion{Fe}{1} and \ion{Ti}{2} lines at 4528 \AA\ and 4529 \AA, respectively, by \citet{Gulliver1994}, yielded a temperature of $9695\pm25$ K.  Our result is particularly close to the those of \citet[][i.e., 9550 K as based on optical colors]{Castelli1994}, \citet[][i.e., 9553 K as based on the bolometric flux and angular diameter]{Ciardi2001}, and \citet[][i.e., 9560 K, which represents the mean across a rapidly rotating and highly gravity-darkened surface]{Hill2010}.  These latter results and ours may represent a real convergence on an apparent temperature for Vega in the neighborhood of 9550 K .  

The surface gravity in Table \ref{tab_RESULTS}, $3.72\pm0.03$, is somewhat lower than most past results. For example, analyses utilizing the shape of the UV--through-optical SED and those using Balmer line profiles generally find results in the range $\log g$ = 3.90--4.00 \citep[e.g.,][]{Castelli1994, Fitzpatrick1999, GarciaGil2005, Hill2010}. On the other hand, \citet{Gulliver1994}'s line profile analysis yielded $3.75\pm0.02$, very similar to our result.  In considering these values, it should be kept in mind that the rapid rotation of Vega, and the subsequent extreme gravity-darkening, make the concept of a single surface gravity (and a single effective temperature) invalid.  All results for Vega reflect some sort of average across the visible surface of the star and it is not obvious that different techniques should necessarily yield the same results.  We will return to this point later in \S \ref{sec_SUMMARY}.

Many analyses of Vega adopt a microturbulence broadening velocity of 2 \kms, often based on the work of Castelli and Kurucz \citep[e.g.,][]{Castelli1994}.  \citet{Fitzpatrick1999} found a similar value, $1.9 \pm 0.4$ \kms, from modeling the low-resolution UV-through-optical SED using \atlas\ models.  Results from elemental abundance analyses generally yield compatible results, ranging between 1.5 and 2 \kms \citep[e.g.,][]{Sadakane1986,Qiu2001,Saffe2004}.  Our result here, $2.04 \pm 0.02$ \kms, is thus consistent with these previous studies.  On the other hand, analyses by \citet{Adelman1990}, \citet{Hill1993}, and \citet{Hill2010} find values of \vturb\ closer to 1.0 \kms.  We have no explanation for the discrepancy among these results but, clearly, a complete understanding of Vega's atmosphere cannot be claimed until they are resolved.  

The radial velocities noted in Table \ref{tab_RESULTS} are mean values computed from the results for the individual spectral orders of the \iue\ SWP and LWP cameras.  The uncertainties listed are actually the sample standard deviations for the 2 cameras.  The small values, roughly 10\% of the camera resolutions, testify to the fidelity of the wavelength calibrations within each camera.  However, a systematic offset of more than 17 \kms\/ between the two cameras is clear, with the SWP result compatible with the expected radial velocity of Vega (see, e.g., the set of measurements recorded in the \simbad\ database).  Our treatment of \vrad\ as a free parameter for each spectral order eliminates such problems as factors in the analysis.  We defer a discussion of the metallicity of Vega, and the individual elemental abundances until \S \ref{sec_COMPOSITION} below.  

The uncertainties listed for FIT A in Table \ref{tab_RESULTS} are (with the exception of those for \vrad\ as noted above) 1-$\sigma$ statistical uncertainties resulting from the $\chi^2$ minimization procedure.  Thus they underestimate the true uncertainty since they fail to take into account errors in the various assumptions built into the analysis and in the fundamental physical validity of the model atmosphere.  In an attempt to gauge the robustness of the analysis, we ran a number of fits to the data with significant changes in the input assumptions.  These results are also shown in Table \ref{tab_RESULTS}, in columns 3--10. The tests involved: 1) changing the base metallicity of the ATLAS9 models by $\pm$0.5 dex from that assumed in the best-fit model (FITS B and C); 2) changing the assumed value of $v \sin i$ by $\pm$3 \kms\ from that in the best-fit model (FITS D and E); forcing $\log g$ to values $\sim$ $\pm$0.2 dex different than the best-fit result (FITS F and G); and 4) forcing $v_{turb}$ to values $\sim$ $\pm$0.5 \kms\ different from that found in the best fit (FITS H and I).  These tests still don't allow us to place strict absolute error bars on our results, but do give a strong indication that the analysis is very stable against uncertainties in values of the assumed parameters.  We will discuss these results in more detail in \S \ref{sec_COMPOSITION}.

\begin{figure}
\figurenum{1}
\epsscale{1.0}
\plotone{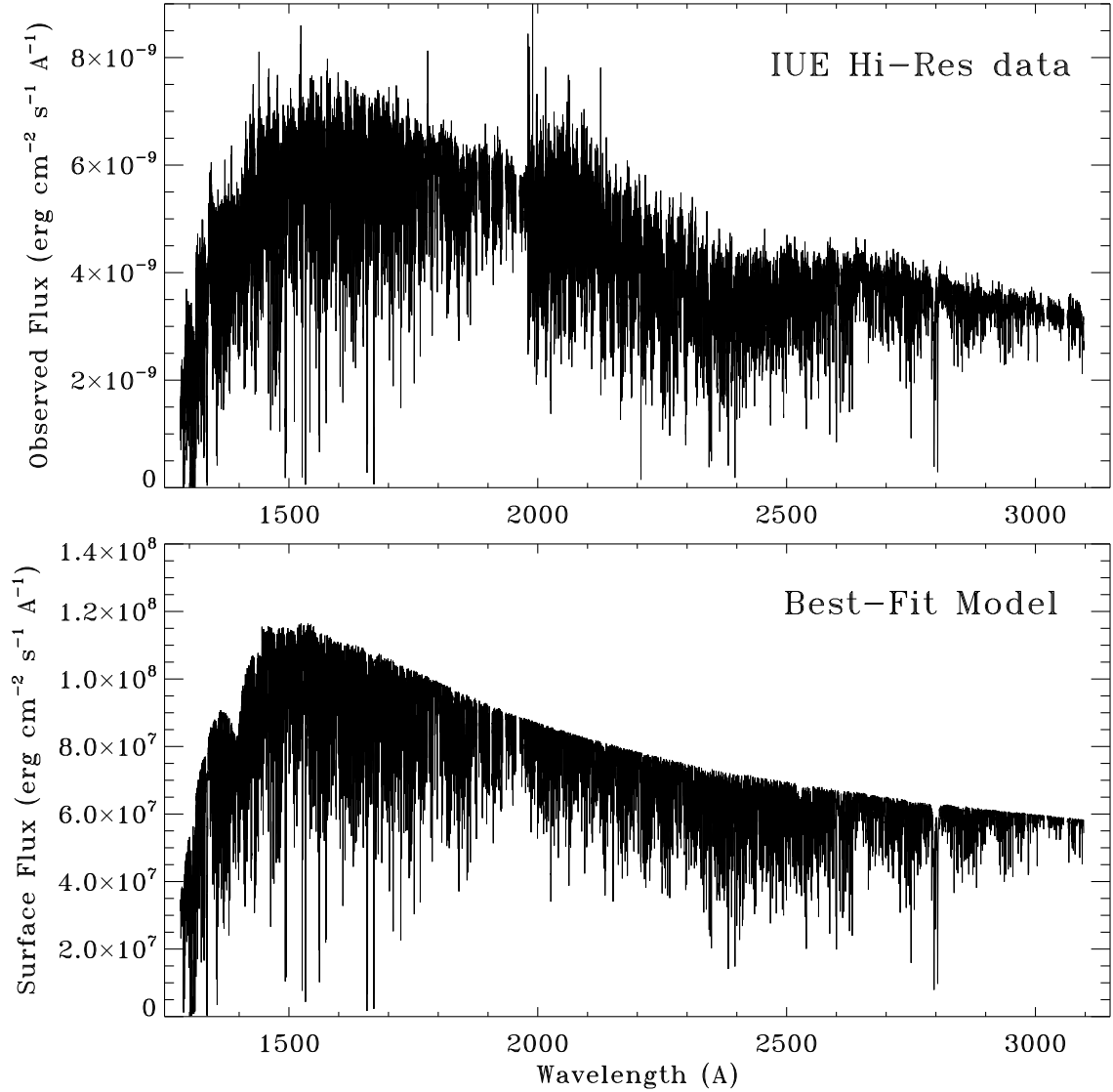}
\caption{\footnotesize {\it Upper Panel:} High-dispersion \iue\ spectrum of Vega, constructed as described in \S \ref{sec_DATA} from the six short-wavelength SWP spectra ($\lambda < 1980$ \AA) and 5 long-wavelength LWP spectra ($\lambda > 1980$ \AA) listed in Table \ref{tab_DATA}.  The apparent noisiness of the data is due to the thousands of individual absorption lines, although the short wavelength end of the LWP data (1980-2300 \AA) has a significantly lower signal-to-noise ratio than other regions.  Small gaps at the long wavelength ends of the SWP and LWP data arise from incomplete overlap between adjacent echelle orders.  Some medium frequency structure in the data, most noticeable around 2200 \AA, arises from incomplete removal of the echelle ``ripple.''  The spectral resolution of the SWP and LWP data are $\sim$23 \kms\ and $\sim$16 \kms, respectively.  {\it Lower Panel:} Best-fitting model of Vega's UV line spectrum, as described in \S \ref{sec_SED}, broadened to the resolution of the \iue\ high-dispersion data.  Apart from the many individual absorption lines, the most notable features in the spectrum are the 1400 \AA\ Lyman-$\alpha$ satellite feature, a \ion{C}{1} edge at 1446 \AA\ due to ionization from the 21648.0 cm$^{-1}$ level, a \ion{Si}{1} edge at 1520 \AA\ due to ionization from the ground state, and the strong \ion{Mg}{2} $h$ and $k$ lines at 2800 \AA.
\label{fig_IUEDATA}}
\end{figure}

An overview of the best-fit model can be gained from the lower panel of Figure \ref{fig_IUEDATA}.  Shown are the surface fluxes of the model, broadened by \vrot\ and the \iue\ spectral resolution, but not mapped to the observed flux levels. The chief features of the model, still visible in the compressed format of Figure \ref{fig_IUEDATA}, are the steeply declining flux in the far-UV (as the wings of the strong \ion{H}{1} Lyman-$\alpha$ line intrude in the observed wavelength window), the 1400 \AA\ Lyman-$\alpha$ satellite feature, a \ion{C}{1} edge at 1446 \AA\ due to ionizations out of the $2s^{2}2p^{2}$ level at 21648.0 cm$^{-1}$, a weaker \ion{Si}{1} edge near 1520 \AA\ due to ionizations out of the ground state, the strong resonance absorption lines of \ion{Mg}{2} at 2800 \AA, and many thousands of other absorption lines spread throughout the observed spectrum.  The 1446 \AA\ and 1520 \AA\ features are difficult to discern in the observed spectrum in the top panel of Figure \ref{fig_IUEDATA} due to calibration deficiencies with the data.  The 1400 \AA\ feature arises from photon absorption during an interaction between an H$^{0}$ atom and a nearby H$^{+}$ ion.  It has been seen in the spectra of DA white dwarfs \citep[e.g.,][]{Greenstein1980}, and was first correctly identified by \citet{Nelan1985} and \citet{Koester1985}.

\begin{figure}
\figurenum{2.1}
\epsscale{1.00}
\plotone{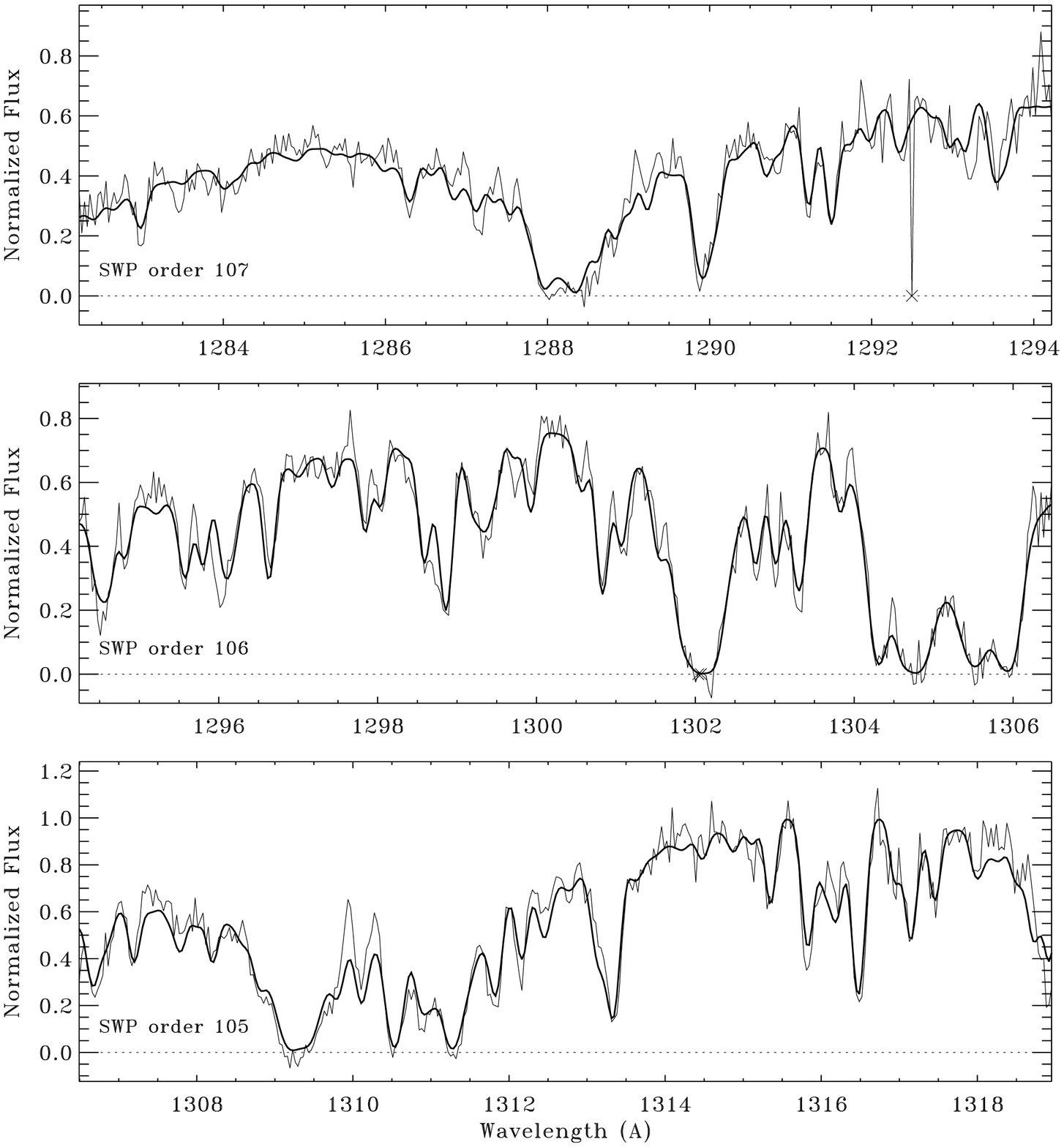}
\caption{{\footnotesize A detailed view of the model fit to Vega's high-resolution \iue\ spectrum.  Shown are the SWP echelle orders 107, 106, and 105, covering the wavelength ranges 1282.22--1294.22 \AA, 1294.24--1306.49 \AA, and 1306.49--1318.95 \AA, respectively.  The model and the data are indicated by the thick and thin curves, respectively.  Data points excluded from the analysis due to detector artifacts are identified with an ``x,'' as at 1292.5 \AA.  Figures 2.1--2.27, which cover the full wavelength range and all 80 SWP and LWP echelle orders included in this analysis, are available in the online version of the Journal.  An ascii file which identifies the various absorption features can be obtained from the author.}
\label{fig_IUEFIT1}}
\end{figure}

A more detailed view of the best-fit model and its level of agreement with the \iue\ high-dispersion data is shown in Figure \ref{fig_IUEFIT1} for the three shortest-wavelength SWP echelle orders (numbers 105, 106, and 107 covering the total wavelength range 1282.22--1318.95 \AA).  The model and the data are indicated by the thick and thin curves, respectively.  Data points excluded from the analysis due to detector artifacts are identified with an ``x,'' as at 1292.5 \AA.  The full set of Figures 2.1--2.27, which cover the entire wavelength range and all 80 SWP and LWP echelle orders included in this analysis, are available in the online version of the Journal.  An ascii file listing all the absorption features that contribute to the model spectrum can be obtained from the author.  A close look at these figures shows that, in general, the quality of the agreement between the model and Vega's UV line spectrum is very good, without any obvious systematic discrepancies.  The range of lines fitted by the model spans weak features, barely visible above the noise level, to very strongly damped features.  The discrepancies that do occur are usually isolated in wavelength space and are generally obviously due to spectral features missing from the atomic line list or to features for which the available atomic data are likely inaccurate, with both issues occurring more frequently at wavelengths shortward of $\sim$1500 \AA. 

\begin{figure}
\figurenum{3.1}
\epsscale{1.00}
\plotone{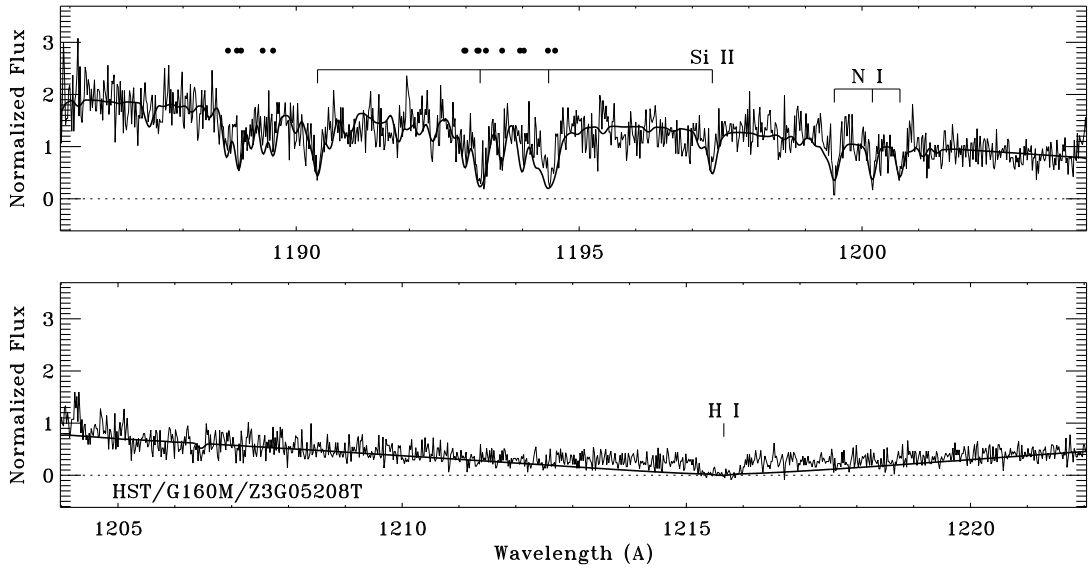}
\caption{A comparison between the best-fit model for Vega (thick curve) and GHRS/G160M spectrum Z3G05208T (thin curve).  The model has been smoothed to match the instrumental resolution of the data and velocity-shifted to align with the observations.  The G160M data were not used in the analysis.  Prominent stellar absorption features are identified.  The \ion{H}{1} Lyman-$\alpha$ line likely contains an interstellar contribution.  Filled circles show the locations of \ion{C}{1} lines arising from the 0, 16.4, and 43.4 cm$^{-1}$ levels.   
\label{fig_GNNNFITS1}}
\end{figure}

\begin{figure}
\figurenum{3.2}
\epsscale{1.00}
\plotone{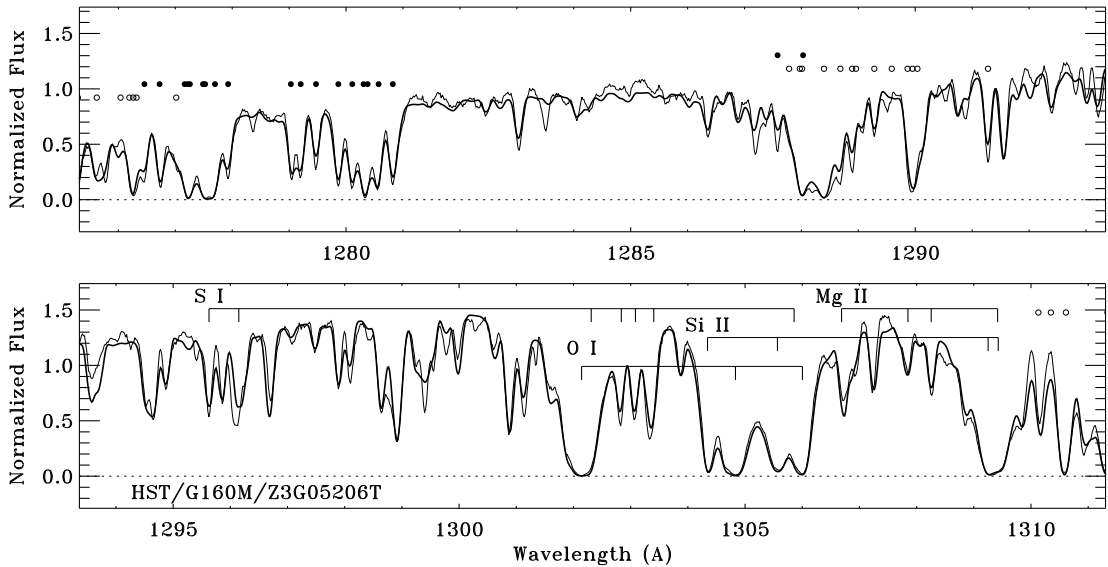}
\caption{Same as Figure 3.1, but for GHRS/G160M spectrum Z3G05206T. Filled circles show the locations of stellar \ion{C}{1} lines arising from the 0, 16.4, and 43.4 cm$^{-1}$ levels. Open circles show \ion{C}{1} lines arising from the 10193.63 cm$^{-1}$ level.
\label{fig_GNNNFITS2}}
\end{figure}

\begin{figure}
\figurenum{3.3}
\epsscale{1.00}
\plotone{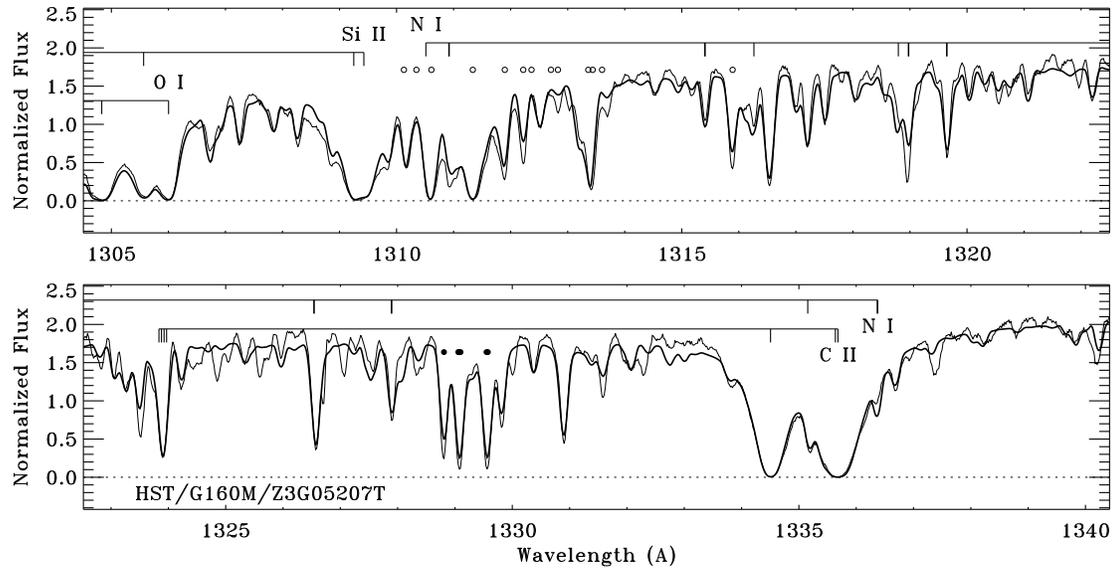}
\caption{Same as Figure 3.1, but for GHRS/G160M spectrum Z3G05207T.
\label{fig_GNNNFITS3}}
\end{figure}

\begin{figure}
\figurenum{3.4}
\epsscale{1.00}
\plotone{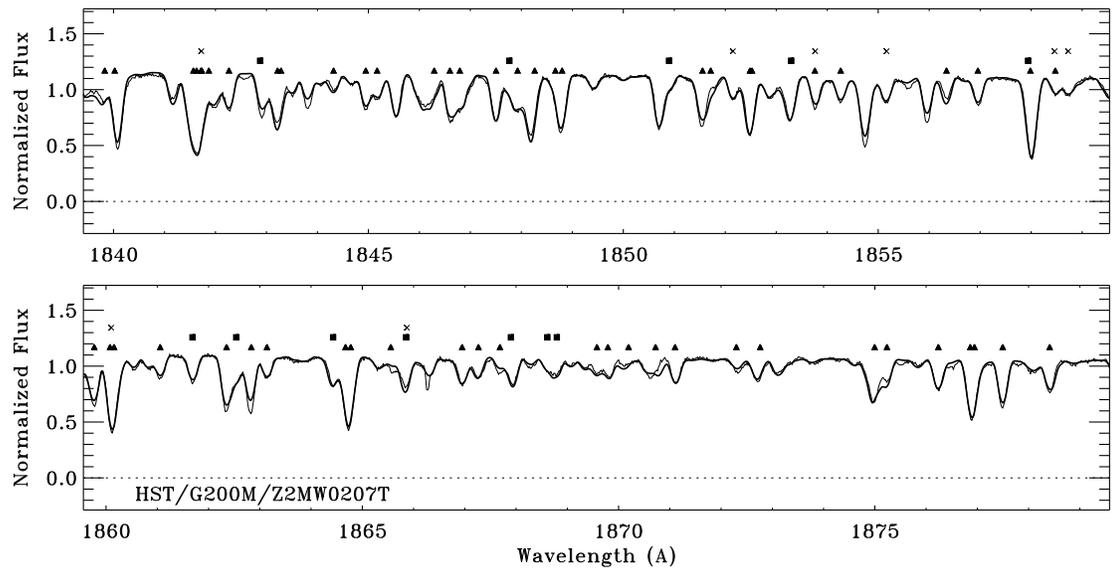}
\caption{Same as Figure 3.1, but for GHRS/G200M spectrum Z2MW0207T.  Triangles, squares, and x's show the locations of \ion{Fe}{2}, \ion{Mn}{2}, and \ion{Cr}{2} lines, respectively.
\label{fig_GNNNFITS4}}
\end{figure}

\begin{figure}
\figurenum{4}
\epsscale{1.00}
\plotone{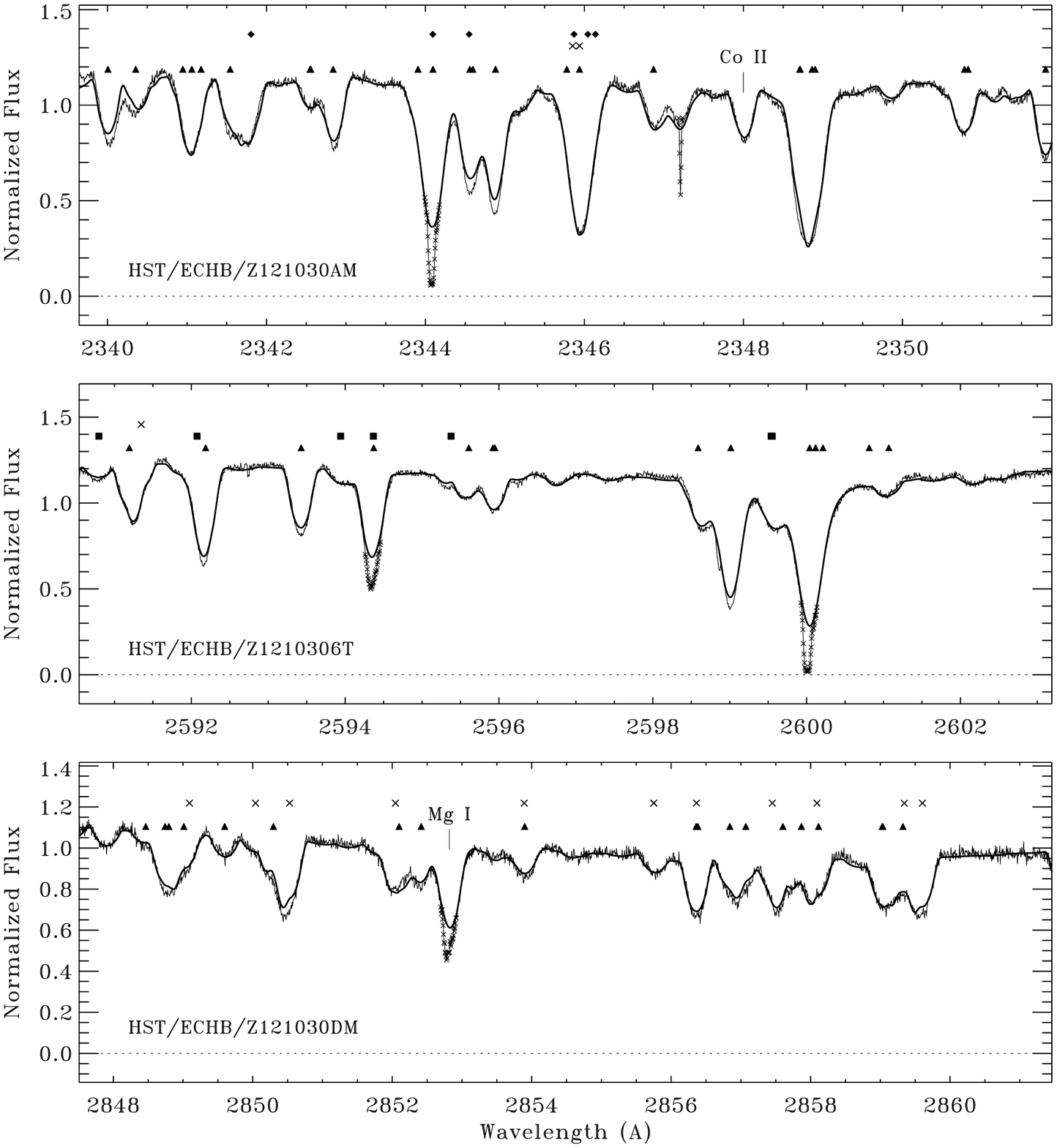}
\caption{A comparison between the best-fit model for Vega (thick curves) and three high-dispersion HST GHRS/ECH-B spectra (thin curves).  The model has been smoothed to match the instrumental resolution of the ECH-B (3 \kms) and velocity-shifted to align with the observations.  Triangles, squares, x's, and diamonds show the locations of \ion{Fe}{2}, \ion{Mn}{2}, \ion{Cr}{2}, and \ion{Ni}{2} lines, respectively.  Several other prominent features are labeled.  The cores of resonance features expected to be contaminated by interstellar absorption are indicated by the small crosses, as are blemishes in the data. 
\label{fig_ECHBFITS}}
\end{figure}

The existence of a small number of \hst/GHRS spectra for Vega (as listed in Table \ref{tab_DATA}) allows a more critical examination of the agreement between theory and observations than possible for the photometrically-inferior \iue\ data --- although only for a limited wavelength range.  This comparison is shown in Figures \ref{fig_GNNNFITS1}--\ref{fig_GNNNFITS4}, for the four G160M and G200M observations, and in Figure \ref{fig_ECHBFITS}, for the three high-resolution ECH-B observations.  For clarity, the medium resolution observations in Figures \ref{fig_GNNNFITS1}--\ref{fig_GNNNFITS4} have been stretched over two panels each.  We computed the best-fit model exactly as prescribed above, using the parameters listed in column 2 of Table \ref{tab_RESULTS}.  We then broadened the model using the appropriate instrumental resolutions and determined best-fit values of the velocity offset (\vrad) and the 5-point ``continuum'' (used to map the model surface fluxes onto the normalized fluxes) for each observation.  The GHRS G160M and G200M spectral resolutions were evaluated from the formulae 
\begin{equation} 
\label{g160M}
(\lambda/\Delta\lambda)_{G160M} = 16.82\lambda - 2986.4    \;\;\;\;\;\;  {\rm and}
\end{equation}
\begin{equation} 
\label{g200M}
(\lambda/\Delta\lambda)_{G200M} = 16.20\lambda - 6940.0    \;\;\;\;\;\; ,
\end{equation}
obtained from Glenn Wahlgren (private communication).  For each observation, the value of $\Delta\lambda$ was evaluated at the central wavelength and adopted for the whole observation.  These FWHM values were 0.070 \AA\ (17 \kms), 0.069 \AA\ (16 \kms), 0.069 \AA\ (16 \kms), and 0.080 \AA\ (13 \kms) for the observations in Figures \ref{fig_GNNNFITS1}, \ref{fig_GNNNFITS2}, \ref{fig_GNNNFITS3}, and \ref{fig_GNNNFITS4}, respectively.  We adopted a Gaussian with FWHM = 3 \kms\ for the ECH-B resolution, as obtained from the GHRS Instrument Handbook.\footnote{The GHRS Handbook was accessed at http://www.stsci.edu/instrument-news/handbooks/ghrs/GHRS\_1.html.}

Figure 3 reinforces the comments above, based on the \iue\ data, that the quality of the fit is good in general, but better at the longer wavelengths.  Note that the $\log gf$ values of the prominent \ion{C}{1} lines near 1289 \AA\ and 1310 \AA, which all arise from the 2s$^2$2p$^2$ level at 10192.63 cm$^{-1}$, were empirically determined by us from the spectra of Vega and Sirius A (see Paper II).  The values obtained from VALD were clearly inaccurate and produced features much weaker than observed.  These lines (indicated by the open circles in Figure 3) were masked off during the current analysis and so did not influence the determination of the C abundance.  A close inspection of model and observations in Figures \ref{fig_GNNNFITS2} and \ref{fig_GNNNFITS3} suggests that the model spectrum may have been slightly over-broadened relative to the data (see, e.g., the region near the strong \ion{C}{1} lines at 1280 \AA).  Such an effect is not obvious in Figure \ref{fig_GNNNFITS1} (at least partly due to the noise level) or in Figure \ref{fig_GNNNFITS4}.  These impressions are seemingly verified by allowing the fitting process to choose optimal values for $v \sin i$ when fitting the GHRS data.  The results of are 18.0 \kms\ for the G160M data of Figures \ref{fig_GNNNFITS2} and \ref{fig_GNNNFITS3} and 20.2 \kms\ for the G200M data of Figure \ref{fig_GNNNFITS4}.

This issue of line widths can be examined further with the three GHRS/ECH-B spectra shown in Figure \ref{fig_ECHBFITS}.  These data allow the most accurate determination of \vrot, since their spectral resolution (3 \kms) is significantly narrower than the rotational broadening.  The only accommodations to the data were the determination of velocity offsets and 5-point ``continua'' to map the flux levels for each spectrum.  The standard parameters clearly provide very good fits to the observed features (mostly due to \ion{Fe}{2}).  If we allow \vrot\ to be determined by the ECH-B data themselves, a value of 20.7 \kms\ is found, which is in excellent agreement with the recent results of \citet[][i.e., \vrot\ = 20.8 $\pm$ 0.2 \kms]{Hill2010}.  This is reasonably consistent with our assumed value of 21.8 \kms\ for the \iue\ data analysis, and broader than suggested by the GHRS G160M spectra.  It may be that the instrumental broadening adopted for the G160M data are slightly overestimated, although we cannot rule out that this is an artifact resulting from inaccuracies in the atomic line list.  This issue has no bearing on the results of our analysis, which were carried out with the \iue\ data.  Further, the value of \vrot\ suggested by the ECH-B data is comfortably close to the assumed value of 21.8 \kms\ and Table \ref{tab_RESULTS} shows that our results are insensitive to the exact value adopted for \vrot\ (due to the fact that the spectral resolution of the \iue\ data themselves are comparable to \vrot). Note the presence of interstellar contamination arising from the local ISM in some of the line cores in Figure \ref{fig_ECHBFITS} (indicated by the x's).  A discussion of this and a detail of the \ion{Fe}{2} line at 2344 \AA\ can be found in \citet{Lallement1995}.

The very high-resolution of the ECH-B data allows for a potentially detailed study of the UV line profiles, akin to the attention received by optical line profiles. Such studies can involve corrections to line wavelengths or oscillators strengths, or both, and (particularly important in the case of Vega) the effects of its high rotation speed and high inclination angle on the resultant line profiles, particularly in their cores \citep[see, e.g.,][]{Gulliver1994,Yoon2008,Takeda2008}.  Unfortunately (or mercifully), the amount of such data available is far too small to contribute significantly to these efforts and the comparison in Figure \ref{fig_ECHBFITS} suggests that only mild adjustments in line position or strengths would be required to achieve excellent agreement.  The relatively low spectral-resolution of our main dataset --- the \iue\ high-resolution spectra --- and their lower photometric quality limit their usefulness for detailed profile analysis.  However, the ability of the current model to reproduce the degree of detail available in the \iue\ high-resolution data, spanning nearly 2000 \AA\ and including thousands of spectral features, as well as the UV through optical SED, is impressive and lends confidence in the derivation of elemental abundances from the UV spectral synthesis.


\section{The Spectral Energy Distribution} \label{sec_SED}

A more global view of our results is shown in Figure \ref{fig_SEDPLOT2} where we compare the observed SED of Vega (thin curve) with the predictions of our best-fit model (thick curve) over the range 1050 to 6700 \AA\ (thick line).  The observed SED of Vega was taken from the CALSPEC database, file ``alpha\_lyr\_stis\_005.ascii,'' and is discussed by \citet{Bohlin2007}\footnote{The CALSPEC database was accessed at http://archive.stsci.edu/hst/observatory/cdbs/calspec.html.}.  This SED is a composite, based on low-resolution \iue\ SWP data, HST/STIS G230LB and G430L data, and Kurucz atmosphere models.  We modified it in two ways.  First, we replaced the spectrum below 1200 \AA\ (based on a model) with the BEFS data described in \S \ref{sec_DATA}.  Second, we replaced the data in the range 1250--1978 \AA\/ (based on \iue\ and HST/STIS) with the mean low-resolution \iue\ SWP spectrum processed as described in \S \ref{sec_DATA}.  We did this because we noticed a large discrepancy in the flux levels between our processed \iue\ spectrum and the CALSPEC fluxes between 1300 and 1800 \AA), in the sense that the CALSPEC fluxes are larger, in a smooth wavelength-dependent manner and by as much as nearly 9\% near 1500 \AA.  Outside the 1300--1800 \AA\ range the fluxes agree very well.  We have never encountered such a discrepancy when modeling low-resolution \iue\ spectra of early-type stars using data processed by the MF00 algorithms and it made us suspicious of the short wavelength end of the CALSPEC SED. Hence, we made the wholesale replacement of the 1250--1978 \AA\ data. (We note that, at longer wavelengths, \iue\ LWP spectra processed with the MF00 algorithms agree well with the CALSPEC SED.)  Ultimately this modification of the CALSPEC SED has no effect on the conclusions of this study, but its motivation suggests that the far-UV fluxes of Vega have not yet been conclusively established.

\begin{figure}
\figurenum{5}
\epsscale{1.00}
\plotone{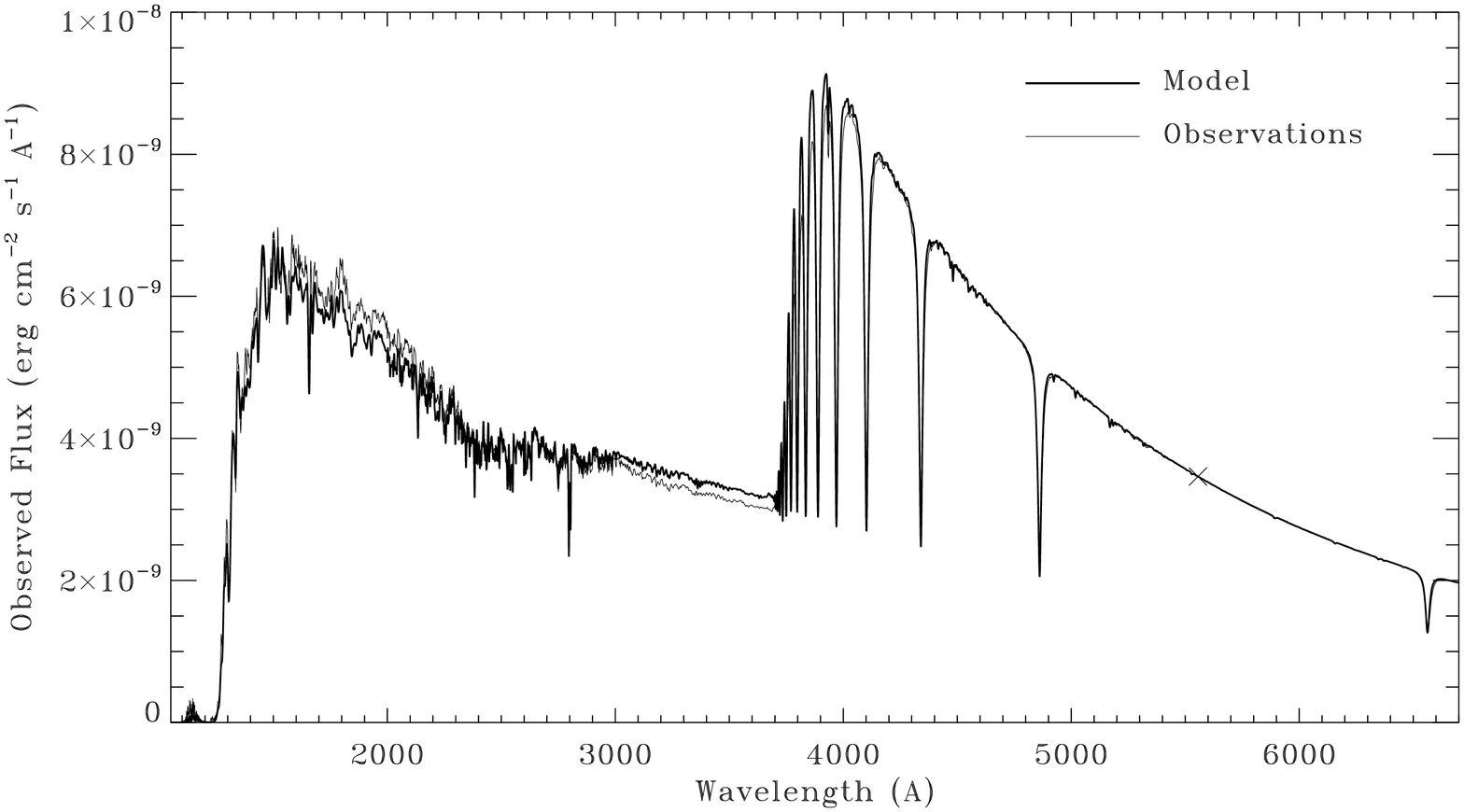}
\caption{A comparison of the observed UV-through-optical SED for Vega (thin curve) and the best-fit model (thick curve).  The model was scaled by a factor of $6.44\times10^{-17}$ (corresponding to an angular diameter of 3.31 mas) to match the observations at 5556 \AA\ (marked with an ``x'').  The observed SED was taken from the CALSPEC database, with some modifications in the far-UV (see \S \ref{sec_SED}).  For this comparison, the model was extended outside the wavelength range over which the fitting procedure operated (1282--3097 \AA) and was broadened to match the spectral resolution of the SED data (typically $\sim$5 \AA\ at $\lambda < 5700$ \AA\ and $\sim$10 \AA\ at $\lambda < 5700$ \AA).
\label{fig_SEDPLOT1}}
\end{figure}

\begin{figure}
\figurenum{6}
\epsscale{1.00}
\plotone{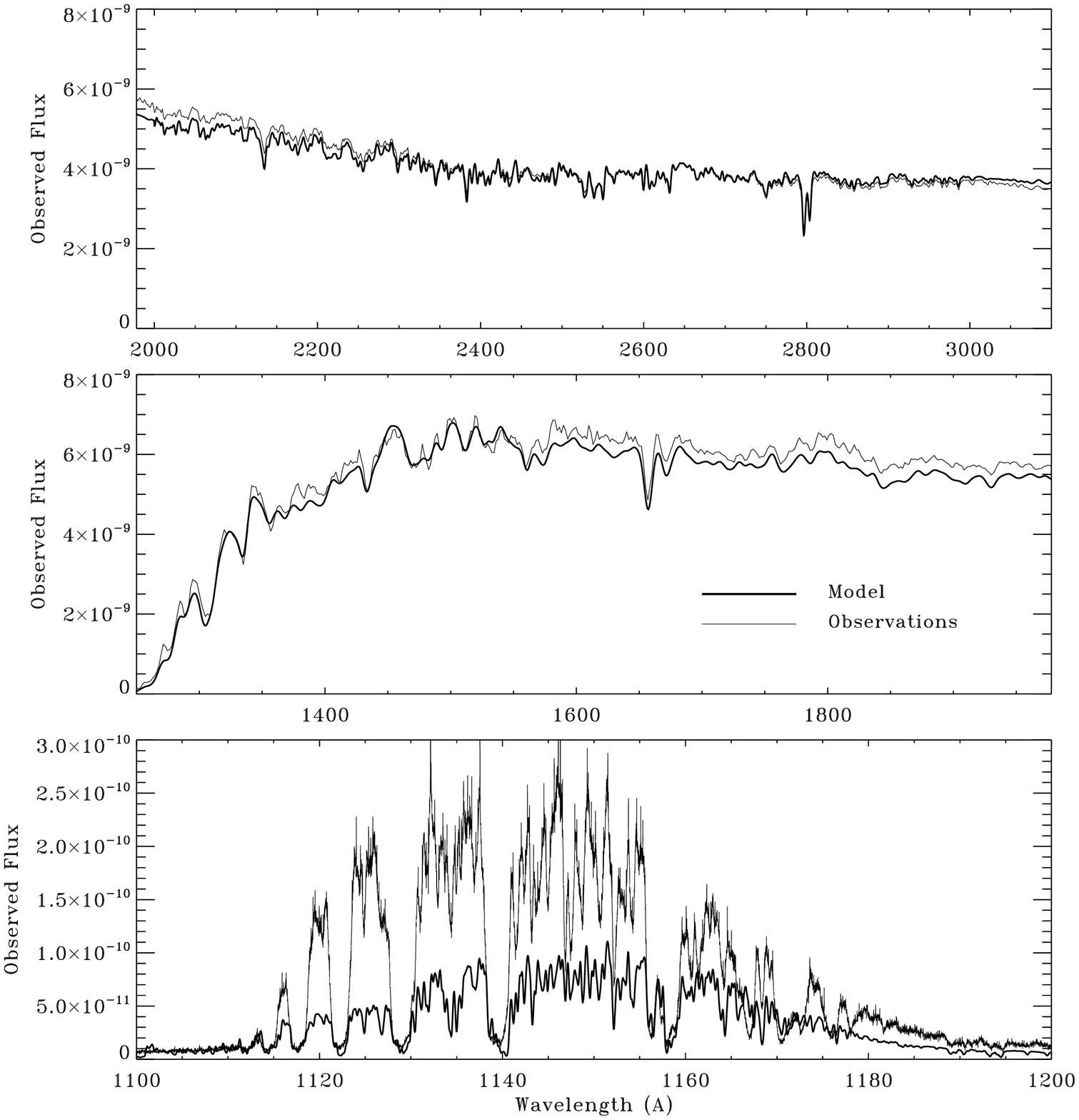}
\caption{A closer look at the observed UV SED for Vega (thin curve) and the best-fit model (thick curve). The data and model are the same as in Figure \ref{fig_SEDPLOT1}; the flux units are ergs cm$^{-2}$ s$^{-1}$ \AA$^{-1}$ and the model was scaled by a factor of $6.44 \times 10^{-17}$ to match the observations at 5556 \AA.  The top two panels show the wavelength region covered by the \iue\ high-resolution data from which the best-fitting model was determined.  The model has been smoothed to a resolution of $\sim$5 \AA\/ in this region to match the observed SED.  The bottom panel shows the small flux island between 1100 and 1200 \AA, which was not included in the analysis.  The strongest features in the observed spectrum, from the BEFS instrument and with a resolution of $\sim$0.23 \AA, are \ion{C}{1} lines approaching the \ion{C}{1} ground-state ionization edge at 1102 \AA.
\label{fig_SEDPLOT2}}
\end{figure}

The model spectrum shown in Figure \ref{fig_SEDPLOT1} was computed using SPECTRUM and the parameters listed in column 2 of Table \ref{tab_RESULTS}, with the spectral range extended outside the fitting region, and then broadened to match the instrumental resolution of the datasets comprising the observed SED.  The resulting surface fluxes were then scaled to match the flux of the observed SED at 5556 \AA, namely $3.46\times10^{-9}$ ergs cm$^{-2}$ \AA$^{-1}$ (marked with an ``x'' on the figure).  The scaling factor of $6.44\times10^{-17}$, which, in the absence of interstellar extinction, is essentially the square of the angular radius, implies an angular diameter for Vega of 3.31 mas.  This is reasonably consistent with measured values of 3.24$\pm$0.07 mas by \citet{HanburyBrown1974}, 3.28$\pm$0.01 mas by \citet{Ciardi2001}, and 3.32 mas by \citet{Aufdenberg2006}, the last of which was based on rapidly rotating models of Vega with a large equator-to-pole temperature range.  As can be seen in Figure \ref{fig_SEDPLOT1}, the model provides an excellent overall representation of the SED, particularly in the relative levels of the optical and UV fluxes.  The major discrepancy is that the Balmer continuum in the model appears less steep than in the observations.  This results in an overestimate of the flux just shortward of the Balmer jump and an underestimate near $\sim$2000 \AA, both by no more than 5--6\%.  The observed and predicted fluxes are consistent between about 1300 and 1500 \AA.  Also, although not obvious in Figure \ref{fig_SEDPLOT1}, the model overestimates the flux between the upper Balmer lines and produces somewhat narrower Balmer lines than seen in the observed SED.  A higher surface gravity, near $\log g = 4.0$, would resolve the Balmer line width discrepancy, but not the Balmer continuum problem.  

In Figure \ref{fig_SEDPLOT2} we zoom in to provide a closer view of the comparison between the model and observed SEDs in the UV.  The Balmer continuum discrepancy is evident in the top panel, which illustrates the wavelengths covered by \iue\/'s LWP camera, although the detailed structure in the model SED matches the observations beautifully.  The middle panel, covering the range of the SWP camera also shows good detailed agreement between model and observations, although the model fluxes tend to run high longward of 1600 \AA.  Recall that the $\lambda < 2000$ \AA\ region is where our processed low-resolution \iue\ SWP spectrum disagrees with the CALSPEC SED and we cannot rule out that some of the differences between model and observed SEDs are due to calibration issues.  Nevertheless, the best-fit model reproduces the UV SED level (as normalized by the flux at 5556 \AA) very well and we conclude that results based on the UV line spectrum and the UV SED are in accord.       

The bottom panel in Figure \ref{fig_SEDPLOT2} shows the small island of flux below Lyman-$\alpha$.  This region was not included in the fitting procedure and it is clear that the model fails to reproduce the observed flux level as revealed by the BEFS data.  The strong absorption features observed in the model and observed spectra are ground state \ion{C}{1} lines.  The discrepancy cannot be remedied simply by invoking a larger value of \teff, since this would require a temperature increase of much more than 100 K (which is clearly inconsistent with the rest of the spectrum) and would not match the shape of the BEFS energy distribution.  The shape of the emergent flux distribution in this region is modulated by the Lyman-$\alpha$ line on the red side and the confluence of strong \ion{C}{1} lines on the blue, culminating in the ground state \ion{C}{1} ionization edge at 1102 \AA\ \citep[see, e.g.,][]{Lemke1996}.  In addition, the region is sensitive to the temperature structure of the outer atmosphere and shows weak emission features of \ion{Fe}{2} and \ion{Cr}{2} that cannot be accounted for by LTE calculations \citep[see, e.g.,][]{vanNoort1998}.  The NLTE calculations of \citeauthor{vanNoort1998} are also able to account for the non-zero flux in the Lyman-$\alpha$ line core, as seen in Figure \ref{fig_GNNNFITS1}.  Even in the absence of Vega's own particular peculiarities, it is perhaps not surprising that our LTE model predictions breakdown in this spectral region.  

We conclude that our best-fitting atmosphere model for Vega --- determined exclusively from fitting the absorption line spectrum in the UV from 1282--3097 \AA\ and without reference to the shape or level of the observed SED --- produces excellent agreement with the emergent flux spectrum over the range $\sim$1200--7000 \AA.  The chief discrepancies noted are the slope of the Balmer continuum, the overestimate of flux between the higher order Balmer lines, and the underestimate of the flux in the 1100--1200 \AA\ spectral island.   It is not immediately clear if these discrepancies can be attributed to the general physical and computational complexities inherent in modeling these spectral regions --- and thus would limit our ability to model the spectra of any A-type star --- or if they represent a particular problem affecting Vega due to its own physical complexity. 

\section{The Composition of Vega's Atmosphere} \label{sec_COMPOSITION}

Given the beautiful agreement between the observed and computed spectra for Vega, and the consistency between spectral line and SED diagnostics, we proceed with some confidence to examine the composition of Vega's atmosphere are revealed by our best-fit model.  The uncertainties listed in Table \ref{tab_RESULTS} for the 17 elements with individually measured abundances reflect only --- as noted earlier for the other parameters --- the statistical uncertainties in the measurement technique.  Additional uncertainty comes from some of the input assumptions in the analysis and we have attempted to gauge the impact of these by running the other fits listed in Table \ref{tab_RESULTS}.  In FITS B and C, we changed the assumed abundance of the Kurucz atmospheric structure model (and of all elements not determined individually) by  $\pm$0.5 dex, respectively and in FITS D and E we changed the assumed value of \vrot\ by $\pm$3.0 \kms, respectively.  To further investigate the stability of the abundances, we also forced changes of $\sim\pm$0.2 dex in the best-fit value of the surface gravity \logg\ (FITS F and G) and about $\pm$0.5 \kms\ in the best-fit value of the microturbulence velocity \vturb\ (FITS H and I).  These latter two sets of tests are rather heavy-handed in that they deliberately skew the fit away from the optimal values, but nevertheless provide some insight into its robustness.  

All in all, as a detailed look at Table \ref{tab_RESULTS} will reveal, the abundance determinations are remarkably stable.  Particularly note the virtual insensitivity to the choice of \vrot\ and the small variations induced by a factor-of-10 change in the composition of the ATLAS9 structure model.  This latter point is important since the scaled solar composition of the ATLAS9 models used for our analysis does not match the composition determined from the UV spectral synthesis.  We used the [m/H] = $-0.5$ models since they are the closest ATLAS9 models to the overall composition inferred from our results, but the detailed element-by-element composition profiles are different.  An obvious follow-up to our study would be to redo the fitting procedure using a grid of atmosphere models computed with a more Vega-like composition, e.g., using the ATLAS12 program \citep{Kurucz1993,Castelli1994}.  However, the insensitivity exhibited by our analysis to large changes in [m/H] suggests that this would have little effect on our results, consistent with the findings of \citet{Castelli1994}.  The largest variations in the abundances (of some elements) occur, not surprisingly, when the microturbulence velocity is disturbed from its best fit-value.  The particular elements most affected should be those whose most important lines (from the point of view of the abundance determination) lie of the ``flat part'' of the curve of growth and are thus most sensitive to changes in \vturb.  Stronger and weaker lines are not affected.  Table \ref{tab_RESULTS} reveals that the elements S, Fe, Ni, and Zn are most affected by {\it ad hoc} modifications in \vturb, which indicates that they are most influential in determining its value.  Considering all the tests in Table \ref{tab_RESULTS}, it is not unreasonable to consider the abundance determinations to be generally stable against the assumptions in the analysis to a level of $\sim\pm0.05-0.07$.

Other fundamental sources of uncertainty in the abundance results, not accounted for in Table \ref{tab_RESULTS}, arise from concerns about the available atomic data and the input physics used in the models.  It is in this regime that the benefits of our approach are most obvious.  Namely, the use of observations in the UV spectral region and the adoption of the latest atomic transition data mean that, for a given element, the analysis generally utilizes spectral features from low-lying atomic levels, includes dominant ionization stages and sometimes more than one ionization stage, and potentially utilizes many different lines arising from a number of different energy levels.  These attributes reduce the sensitivity of the results to: 1) NLTE effects (which are generally reduced for low-lying levels, see below), 2) complications due to the rotational broadening formulation \citep{Yoon2008}, and 3) uncertainties in the atomic parameters for specific transitions.

\begin{figure}
\figurenum{7}
\epsscale{1.00}
\plotone{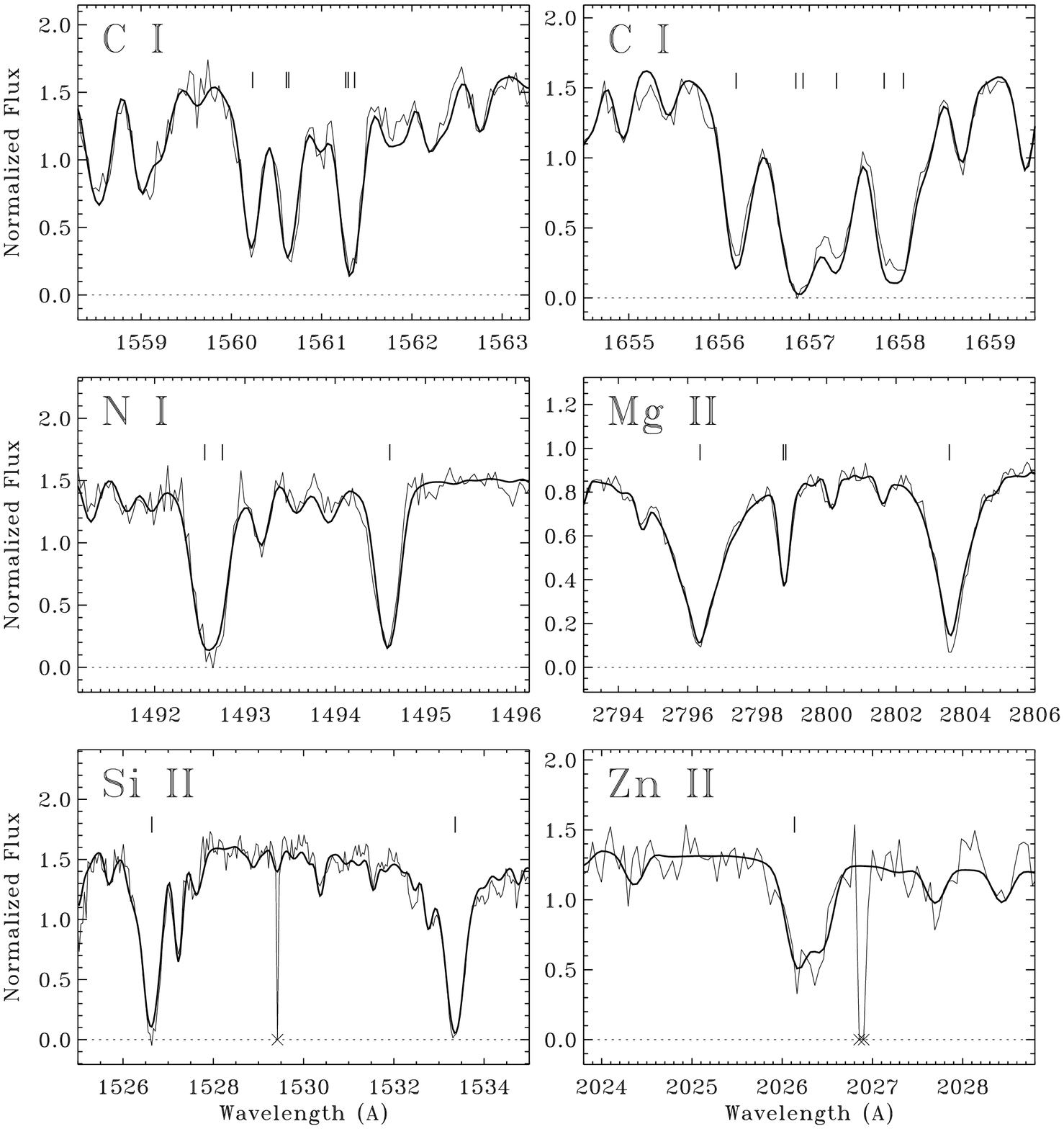}
\caption{A detailed view of the high-dispersion \iue\ spectrum (thin curves) and the best-fit model (thick curves) in the vicinity of selected stellar absorption lines.  The featured chemical species are identified in the upper left of each panel and the location(s) of the line(s) of that species indicated by the vertical tick marks.  Data points excluded from the analysis due to detector artifacts are identified with an ``x,'' as near 2027 \AA\ in \ion{Zn}{2} panel. These lines are discussed in \S \ref{sec_COMPOSITION}.
\label{fig_DETAILS}}
\end{figure}

Before examining the composition of Vega's atmosphere, we first comment on the abundance determinations for several specific elements:

\begin{itemize}
\item \underline{CARBON} The abundance of C is particularly stable and is based on numerous \ion{C}{1} lines and two strong \ion{C}{2} lines near 1335 \AA\ (see Figure \ref{fig_GNNNFITS3}).  The \ion{C}{1} and \ion{C}{2} results are compatible, i.e., excluding the \ion{C}{2} lines from the analysis produces nearly no change in the derived C abundance.  When first using the original $\log gf$ values obtained from VALD, we saw that \ion{C}{1} lines from the 10192.63 cm$^{-1}$ level produced large discrepancies between the model and the observations.  NLTE effects are not expected to be strong for this low-lying level \citep{Stuerenburg1990} and poorly-determined atomic parameters seemed a likely source for the discrepancy, which varies in severity among the different lines.  Lines from this level were excluded from the analysis by setting the data weights to zero in their vicinity and empirical estimates of their $\log gf$ values were made, as detailed in Paper II.  The fits shown in Figures \ref{fig_IUEFIT1}, \ref{fig_GNNNFITS2}, and \ref{fig_GNNNFITS3} were performed with these empirically-determined values although, again, these features did not contribute to the determination of the C abundance.  The ground state \ion{C}{1} lines (i.e., from the 0, 16.4, and 43.4 cm$^{-1}$ levels) are well-fit by our model, as has been seen in Figures \ref{fig_GNNNFITS2} and \ref{fig_GNNNFITS3} and also in the top panels of Figure \ref{fig_DETAILS} where we show closeups of the 1560 \AA\ and 1657 \AA\ complexes.  These panels can be compared with Figure 9 of \citet{GarciaGil2005}.  Our results suggest that the difficulty in fitting those lines encountered by \citet{GarciaGil2005} was due to the adoption of a too-high C abundance, and was not an NLTE effect.  



\item \underline{NITROGEN} The N abundance is also very stable, as can be seen in Table \ref{tab_RESULTS}.  The results are based on a number of strong \ion{N}{1} lines from the $2s^22p^3$ levels at 19224, 19233, 22839 cm$^{-1}$.  The good fit to the strongest \ion{N}{1} lines, at 1492.625, 1492.820, and 1494.675 \AA, can be seen in Figure \ref{fig_DETAILS}.  The resultant abundance is also compatible with --- although was not based on --- the resonance triplet near 1200 \AA, as can be seen in Figure \ref{fig_GNNNFITS1}.  NLTE effects are not expected to be significant for the low-lying levels used here \citep{RentzschHolm1996}.

\item \underline{OXYGEN} The O abundance is strongly influenced by the very strong lines at 1302.168, 1304.858, and 1306.029 \AA, arising from levels at 0, 158, and 227 cm$^{-2}$, respectively.  These lines can be seen in Figure \ref{fig_GNNNFITS2}.  The latter two lines are blended with strong \ion{Si}{2} features at 1304.370 and 1305.592 \AA.   Much weaker O I lines at 1355.598 and 1358.512 \AA\/ are coincident with \ion{C}{1} transitions from the 10192.63 cm$^{-1}$ level and are masked from the analysis.

\item \underline{MAGNESIUM} The Mg abundance is based on a large number of \ion{Mg}{1} and \ion{Mg}{2} lines.  The most prominent Mg features are \ion{Mg}{2} 2797.929 and 2799.998 \AA\ and the \ion{Mg}{2} $h$ and $k$ resonance lines at 2795.528 and 2802.705 \AA. The region surrounding these features is shown in Figure \ref{fig_DETAILS}.  The broadening of the $h$ and $k$ lines was handled by SPECTRUM using the Anstee-O'Mara theory of line broadening by collisions with neutral H atoms, with coefficients from \citet{Barklem1998}.

\item \underline{SILICON}  The Si abundance is based on numerous lines of \ion{Si}{1} and \ion{Si}{2}, including strong \ion{Si}{2} resonance and fine structure features at 1304.370 and 1309.276 \AA\ (see Figures \ref{fig_GNNNFITS2} and \ref{fig_GNNNFITS3}) and at 1526.707 and 1533.431 \AA\ (see the closeup view in Figure \ref{fig_DETAILS}).  The derived abundance produces good agreement with the strong resonance features near 1195 \AA\ (see Figure \ref{fig_GNNNFITS1}), although this wavelength range was not included in the fit.  Our original line list produced large discrepancies between model and observations for a number of \ion{Si}{1} features, mainly between 1539 and 1563 \AA, arising from low-lying levels.  As in the case of \ion{C}{1}, these levels are not expected to be susceptible to NLTE effects and the varying level of the discrepancy from feature-to-feature suggests deficiencies in the transition strengths. We empirically measured $\log gf$ values for many of these features (see Paper II) and utilized these results in producing the spectra seen in Figure 2 --- although the lines themselves did not contribute to the Si abundance.   

\item \underline{PHOSPHORUS}  The element P stands out in our analysis as showing the largest under-abundance relative to the solar standard (see below).  The abundance is based on a number of weak lines arising from the ground state and low-lying levels of \ion{P}{1} and \ion{P}{2}, all located between 1301 and 1787 \AA.  P has one of the largest statistical uncertainties among the abundance determinations --- reflective of the weakness of the lines and their frequent blending with other features --- but not large enough to explain its apparent depletion relative to other elements.  

\item \underline{ZINC} The element Zn stands out in our analysis as showing the only over-abundance relative to the solar standard (see below).  The abundance is strongly influenced by the two strong \ion{Zn}{2} resonance lines at 2025.483 and 2062.005 \AA.  The stronger of these is shown in closeup in the bottom row of Figure \ref{fig_DETAILS}.  Both features are located in noisy regions of the spectrum and are blended with nearby lines, resulting in the largest statistical uncertainty among the abundance measurements.  There is no reason to question the accuracy of the oscillator strengths for the these transitions since interstellar medium studies utilizing these lines typically find the Zn abundance to be comparable to the solar standard \citep[e.g.,][]{Jenkins2009}.
\end{itemize}

\begin{figure}
\figurenum{8}
\epsscale{1.00}
\plotone{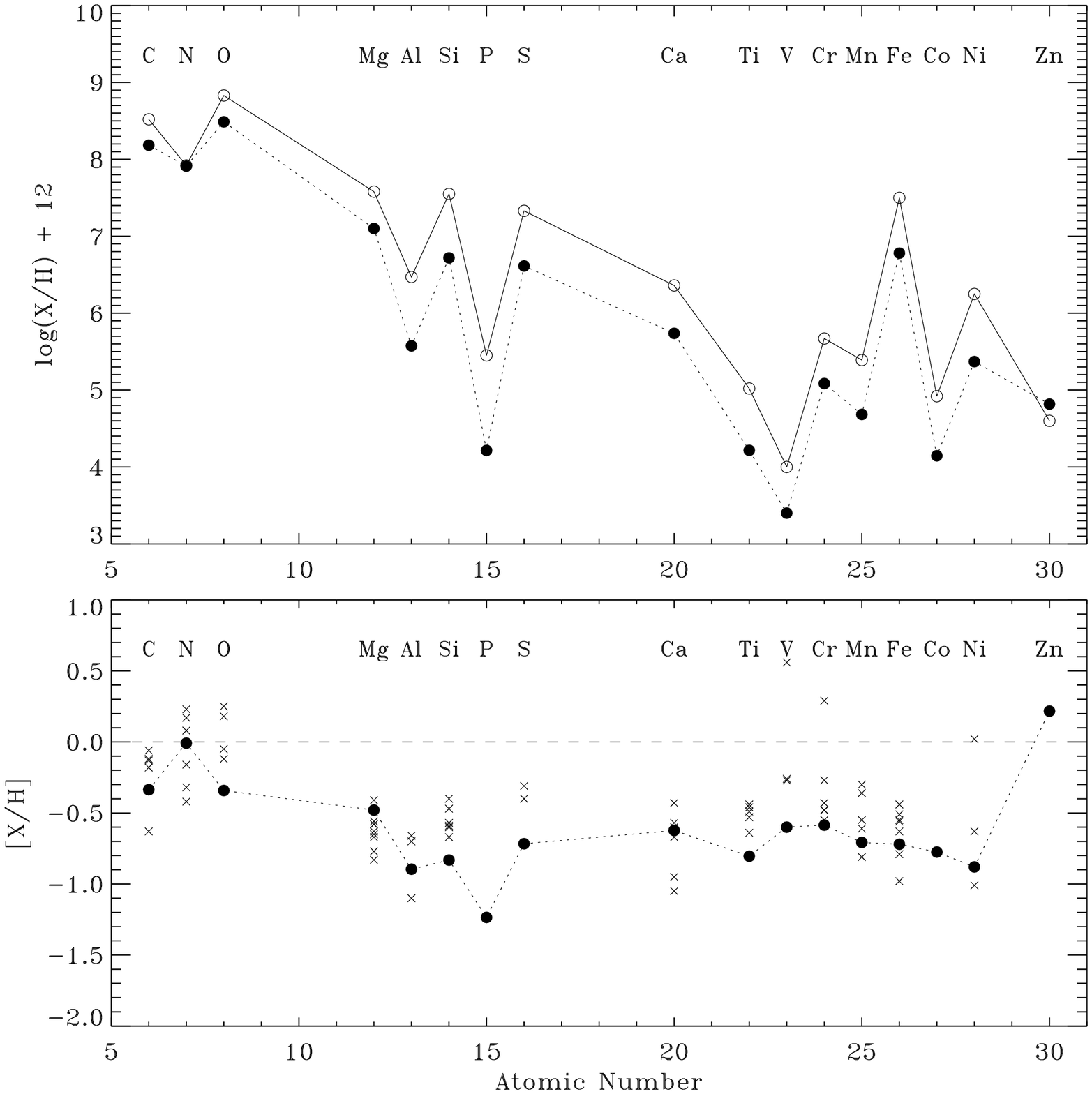}
\caption{The surface abundances of Vega's atmosphere.  The top panel shows the abundances (relative to H) derived for 17 elements from our analysis of the high-dispersion \iue\ spectrum (filled circles).  The solar composition of \citet{Grevesse1998} is shown for comparison (open circles).  The lower panel shows the ``metallicity'' for our 17 elements, relative to the \citet{Grevesse1998} scale (filled circles)  The small x's in the lower panel show results from other investigations of Vega's composition, including \citet{Ferrero1983}, \citet{Gigas1986}, \citet{Gigas1988}, \citet{Stuerenburg1990}, \citet{Adelman1990}, \citet{Lemke1996}, \citet{RentzschHolm1996}, \citet{Ilijic1998}, \citet{Wedemeyer2001}, \citet{Qiu2001}, \citet{Przybilla2001}, \citet{Saffe2004}, \citet{GarciaGil2005}, and \citet{Yoon2008}.   
\label{fig_ABUND}}
\end{figure}

Given the above background, we present in Figure \ref{fig_ABUND} the composition results from our analysis of Vega's UV spectrum.  The top panel shows the abundances (filled circles) in the standard form relative to the abundance of H, i.e., as listed in Table \ref{tab_RESULTS}.  Open circles show the \citet[][hereafter GS98]{Grevesse1998} solar composition for comparison.  The abundance values are plotted against atomic number and the elements are identified along the top of the figure.  The bottom panel of Figure \ref{fig_ABUND} plots the difference between the Vega and the GS98 solar composition in the form ${\rm [X/H]} \equiv \log{\rm(N_X/N_H)}_{Vega} - \log{\rm(N_X/N_H)}_{GS98}$.  The small x's in the panel show the results from other investigations of Vega's atmospheric composition, and include results from \citet{Ferrero1983}, \citet{Gigas1986}, \citet{Gigas1988}, \citet{Stuerenburg1990}, \citet{Adelman1990}, \citet{Lemke1996}, \citet{RentzschHolm1996}, \citet{Ilijic1998}, \citet{Wedemeyer2001}, \citet{Qiu2001}, \citet{Przybilla2001}, \citet{Saffe2004}, \citet{GarciaGil2005}, and \citet{Yoon2008}.  

Our results are not sufficiently different from the generally accepted pattern of abundances to warrant a detailed discussion of Vega's composition, but a number of significant points can be made:

\begin{enumerate}
\item Figure \ref{fig_ABUND} shows that the \iue\ high-resolution data and our analysis technique allow access to an important range of elements, including CNO, the light metals (e.g, Mg, Si, etc.) and the Fe-group.

\item The established abundance pattern of Vega (i.e., significant depletion of the elements heavier than Mg and more solar-like CNO), which lead \citet{Baschek1988} to note its resemblance to the $\lambda$ Boo stars, is evident in our results.

\item The depletions of Mg and heavier elements show good element-to-element coherence, with the exceptions of P and Zn, as discussed above.  The mean depletion of these elements (excluding P and Zn) is [X/H] $= -0.72 \pm 0.13$ (s.d.), i.e., a factor of five subsolar as compared to the GS98 solar scale.  The scatter is larger than can be accounted for by the uncertainties in the analysis.

\item The slightly super-solar abundance of Zn is not inconsistent with the hypothesis that the $\lambda$ Boo (and Vega) abundance phenomenon arises from the accretion of metal-deficient circumstellar gas, with the deficiency arising from the condensation of some elements onto dust grains \citep{Venn1990}.  In the interstellar medium, Zn is seen to resist incorporation into grains and generally maintains a solar-level abundance \citep[e.g.,][]{Jenkins2009}.

\item The enhancement of N relative to C and O is consistent with the hypothesis that processed material may have been mixed to the surface, as a result of Vega's rapid rotation. \citep[See the discussion in][]{Yoon2008}.  On the other hand, the depletion characteristics of N in the ISM differ from those of C and O \citep{Jenkins2009} and accretion of depleted circumstellar gas might produce such a pattern.  Both processes may be occurring simultaneously.

\end{enumerate}

\section{FINAL COMMENTS} \label{sec_SUMMARY}
We have shown that the emergent UV spectrum of Vega, as recorded by the \iue\ satellite in both high-dispersion and low-dispersion, can be fit remarkably well --- both in detail and in its broad properties --- by a single-temperature synthetic spectrum based on the LTE, plane-parallel ATLAS9 models of Kurucz and the LTE spectral synthesis program SPECTRUM of Gray, with an updated atomic line list.  The chief discrepancies between the model and the observations --- i.e., those which cannot be eliminated by adjustments in the model parameters --- are an underestimate of the slope of the Balmer continuum, producing a model flux excess near 3500 \AA\ and a deficit near 2000 \AA\ (both by up to 5--6\%) and a deficit of model flux in the spectral ``island'' between 1100 and 1200 \AA.  Since few A-type stars have UV-through-optical data approaching the extent and quality of Vega's, it is not clear whether these deficiencies are Vega-related, and result from its high rotational velocity and subsequently large gravity-darkening, or are general deficiencies which would plague modeling attempts of any A-type star.  Studies of additional, more normal, stars are needed to address this issue.  At the detailed level, small discrepancies at discrete wavelengths point to incompleteness and to deficiencies in the atomic line list used to generate the model.  The most notable discrepancies involve transitions from the 10193.63 cm$^{-1}$ level of \ion{C}{1} and ground-state and fine-structure levels of \ion{Si}{1}.  In conjunction with a concurrent study of the UV spectrum of Sirius A, we have produced empirical estimates of the $\log gf$ values of these lines (see Paper II). Attempting to improve the quality of the UV transition data will be an ongoing project.

If Vega were a normal, equator-on, slow-rotating star, then its emergent spectrum and our analysis --- which is based strictly on the high-dispersion UV line spectrum, but consistent with the UV-through-optical SED --- would indicate that it has a surface temperature \teff\ $\simeq 9550$ K, surface gravity \logg\ $\simeq 3.7$, a general surface metallicity [m/H] $\simeq -0.5$, and a microturbulence velocity \vturb\ $\simeq 2.0$ \kms.  The Balmer lines, which we examined only cursorily, seem to require a larger \logg\ to explain their widths, as has been found in other studies.  Since Vega is not a normal star but, rather, a rapid rotator, these properties must be treated with caution since the emergent spectrum will be dependent on viewing angle.  Our results must be regarded as representing some sort of average across the observed hemisphere.  In general, the simple spectroscopically-derived properties are not the values that one would adopt in stellar structure calculations to understand Vega's evolutionary state.  


Modeling the complex UV line spectrum of Vega, as revealed by the \iue\ high-dispersion data, has allowed us to determine the specific surface abundances for 17 different chemical elements, including CNO, the light metals, and the iron group elements.  The resultant abundance pattern agrees in general with previous results, although there is considerable scatter in the literature.  The success of the modeling technique, coupled with an {\it a prior} appreciation of the advantages to modeling transitions from low-lying levels, transitions from multiple ionization states of the same element, and, in some cases, hundreds or more transitions per element, suggest that these are the most reliable measurements to date for the respective elements.  

Despite its peculiarities, Vega has turned out to provide a powerful test of the extent of our abilities to model the atmospheric properties of the early A-type stars, particularly the detailed UV line spectrum.  The value of the measurements from this pilot study will increase as this analysis is extended to more objects in the rich high-dispersion \iue\ data archive, including both normal and peculiar objects.    

\begin{acknowledgments}
The author acknowledges support from NASA grant NNX08AJ62G.  Special thanks to Richard Gray of Appalachian State University for making his SPECTRUM program available and for many quick and kind responses to a neophyte's questions. Some of the data presented in this paper were obtained from the Multimission Archive at the Space Telescope Science Institute (MAST). STScI is operated by the Association of Universities for Research in Astronomy, Inc., under NASA contract NAS5-26555. Support for MAST for non-HST data is provided by the NASA Office of Space Science via grant NAG5-7584 and by other grants and contracts. 
\end{acknowledgments}



\clearpage
\begin{deluxetable}{clccc}
\tabletypesize{\scriptsize}
\tablewidth{0pc}
\tablecaption{High-Resolution UV Spectra of Vega\label{tab_DATA}}
\tablehead{
\colhead{Instrument}  &
\colhead{Observation} &
\colhead{Program}     &
\colhead{Wavelength}  &
\colhead{Spectral}    \\
\colhead{}            &
\colhead{ID}          &
\colhead{ID}          &
\colhead{Range}       &
\colhead{Resolution}   }
\startdata
\iue/SWP        & SWP32870      &  PHCAL    & 1150--2000 \AA & 23 \kms \\
\iue/SWP        & SWP32887      &  PHCAL    & 1150--2000 \AA & 23 \kms \\
\iue/SWP        & SWP33633      &  STJRP    & 1150--2000 \AA & 23 \kms \\
\iue/SWP        & SWP38411      &  PHCAL    & 1150--2000 \AA & 23 \kms \\
\iue/SWP        & SWP42521      &  PHCAL    & 1150--2000 \AA & 23 \kms \\
\iue/SWP        & SWP45285      &  PHCAL    & 1150--2000 \AA & 23 \kms \\
\iue/LWP        & LWP07490      &  STHRP    & 1900--3200 \AA & 16 \kms \\
\iue/LWP        & LWP12617      &  PHCAL    & 1900--3200 \AA & 16 \kms \\
\iue/LWP        & LWP17648      &  PHCAL    & 1900--3200 \AA & 16 \kms \\
\iue/LWP        & LWP21295      &  PHCAL    & 1900--3200 \AA & 16 \kms \\
\iue/LWP        & LWP23642      &  PHCAL    & 1900--3200 \AA & 16 \kms \\
\hst/GHRS/G160M & Z3G05208T     &  6828     & 1186--1122 \AA & 16 \kms \\
\hst/GHRS/G160M & Z3G05206T     &  6828     & 1276--1311 \AA & 16 \kms \\
\hst/GHRS/G160M & Z3G05207T     &  6828     & 1305--1340 \AA & 16 \kms \\
\hst/GHRS/G200M & Z2MW0207T     &  5673     & 1840--1879 \AA & 13 \kms \\
\hst/GHRS/ECH-B & Z1210306T     &  2461     & 2590--2603 \AA & 3 \kms \\
\hst/GHRS/ECH-B & Z121030AM     &  2461     & 2339--2352 \AA & 3 \kms  \\
\hst/GHRS/ECH-B & Z121030DM     &  2461     & 2847--2861 \AA & 3 \kms \\
BEFS\tablenotemark{a} & BEFS2094  & \nodata &  900--1220 \AA & $\sim$60 \kms \\
\enddata
\tablenotetext{a}{The Berkeley Extreme and Far-Ultraviolet Spectrometer (BEFS) was flown aboard the ORPHEUS telescope on the {\it ORPHEUS-SPAS} space shuttle missions in 1993 and 1996 \citep[see][]{Dixon2002}}
\end{deluxetable}


\begin{deluxetable}{lrrrrrrrrrc} 
\tabletypesize{\scriptsize}
\tablewidth{0pc} 
\tablecaption{Spectral Synthesis Results for Vega}
\tablehead{ 
\colhead{Property}   &
\colhead{\textbf{FIT A}}      & 
\colhead{FIT B}      & 
\colhead{FIT C}      &
\colhead{FIT D}      & 
\colhead{FIT E}      & 
\colhead{FIT F}      &
\colhead{FIT G}      & 
\colhead{FIT H}      & 
\colhead{FIT I}      &    
\colhead{No. of}     \\
\colhead{}      &
\colhead{}      &     
\colhead{}      & 
\colhead{}      &
\colhead{}      & 
\colhead{}      & 
\colhead{}      &
\colhead{}      & 
\colhead{}      & 
\colhead{}      &
\colhead{Lines\tablenotemark{a}}      }
\startdata
$T_{eff}$ (K) & $\textbf{  9547 $\pm$    17}$ & $  9546$ & $  9546$ & $  9543$ & $  9535$ & $  9448$ & $  9549$ & $  9545$ & $  9541$ & \nodata  \\
$\log g$ & $\textbf{  3.72 $\pm$  0.03}$ & $  3.69$ & $  3.83$ & $  3.71$ & $  3.71$ & $\underline{  3.60}$ & $\underline{  4.00}$ & $  3.73$ & $  3.73$ & \nodata  \\
$[$m/H$]$\tablenotemark{b} & $\underline{\textbf{ -0.50}}$ & $\underline{ -1.00}$ & $\underline{  0.00}$ & $\underline{ -0.50}$ & $\underline{ -0.50}$ & $\underline{ -0.50}$ & 
$\underline{ -0.50}$ & $\underline{ -0.50}$ & $\underline{ -0.50}$ & \nodata  \\
$v_{turb}$ (km s$^{-1}$) & $\textbf{  2.04 $\pm$  0.02}$ & $  2.06$ & $  1.94$ & $  1.80$ & $  2.21$ & $  2.00$ & $  2.00$ & $\underline{  1.50}$ & $\underline{  2.50}$ & \nodata
  \\
$v \sin i$ (km s$^{-1}$) & $\underline{\textbf{ 21.80}}$ & $\underline{ 21.80}$ & $\underline{ 21.80}$ & $\underline{ 18.80}$ & $\underline{ 24.80}$ & $\underline{ 21.80}$ & 
$\underline{ 21.80}$ & $\underline{ 21.80}$ & $\underline{ 21.80}$ & \nodata  \\
$v_{rad}$(SWP)\tablenotemark{c} (km s$^{-1}$) & $\textbf{-16.14 $\pm$  2.46}$ & $-16.10$ & $-16.12$ & $-16.03$ & $-16.18$ & $-16.12$ & $-16.09$ & $-16.16$ & $-16.05$ & \nodata  \\
$v_{rad}$(LWP)\tablenotemark{d} (km s$^{-1}$) & $\textbf{  1.37 $\pm$  1.88}$ & $  1.35$ & $  1.38$ & $  1.45$ & $  1.32$ & $  1.36$ & $  1.36$ & $  1.32$ & $  1.41$ & \nodata  \\
C/H\tablenotemark{e} & $\textbf{  8.18 $\pm$  0.02}$ & $  8.15$ & $  8.22$ & $  8.18$ & $  8.18$ & $  8.14$ & $  8.11$ & $  8.19$ & $  8.16$ & 196  \\
N/H\tablenotemark{e} & $\textbf{  7.91 $\pm$  0.03}$ & $  7.88$ & $  7.95$ & $  7.91$ & $  7.92$ & $  7.88$ & $  7.86$ & $  7.97$ & $  7.83$ & 46  \\
O/H\tablenotemark{e} & $\textbf{  8.48 $\pm$  0.03}$ & $  8.45$ & $  8.52$ & $  8.47$ & $  8.49$ & $  8.44$ & $  8.47$ & $  8.48$ & $  8.49$ & 8  \\
Mg/H\tablenotemark{e} & $\textbf{  7.10 $\pm$  0.02}$ & $  7.07$ & $  7.12$ & $  7.08$ & $  7.09$ & $  7.03$ & $  7.11$ & $  7.10$ & $  7.08$ & 327  \\
Al/H\tablenotemark{e} & $\textbf{  5.57 $\pm$  0.02}$ & $  5.55$ & $  5.61$ & $  5.55$ & $  5.57$ & $  5.52$ & $  5.62$ & $  5.60$ & $  5.54$ & 84  \\
Si/H\tablenotemark{e} & $\textbf{  6.71 $\pm$  0.01}$ & $  6.69$ & $  6.74$ & $  6.70$ & $  6.72$ & $  6.68$ & $  6.68$ & $  6.75$ & $  6.67$ & 388  \\
P/H\tablenotemark{e} & $\textbf{  4.21 $\pm$  0.05}$ & $  4.18$ & $  4.26$ & $  4.19$ & $  4.22$ & $  4.15$ & $  4.18$ & $  4.23$ & $  4.19$ & 78  \\
S/H\tablenotemark{e} & $\textbf{  6.61 $\pm$  0.04}$ & $  6.56$ & $  6.65$ & $  6.60$ & $  6.62$ & $  6.57$ & $  6.51$ & $  6.71$ & $  6.47$ & 75  \\
Ca/H\tablenotemark{e} & $\textbf{  5.73 $\pm$  0.03}$ & $  5.71$ & $  5.78$ & $  5.74$ & $  5.75$ & $  5.71$ & $  5.68$ & $  5.83$ & $  5.67$ & 91  \\
Ti/H\tablenotemark{e} & $\textbf{  4.21 $\pm$  0.02}$ & $  4.18$ & $  4.27$ & $  4.20$ & $  4.23$ & $  4.16$ & $  4.27$ & $  4.27$ & $  4.17$ & 514  \\
V/H\tablenotemark{e} & $\textbf{  3.40 $\pm$  0.03}$ & $  3.37$ & $  3.45$ & $  3.36$ & $  3.42$ & $  3.34$ & $  3.44$ & $  3.42$ & $  3.38$ & 462  \\
Cr/H\tablenotemark{e} & $\textbf{  5.08 $\pm$  0.01}$ & $  5.06$ & $  5.12$ & $  5.06$ & $  5.09$ & $  5.04$ & $  5.13$ & $  5.15$ & $  5.01$ & 2933  \\
Mn/H\tablenotemark{e} & $\textbf{  4.68 $\pm$  0.02}$ & $  4.65$ & $  4.72$ & $  4.65$ & $  4.69$ & $  4.64$ & $  4.73$ & $  4.73$ & $  4.63$ & 1690  \\
Fe/H\tablenotemark{e} & $\textbf{  6.78 $\pm$  0.01}$ & $  6.76$ & $  6.82$ & $  6.77$ & $  6.78$ & $  6.74$ & $  6.83$ & $  6.91$ & $  6.65$ & 7416  \\
Co/H\tablenotemark{e} & $\textbf{  4.14 $\pm$  0.03}$ & $  4.11$ & $  4.19$ & $  4.13$ & $  4.14$ & $  4.09$ & $  4.19$ & $  4.21$ & $  4.08$ & 637  \\
Ni/H\tablenotemark{e} & $\textbf{  5.37 $\pm$  0.02}$ & $  5.36$ & $  5.44$ & $  5.37$ & $  5.38$ & $  5.34$ & $  5.44$ & $  5.53$ & $  5.25$ & 738  \\
Zn/H\tablenotemark{e} & $\textbf{  4.81 $\pm$  0.10}$ & $  4.80$ & $  4.87$ & $  4.83$ & $  4.82$ & $  4.78$ & $  4.86$ & $  4.93$ & $  4.72$ & 12  \\
\enddata
\tablecomments{The bold-faced model in the second column (FIT A) is the ``best-fit' model discussed in the text.  The other fits are described in \S \ref{sec_BESTFITMODEL} and \S \ref{sec_COMPOSITION}. The underlined parameters were held fixed at the indicated values in the various analyses;  all others were determined by the $\chi^2$ minimization procedure.}
\tablenotetext{a}{The approximate number of spectral features contributing to the abundance measurement for each element.}
\tablenotetext{b}{The quantity [X/H] refers to the general metallicity of the Kurucz \atlas\ models on which the spectral synthesis calculations were based.}
\tablenotetext{c}{The quantity $v_{rad}$(SWP) refers to the mean value for the velocity offset between the model spectrum and the 38 \iue\ SWP orders.  The uncertainty listed is the standard deviation of the sample.}
\tablenotetext{d}{The quantity $v_{rad}$(LWP) refers to the mean value for the velocity offset between the model spectrum and the 42 \iue\ LWP orders.  The uncertainty listed is the standard deviation of the sample.}
\tablenotetext{e}{The individual elemental abundances determined from the spectral synthesis analyses expressed in the form ${\rm X/H} \equiv \log$(N$_{\rm X}$/N$_{\rm H}$) + 12.0, where N$_{\rm X}$/N$_{\rm H}$ is the number abundance of element X relative to that of hydrogen.} 
\label{tab_RESULTS}
\end{deluxetable}


\renewcommand{\baselinestretch}{1}

\begin{figure}
\figurenum{Online 2.1}
\epsscale{1.00}
\plotone{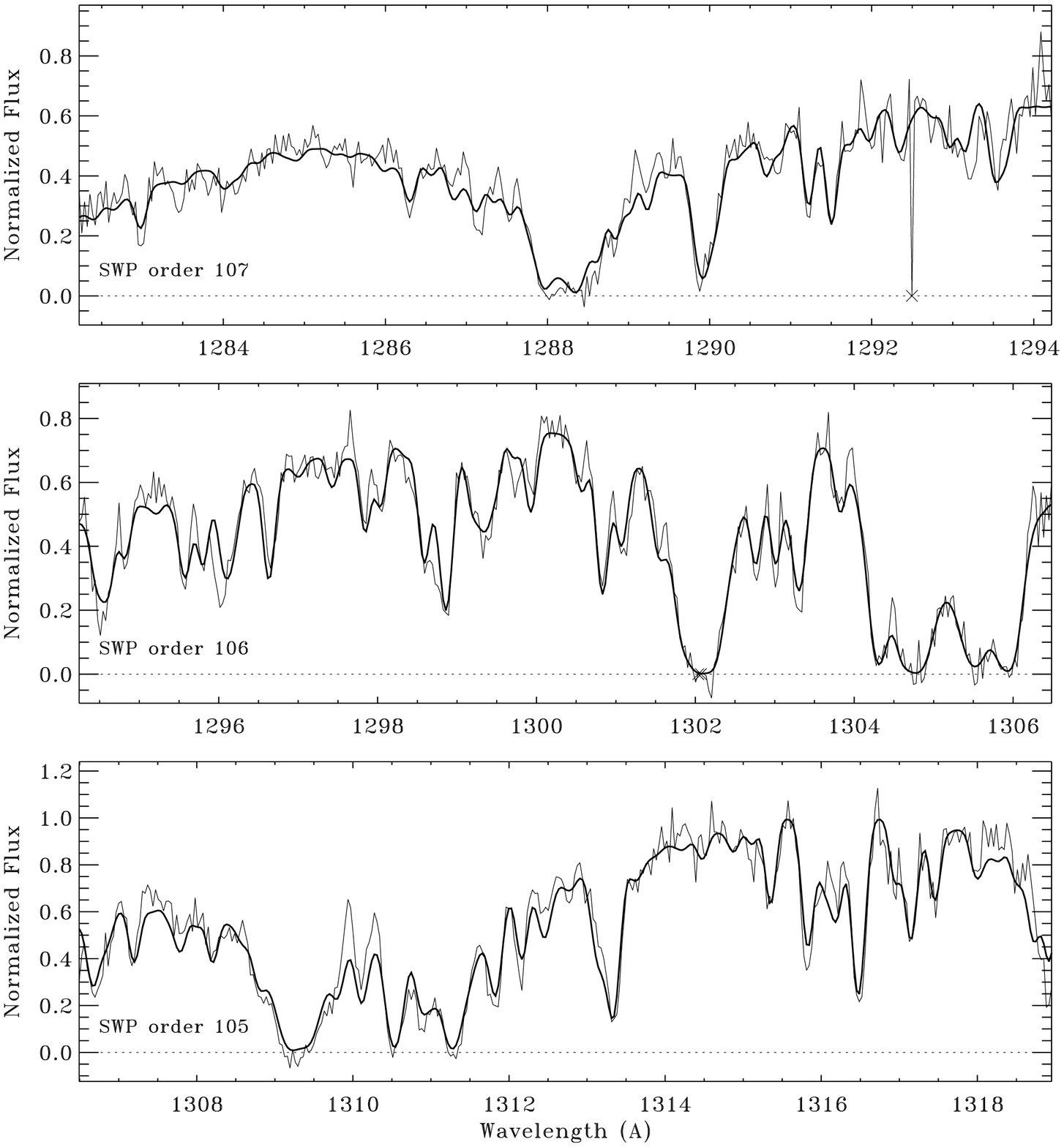}
\caption{\footnotesize A detailed view of the model fit to Vega's high-resolution \iue\ spectrum, for SWP echelle orders 107, 106, and 105, covering the wavelength ranges 1282.22--1294.22 \AA, 1294.24--1306.49 \AA, and 1306.49--1318.95 \AA, respectively.  The model and the data are indicated by the thick and thin curves, respectively.  Data points excluded from the analysis due to detector artifacts are identified with an ``x,'' as at 1292.5 \AA.  An ascii file which identifies the various absorption features can be obtained from the author. 
\label{fig_ONLINE1}}
\end{figure}

\begin{figure}
\figurenum{Online 2.2}
\epsscale{1.00}
\plotone{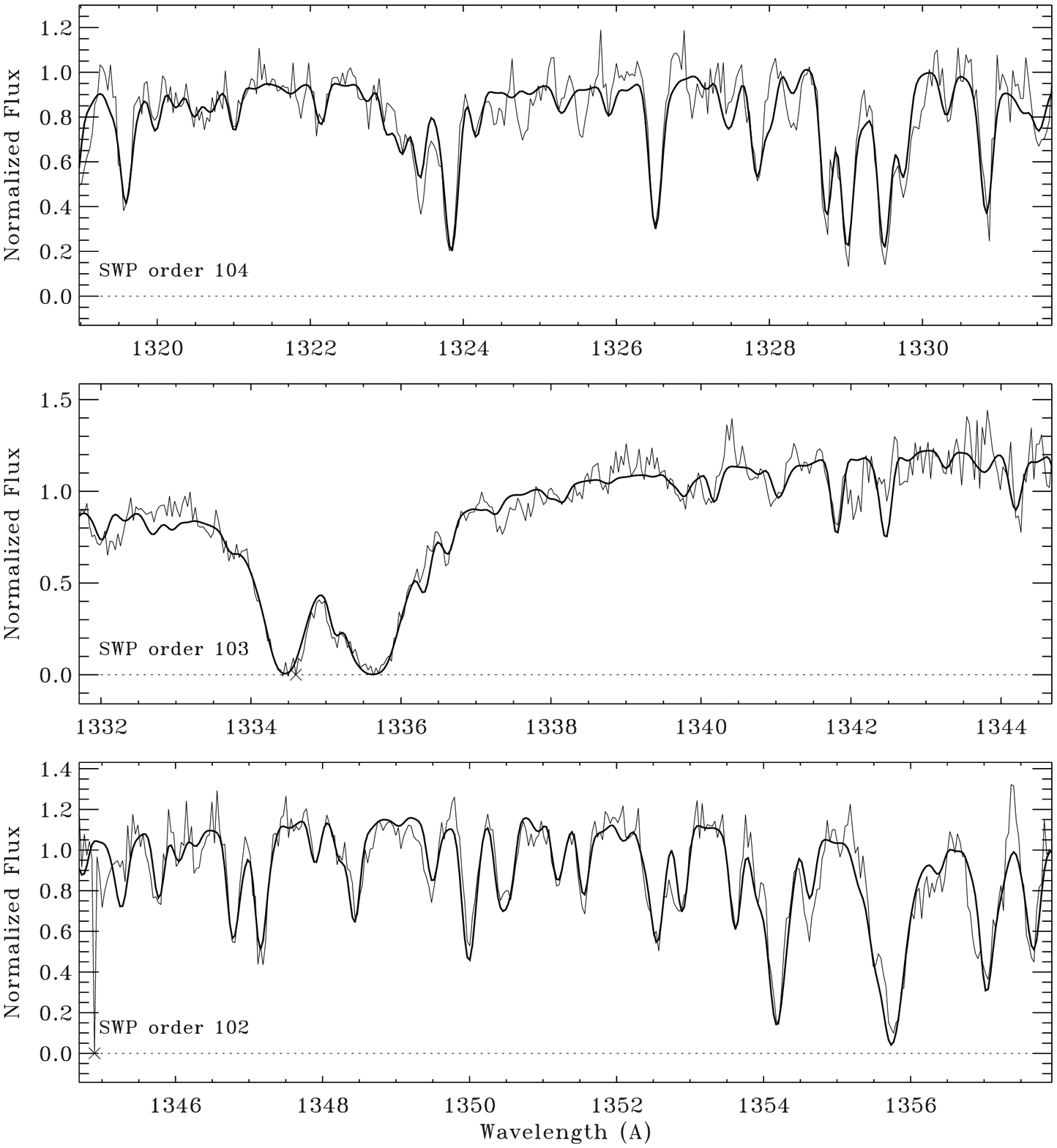}
\caption{Same as Figure \ref{fig_ONLINE1}, but for an additional set of high-dispersion \iue\ echelle orders, identified in the lower left of each panel.}
\end{figure}

\begin{figure}
\figurenum{Online 2.3}
\epsscale{1.00}
\plotone{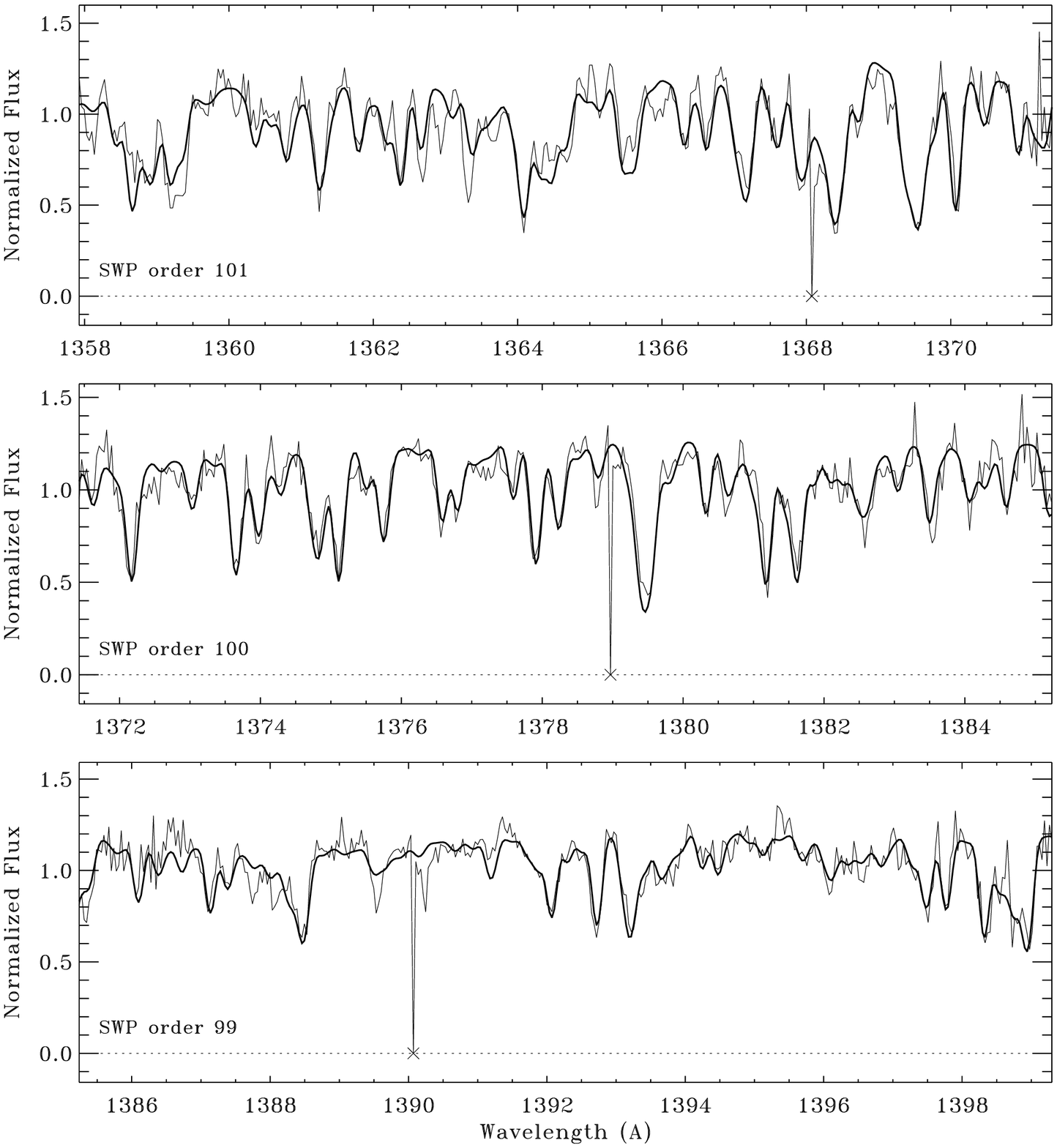}
\caption{Same as Figure \ref{fig_ONLINE1}, but for an additional set of high-dispersion \iue\ echelle orders, identified in the lower left of each panel.}
\end{figure}

\clearpage

\begin{figure}
\figurenum{Online 2.4}
\epsscale{1.00}
\plotone{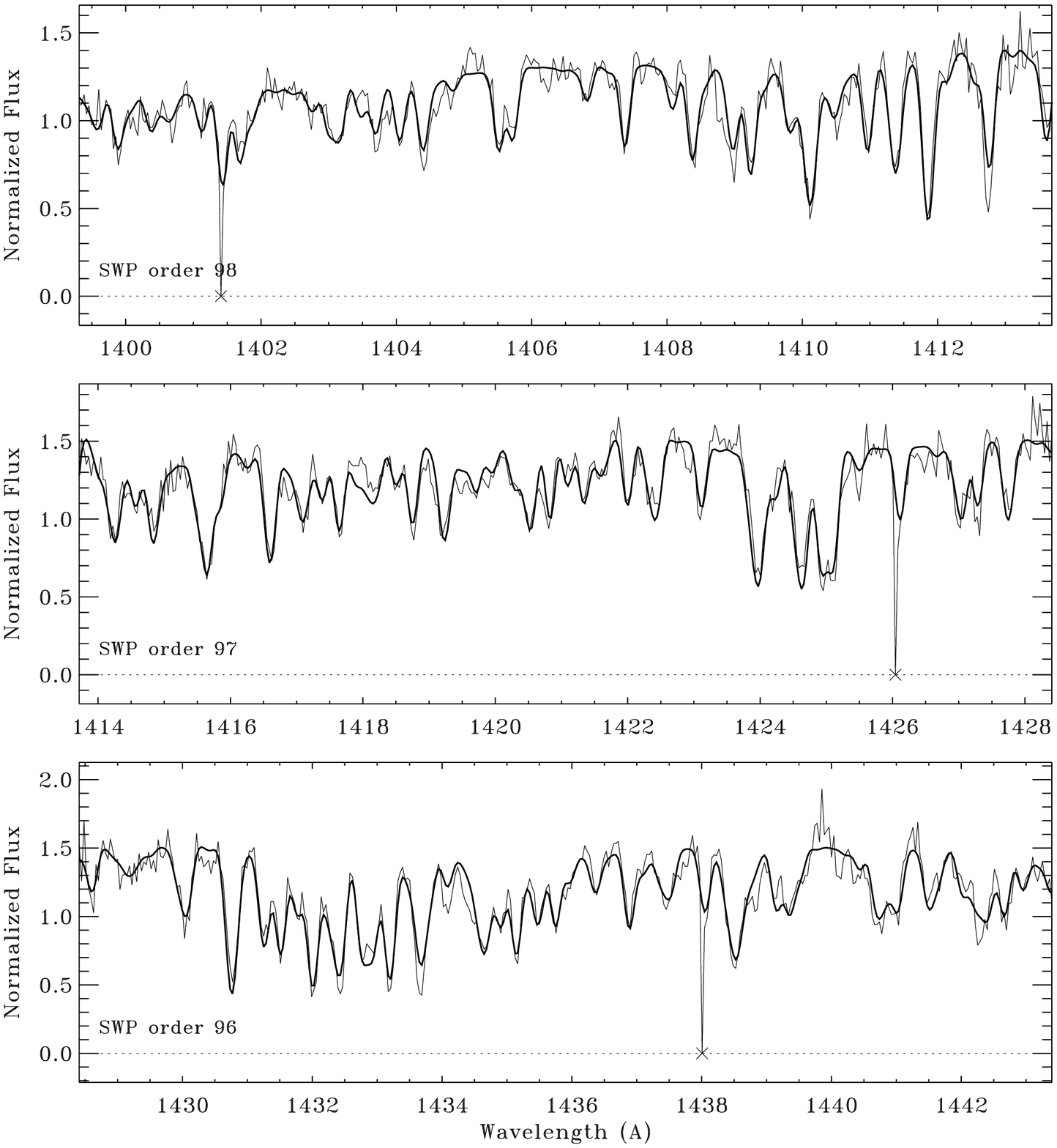}
\caption{Same as Figure \ref{fig_ONLINE1}, but for an additional set of high-dispersion \iue\ echelle orders, identified in the lower left of each panel.}
\end{figure}

\begin{figure}
\figurenum{Online 2.5}
\epsscale{1.00}
\plotone{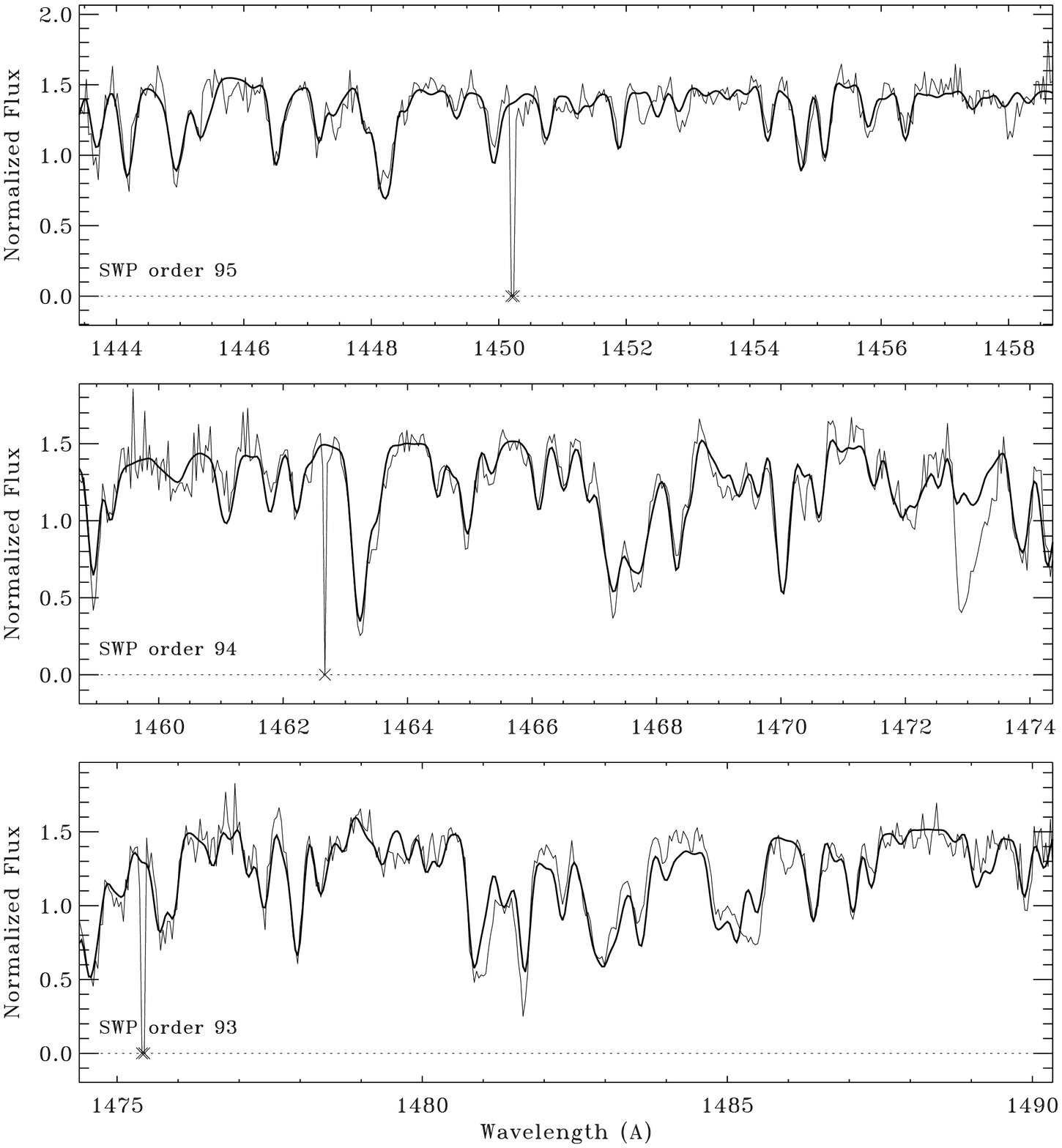}
\caption{Same as Figure \ref{fig_ONLINE1}, but for an additional set of high-dispersion \iue\ echelle orders, identified in the lower left of each panel.}
\end{figure}

\begin{figure}
\figurenum{Online 2.6}
\epsscale{1.00}
\plotone{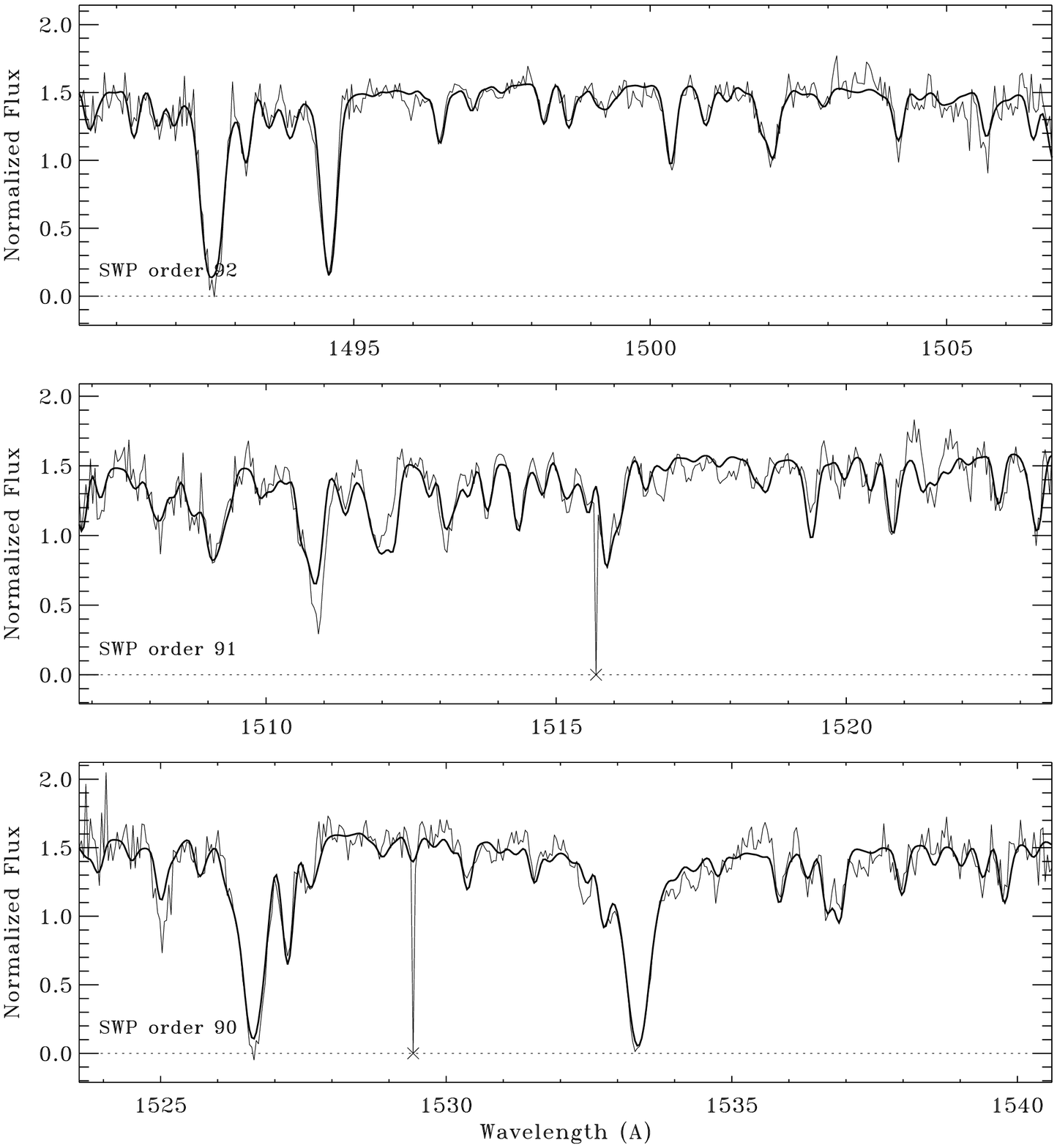}
\caption{Same as Figure \ref{fig_ONLINE1}, but for an additional set of high-dispersion \iue\ echelle orders, identified in the lower left of each panel.}
\end{figure}

\clearpage

\begin{figure}
\figurenum{Online 2.7}
\epsscale{1.00}
\plotone{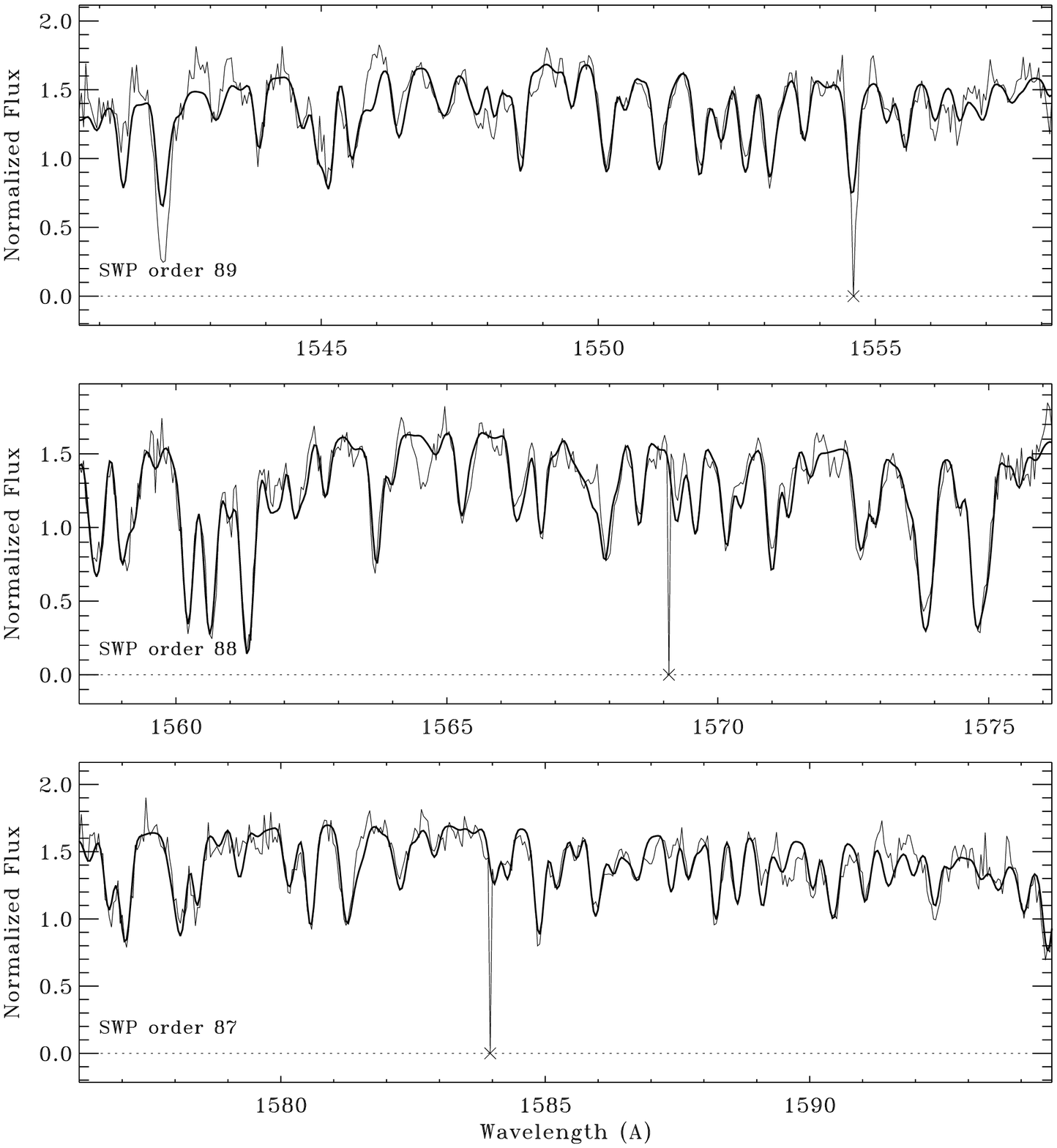}
\caption{Same as Figure \ref{fig_ONLINE1}, but for an additional set of high-dispersion \iue\ echelle orders, identified in the lower left of each panel.}
\end{figure}

\begin{figure}
\figurenum{Online 2.8}
\epsscale{1.00}
\plotone{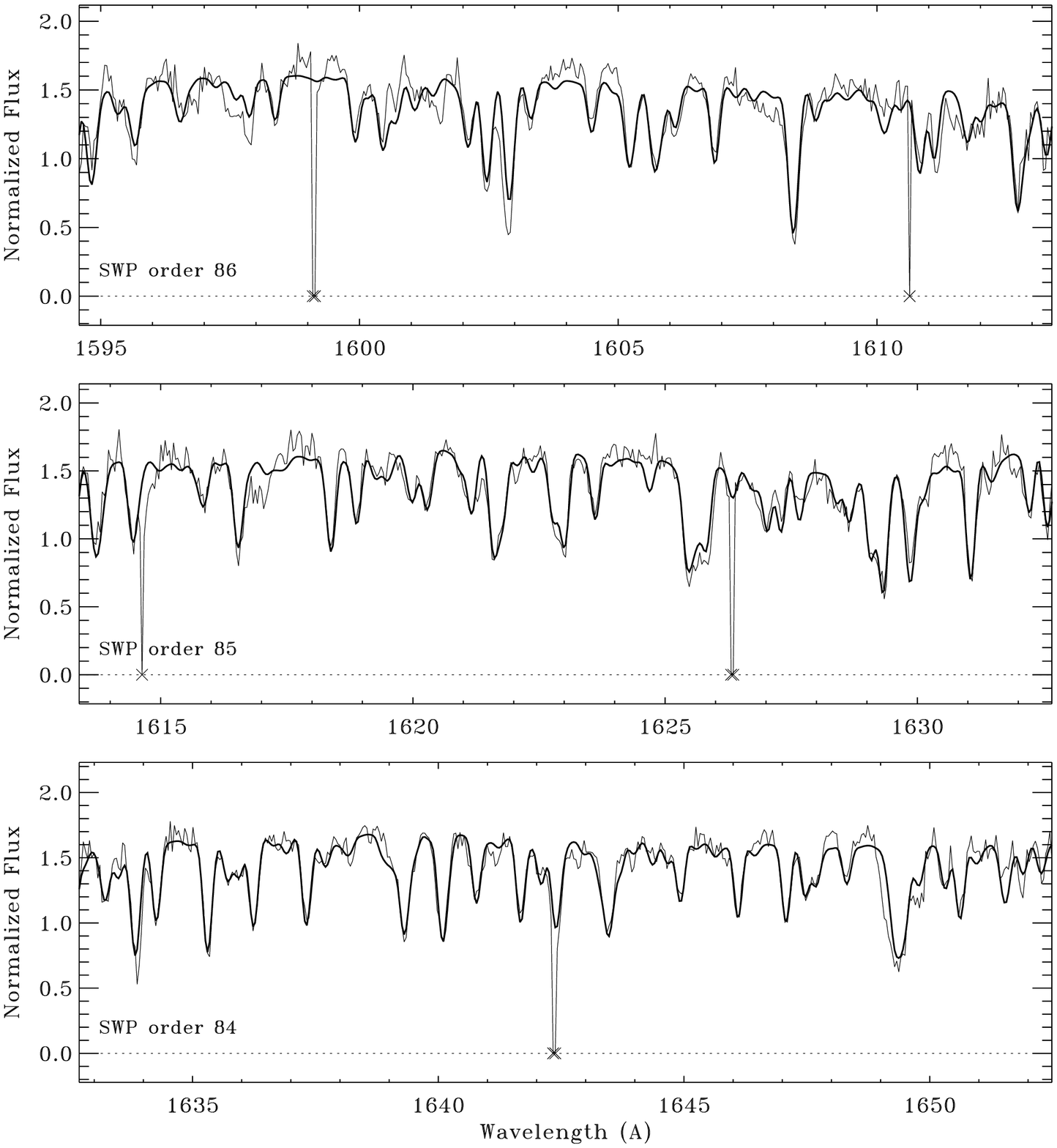}
\caption{Same as Figure \ref{fig_ONLINE1}, but for an additional set of high-dispersion \iue\ echelle orders, identified in the lower left of each panel.}
\end{figure}

\begin{figure}
\figurenum{Online 2.9}
\epsscale{1.00}
\plotone{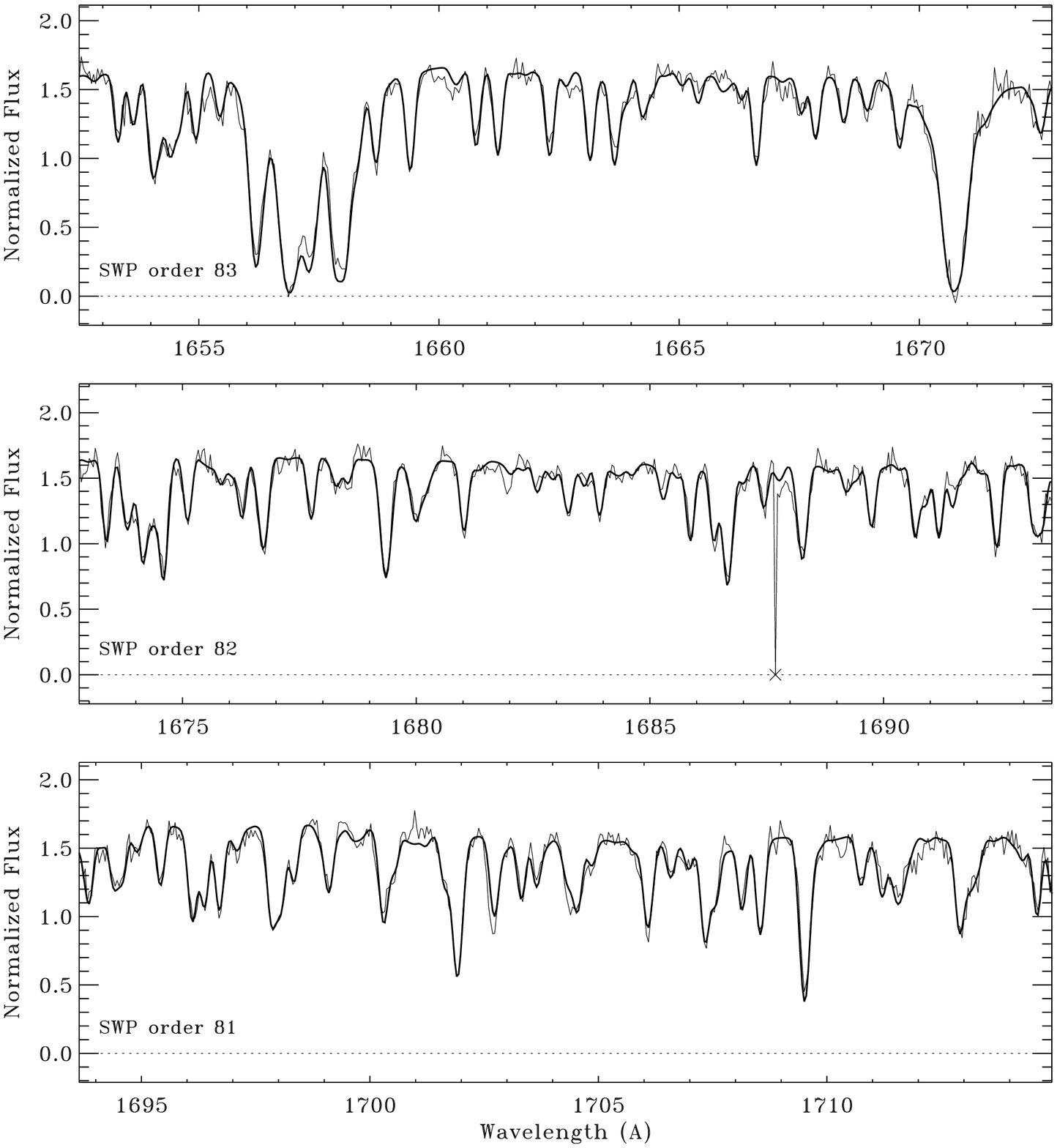}
\caption{Same as Figure \ref{fig_ONLINE1}, but for an additional set of high-dispersion \iue\ echelle orders, identified in the lower left of each panel.}
\end{figure}

\clearpage

\begin{figure}
\figurenum{Online 2.10}
\epsscale{1.00}
\plotone{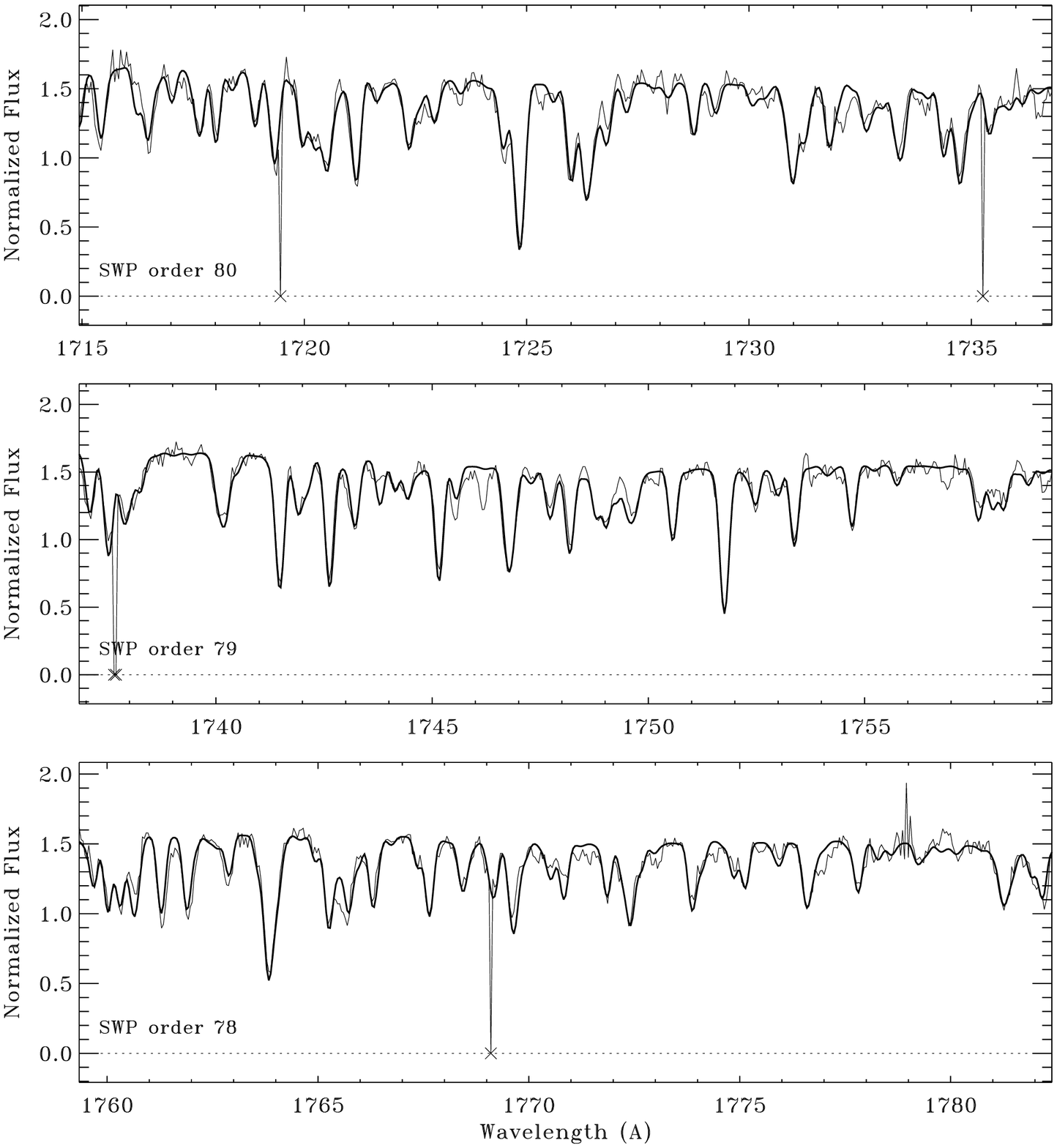}
\caption{Same as Figure \ref{fig_ONLINE1}, but for an additional set of high-dispersion \iue\ echelle orders, identified in the lower left of each panel.}
\end{figure}

\begin{figure}
\figurenum{Online 2.11}
\epsscale{1.00}
\plotone{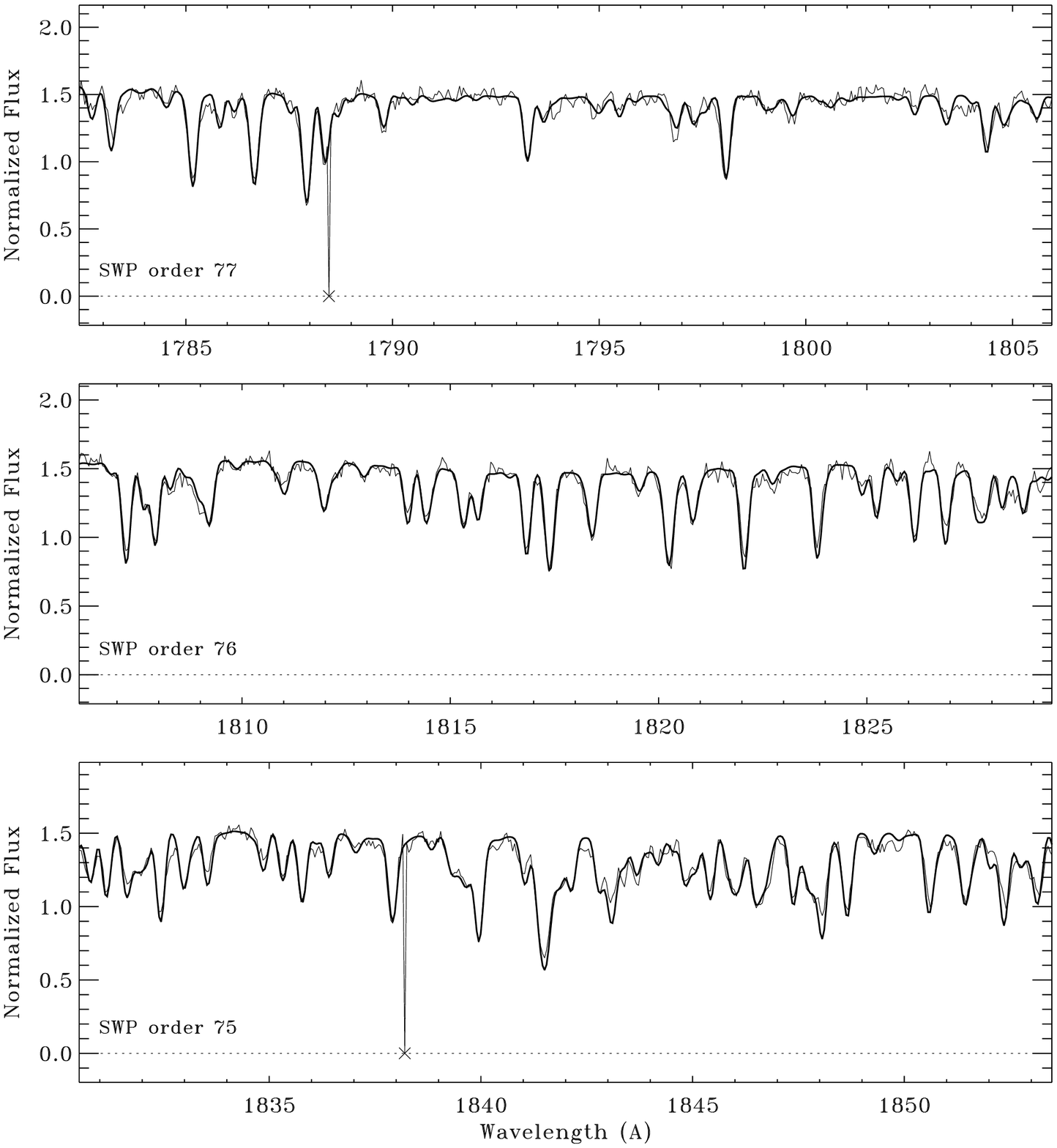}
\caption{Same as Figure \ref{fig_ONLINE1}, but for an additional set of high-dispersion \iue\ echelle orders, identified in the lower left of each panel.}
\end{figure}

\begin{figure}
\figurenum{Online 2.12}
\epsscale{1.00}
\plotone{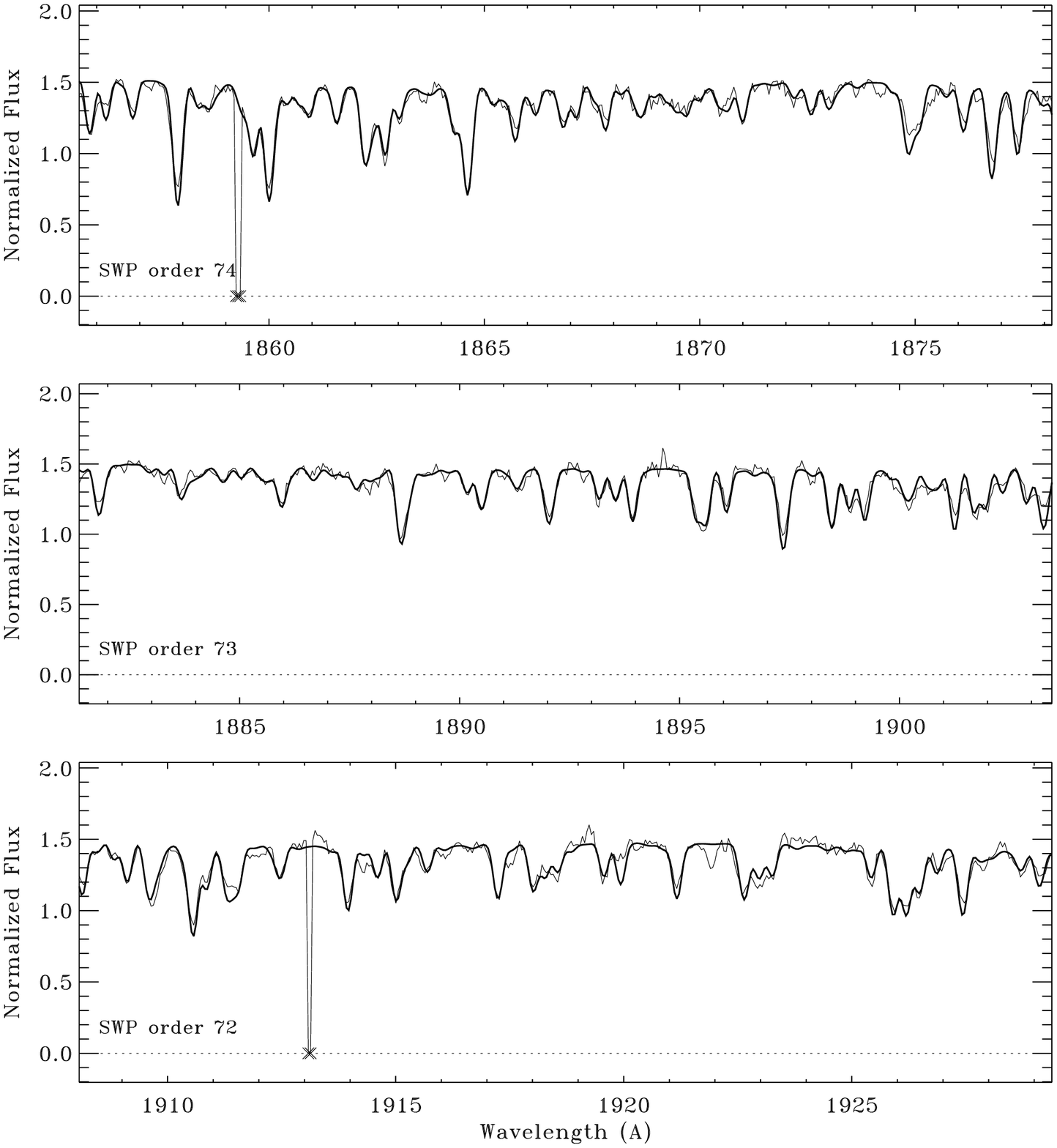}
\caption{Same as Figure \ref{fig_ONLINE1}, but for an additional set of high-dispersion \iue\ echelle orders, identified in the lower left of each panel.}
\end{figure}

\clearpage

\begin{figure}
\figurenum{Online 2.13}
\epsscale{1.00}
\plotone{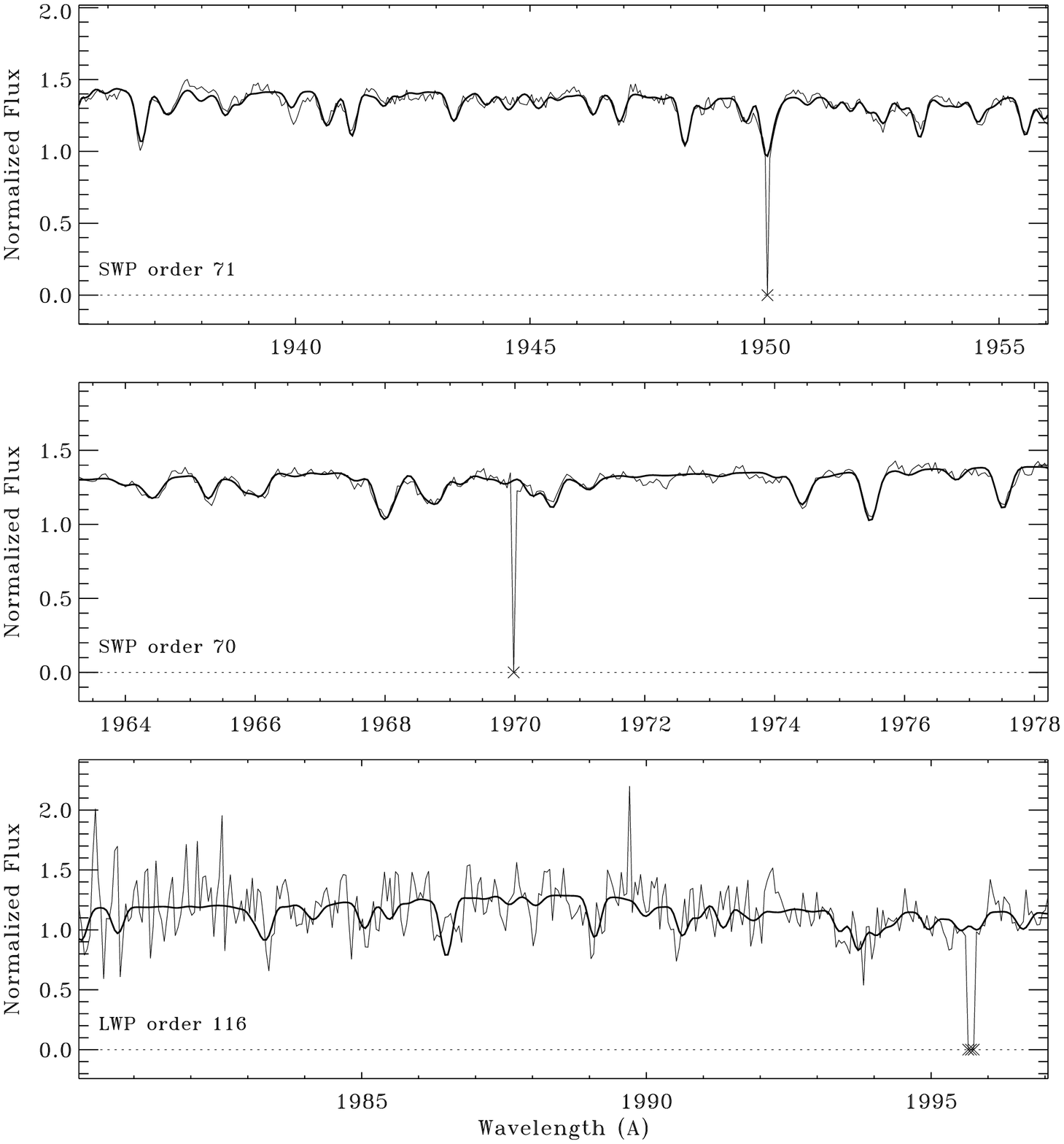}
\caption{Same as Figure \ref{fig_ONLINE1}, but for an additional set of high-dispersion \iue\ echelle orders, identified in the lower left of each panel.}
\end{figure}

\begin{figure}
\figurenum{Online 2.14}
\epsscale{1.00}
\plotone{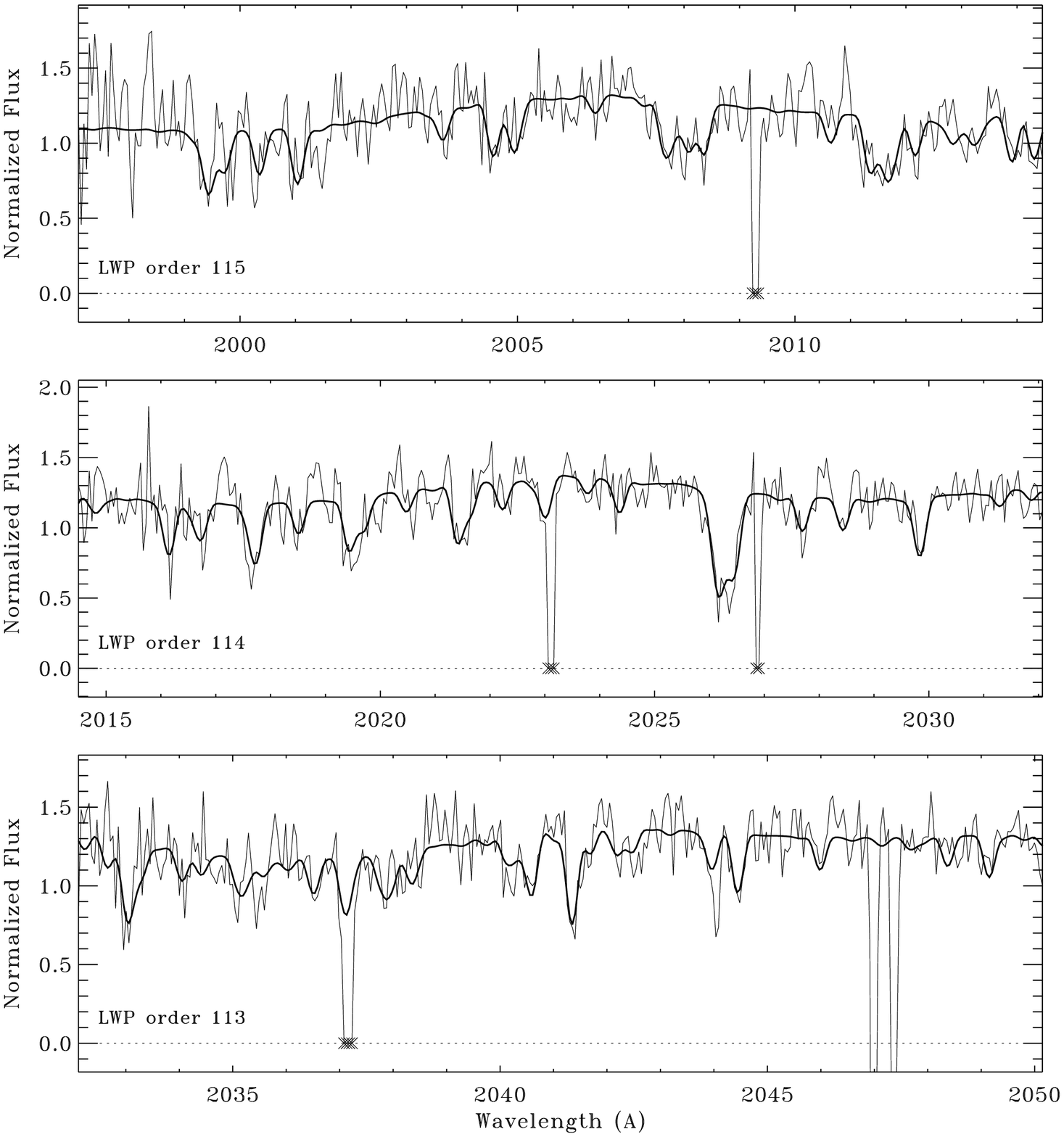}
\caption{Same as Figure \ref{fig_ONLINE1}, but for an additional set of high-dispersion \iue\ echelle orders, identified in the lower left of each panel.}
\end{figure}

\begin{figure}
\figurenum{Online 2.15}
\epsscale{1.00}
\plotone{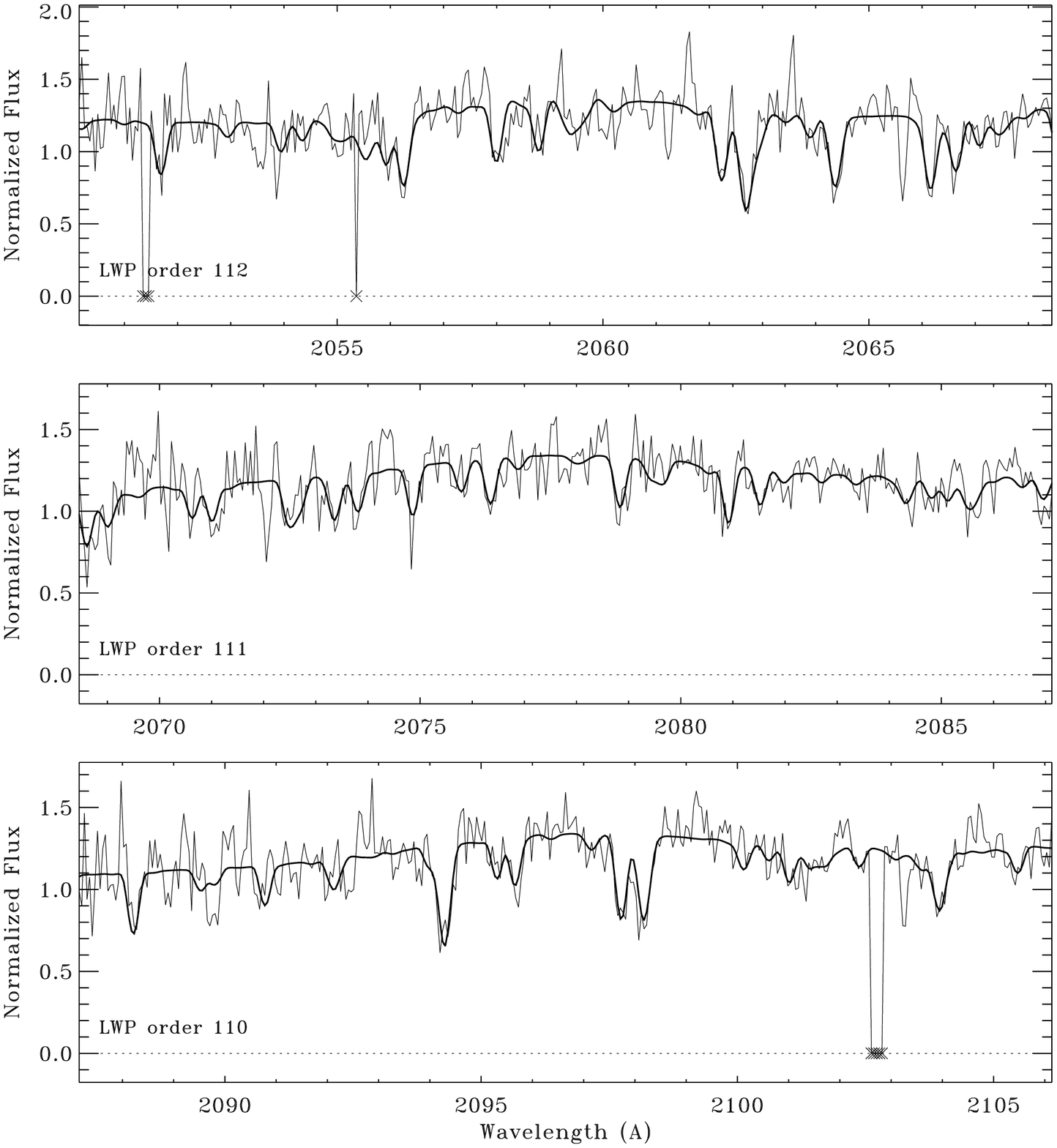}
\caption{Same as Figure \ref{fig_ONLINE1}, but for an additional set of high-dispersion \iue\ echelle orders, identified in the lower left of each panel.}
\end{figure}

\clearpage

\begin{figure}
\figurenum{Online 2.16}
\epsscale{1.00}
\plotone{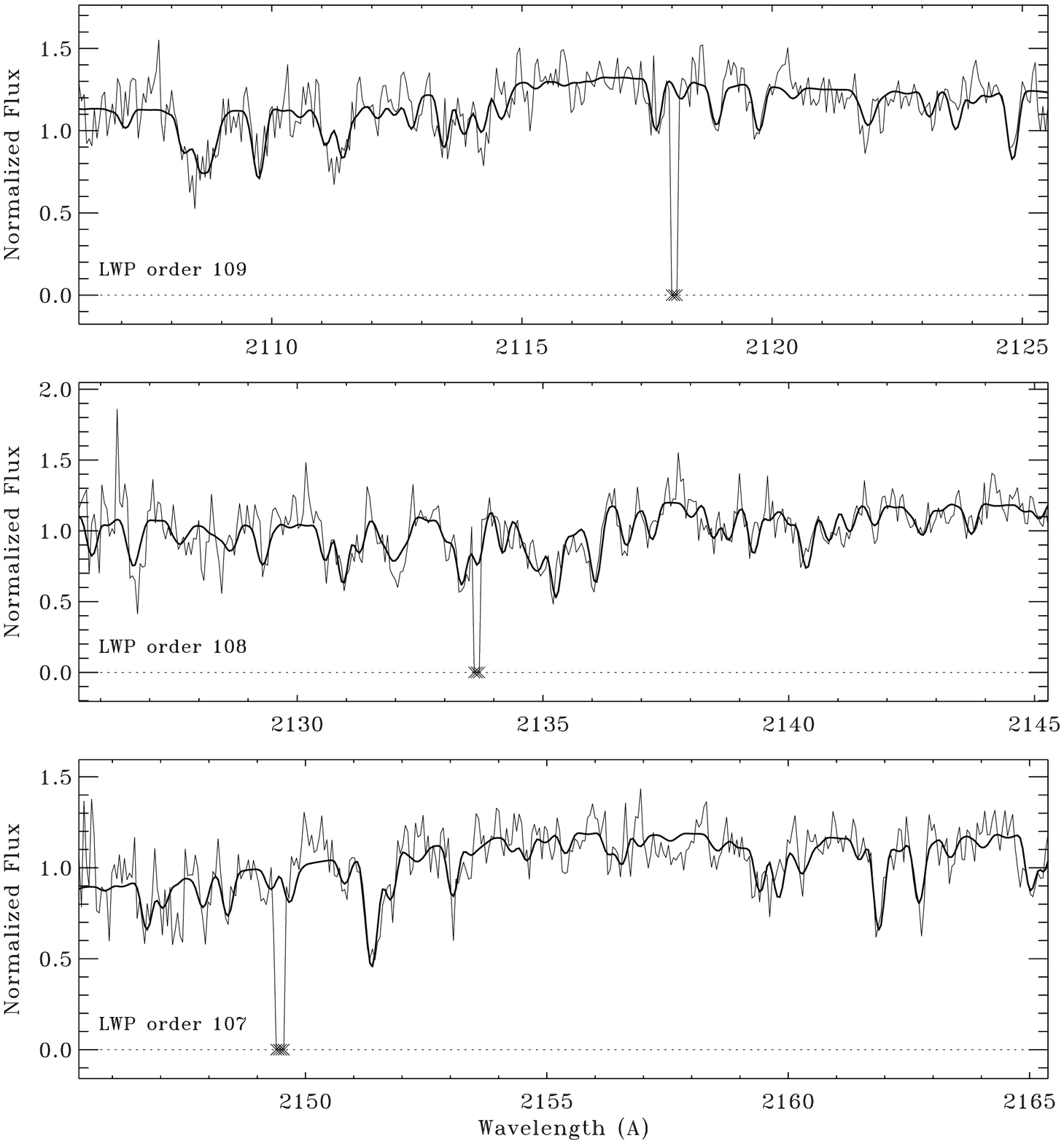}
\caption{Same as Figure \ref{fig_ONLINE1}, but for an additional set of high-dispersion \iue\ echelle orders, identified in the lower left of each panel.}
\end{figure}

\begin{figure}
\figurenum{Online 2.17}
\epsscale{1.00}
\plotone{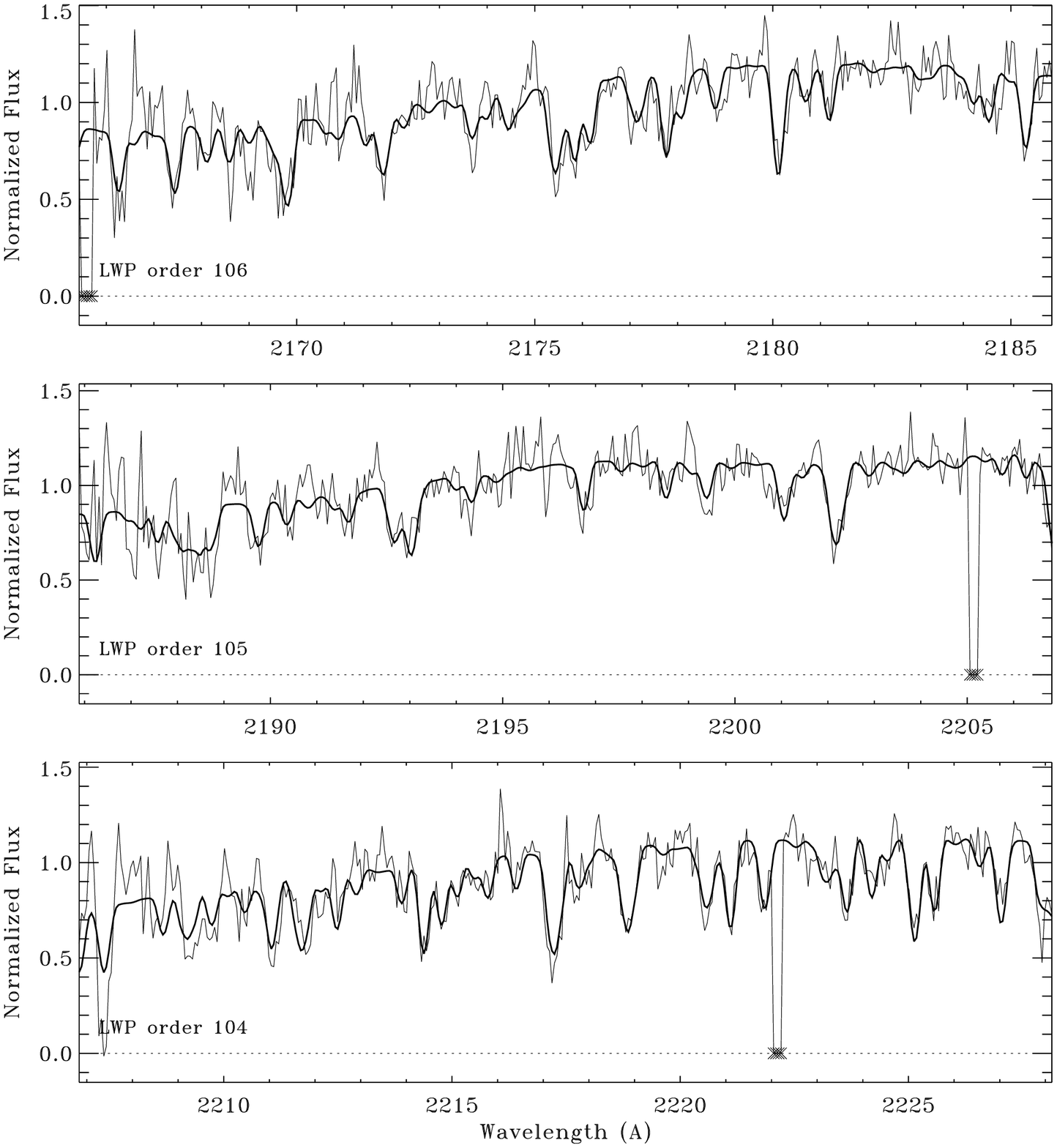}
\caption{Same as Figure \ref{fig_ONLINE1}, but for an additional set of high-dispersion \iue\ echelle orders, identified in the lower left of each panel.}
\end{figure}

\begin{figure}
\figurenum{Online 2.18}
\epsscale{1.00}
\plotone{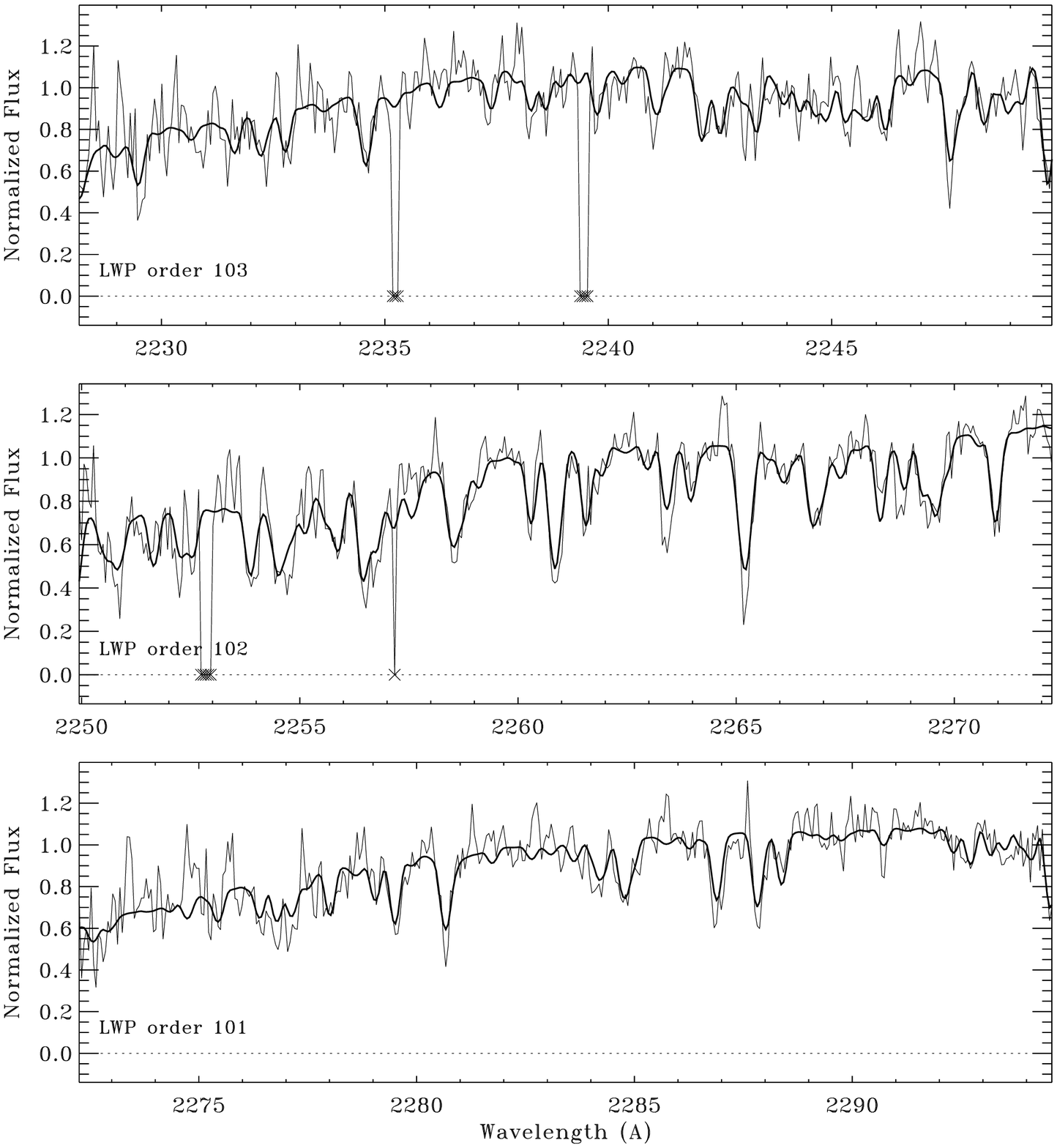}
\caption{Same as Figure \ref{fig_ONLINE1}, but for an additional set of high-dispersion \iue\ echelle orders, identified in the lower left of each panel.}
\end{figure}

\clearpage

\begin{figure}
\figurenum{Online 2.19}
\epsscale{1.00}
\plotone{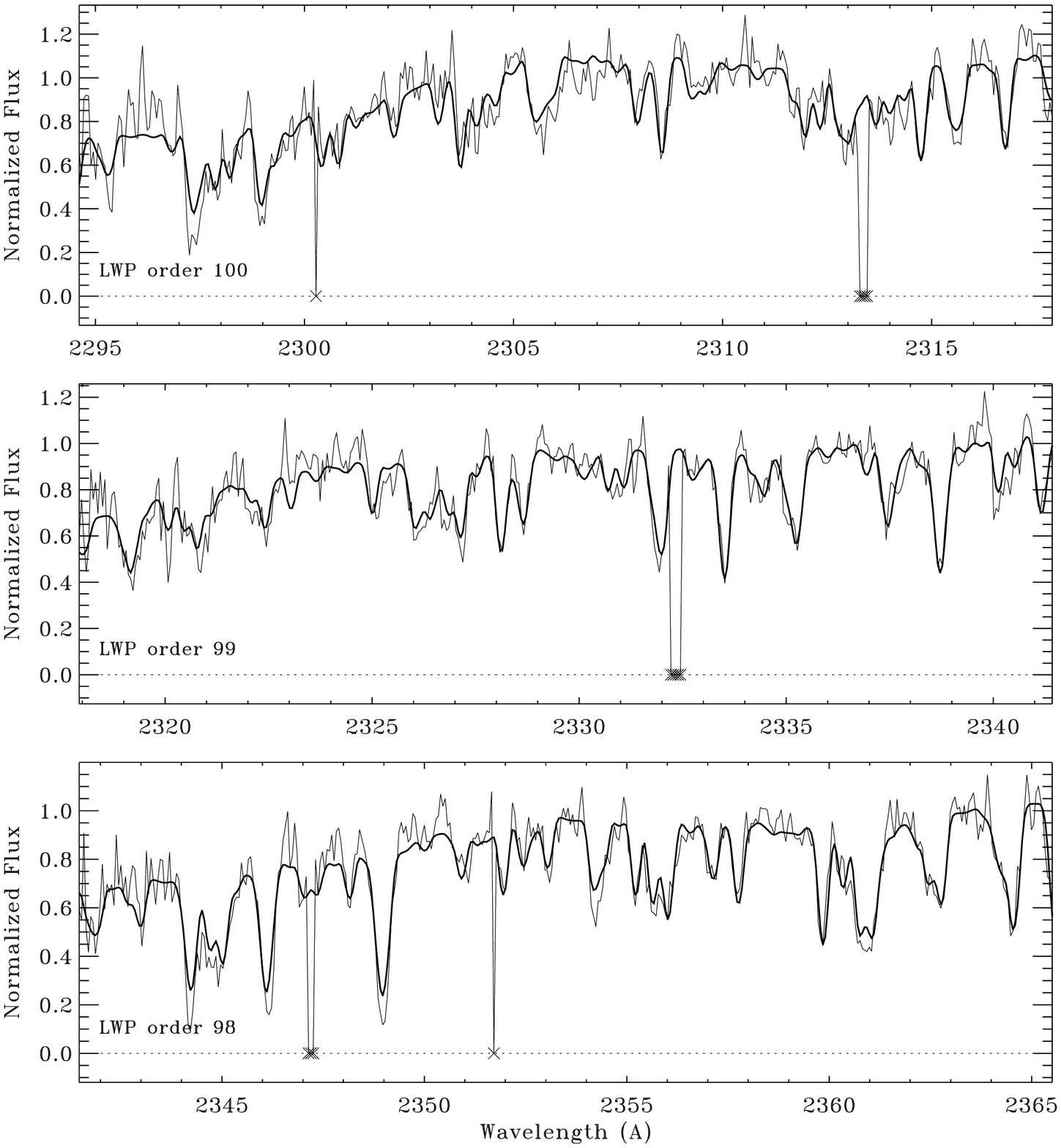}
\caption{Same as Figure \ref{fig_ONLINE1}, but for an additional set of high-dispersion \iue\ echelle orders, identified in the lower left of each panel.}
\end{figure}

\begin{figure}
\figurenum{Online 2.20}
\epsscale{1.00}
\plotone{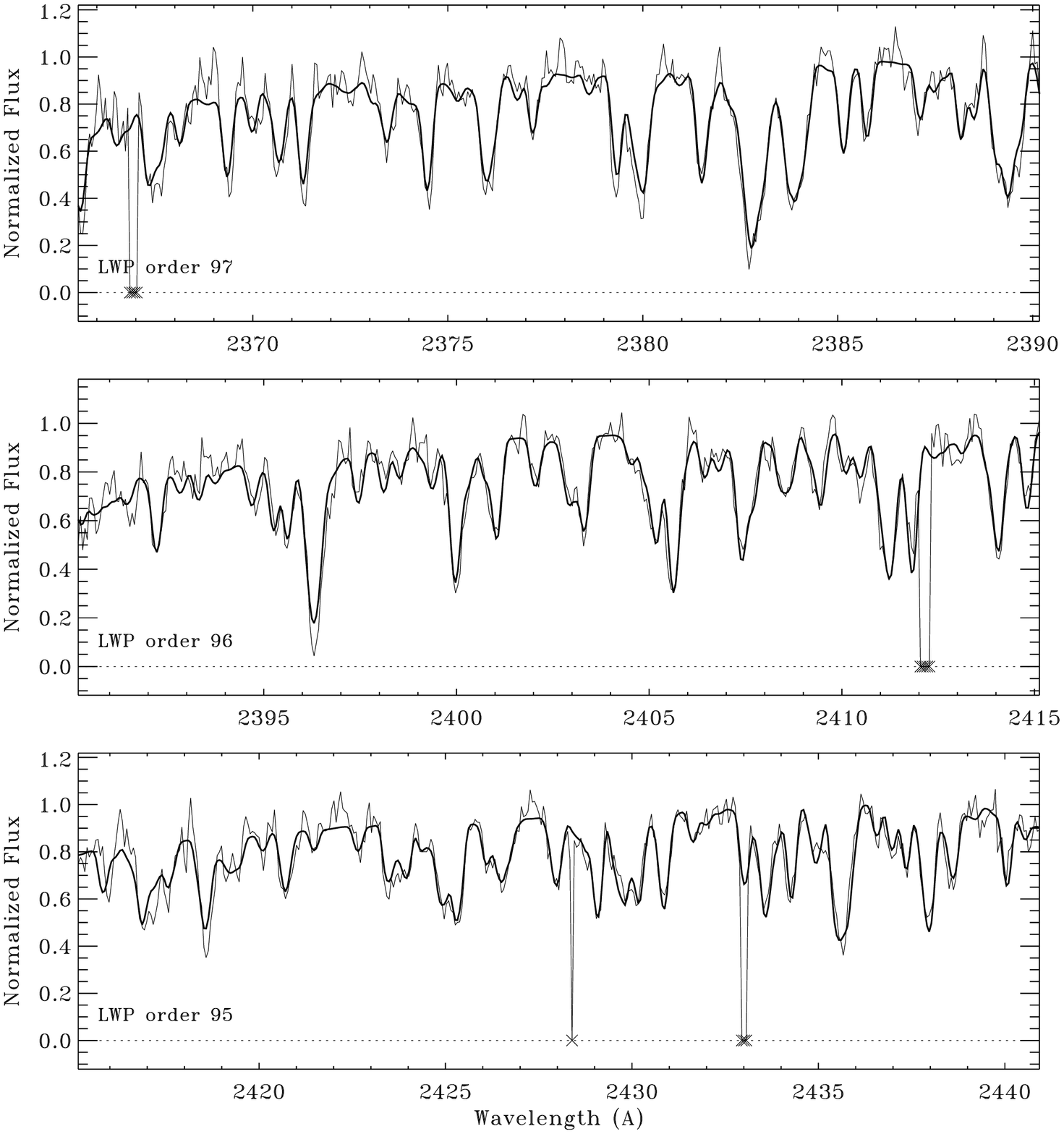}
\caption{Same as Figure \ref{fig_ONLINE1}, but for an additional set of high-dispersion \iue\ echelle orders, identified in the lower left of each panel.}
\end{figure}

\begin{figure}
\figurenum{Online 2.21}
\epsscale{1.00}
\plotone{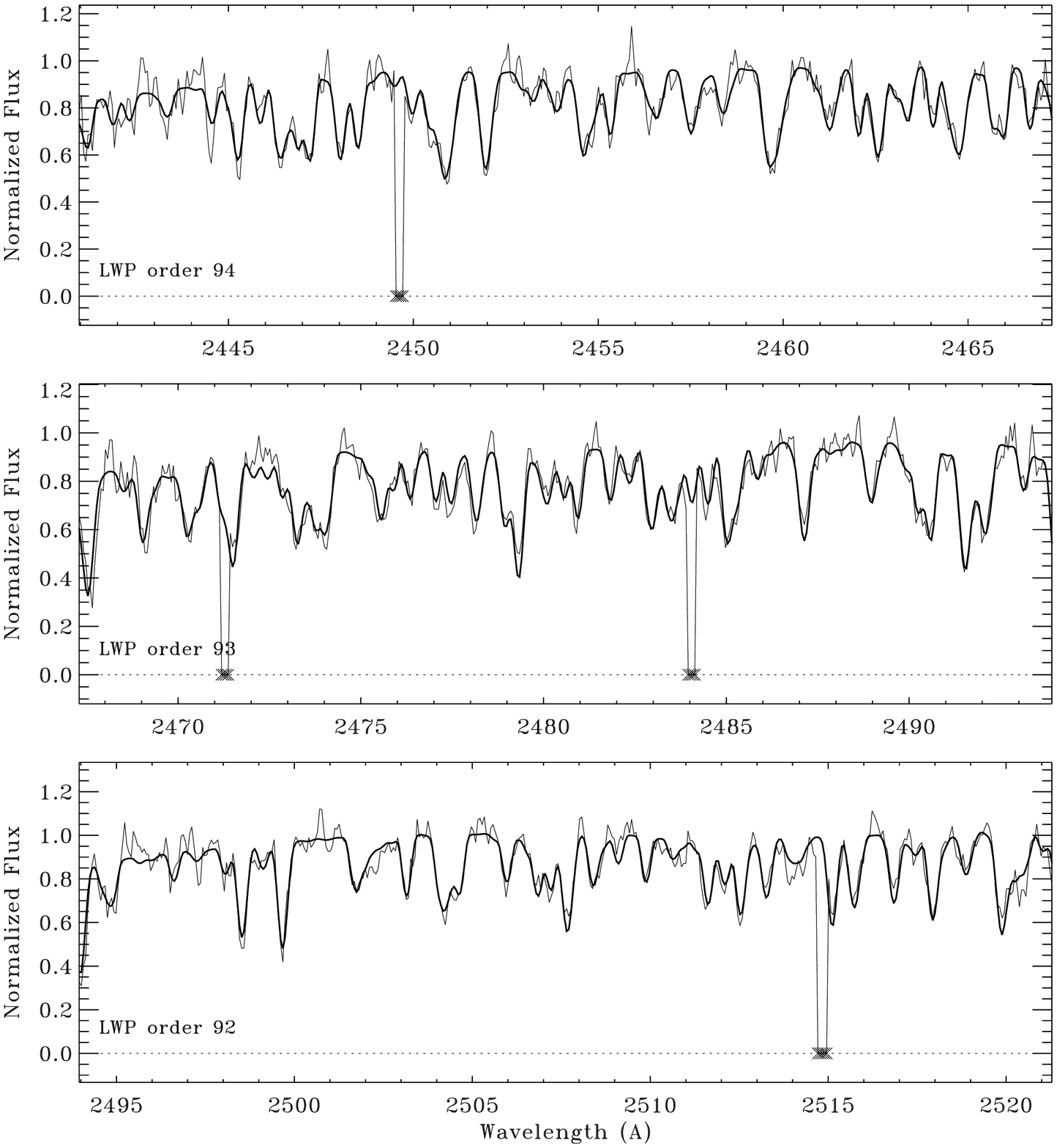}
\caption{Same as Figure \ref{fig_ONLINE1}, but for an additional set of high-dispersion \iue\ echelle orders, identified in the lower left of each panel.}
\end{figure}

\clearpage

\begin{figure}
\figurenum{Online 2.22}
\epsscale{1.00}
\plotone{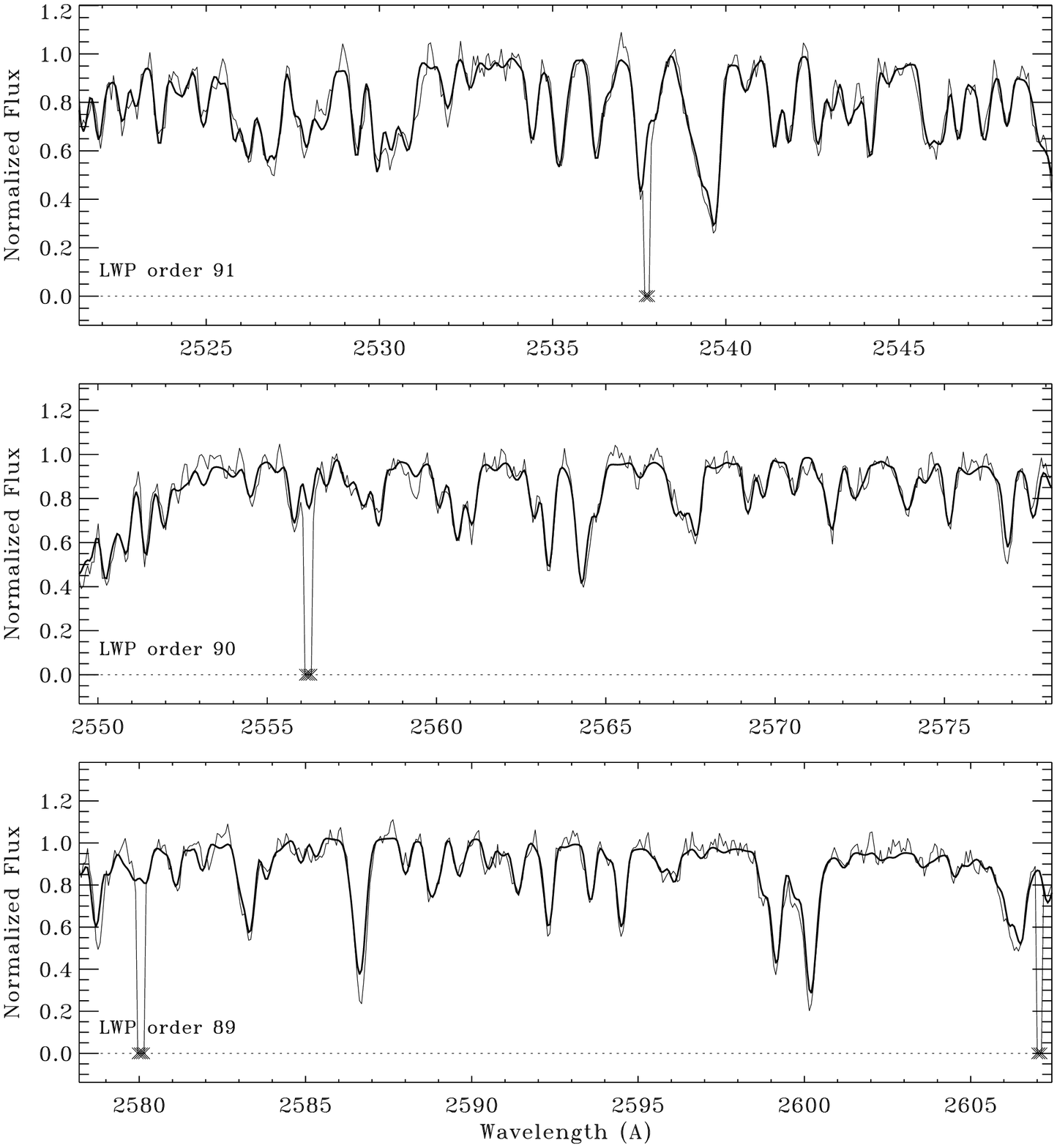}
\caption{Same as Figure \ref{fig_ONLINE1}, but for an additional set of high-dispersion \iue\ echelle orders, identified in the lower left of each panel.}
\end{figure}

\begin{figure}
\figurenum{Online 2.23}
\epsscale{1.00}
\plotone{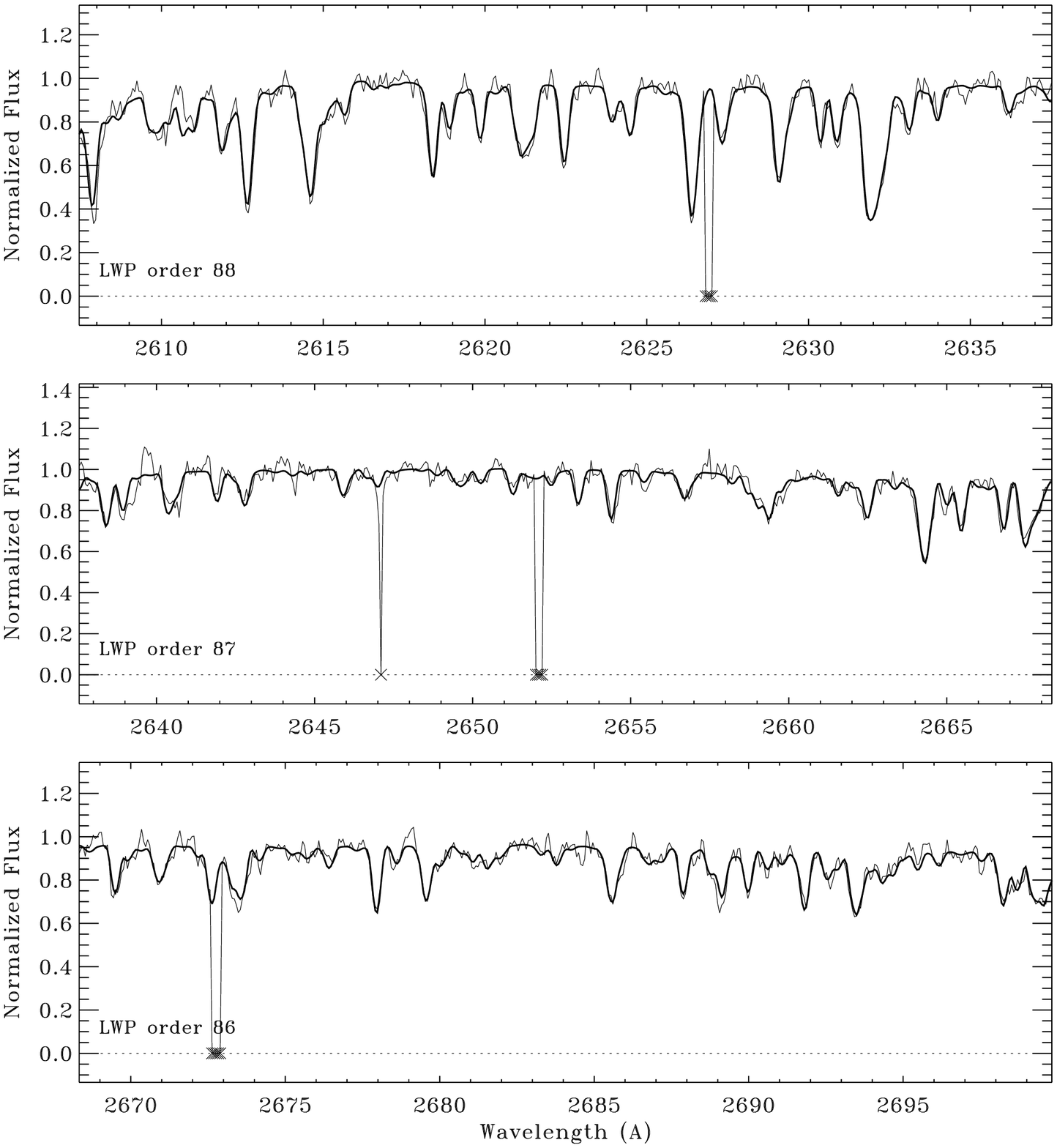}
\caption{Same as Figure \ref{fig_ONLINE1}, but for an additional set of high-dispersion \iue\ echelle orders, identified in the lower left of each panel.}
\end{figure}

\begin{figure}
\figurenum{Online 2.24}
\epsscale{1.00}
\plotone{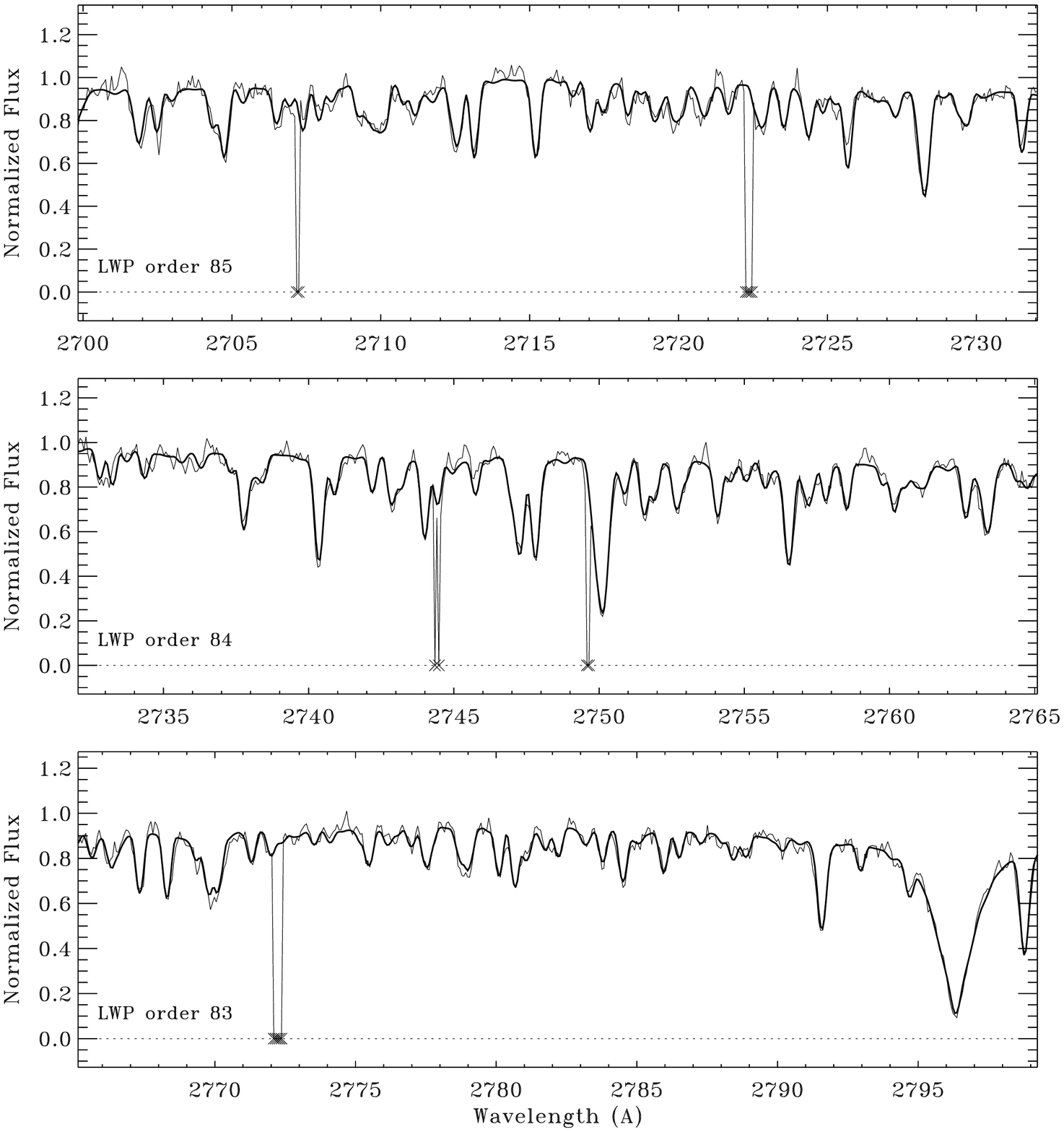}
\caption{Same as Figure \ref{fig_ONLINE1}, but for an additional set of high-dispersion \iue\ echelle orders, identified in the lower left of each panel.}
\end{figure}

\clearpage

\begin{figure}
\figurenum{Online 2.25}
\epsscale{1.00}
\plotone{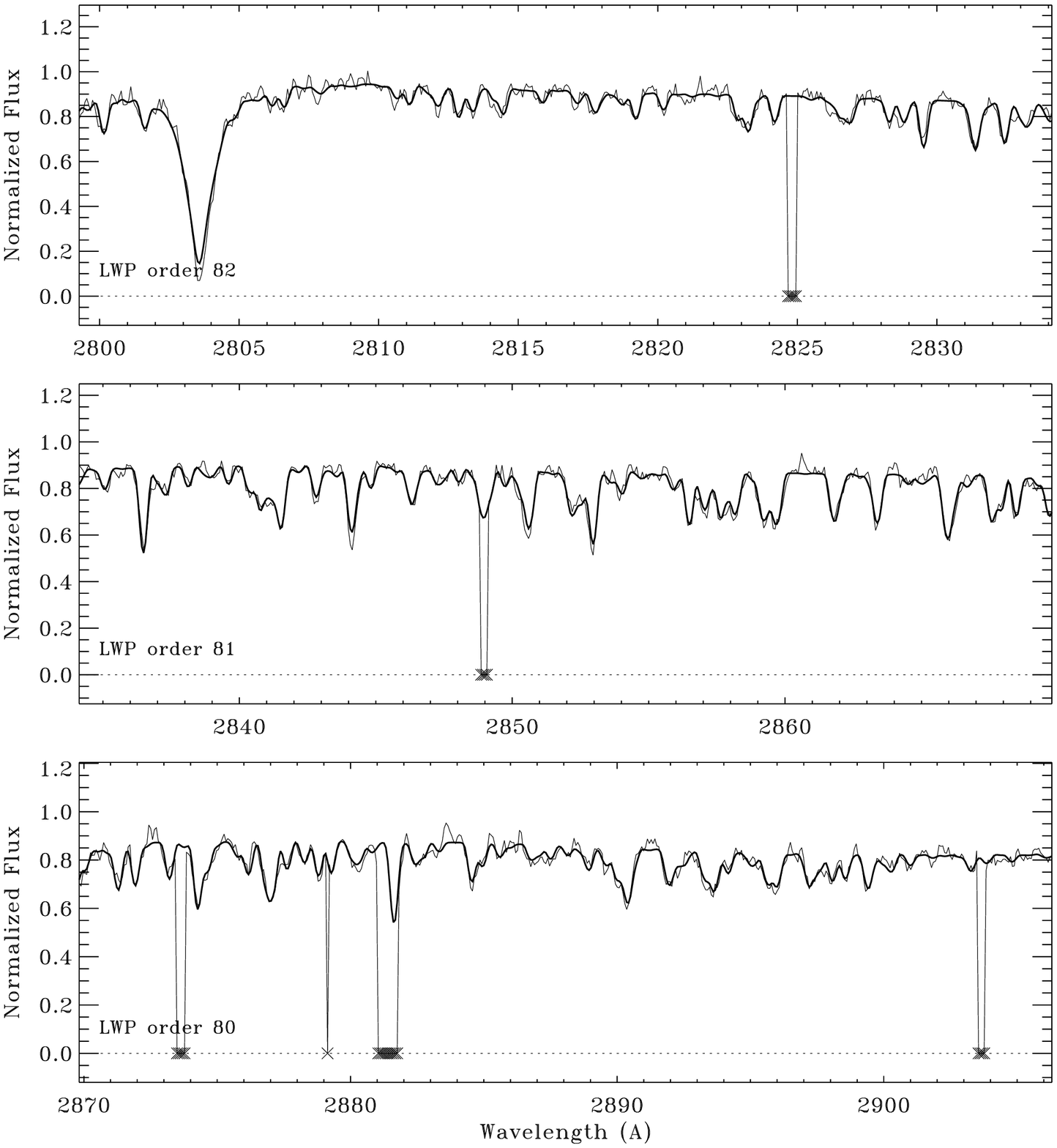}
\caption{Same as Figure \ref{fig_ONLINE1}, but for an additional set of high-dispersion \iue\ echelle orders, identified in the lower left of each panel.}
\end{figure}

\begin{figure}
\figurenum{Online 2.26}
\epsscale{1.00}
\plotone{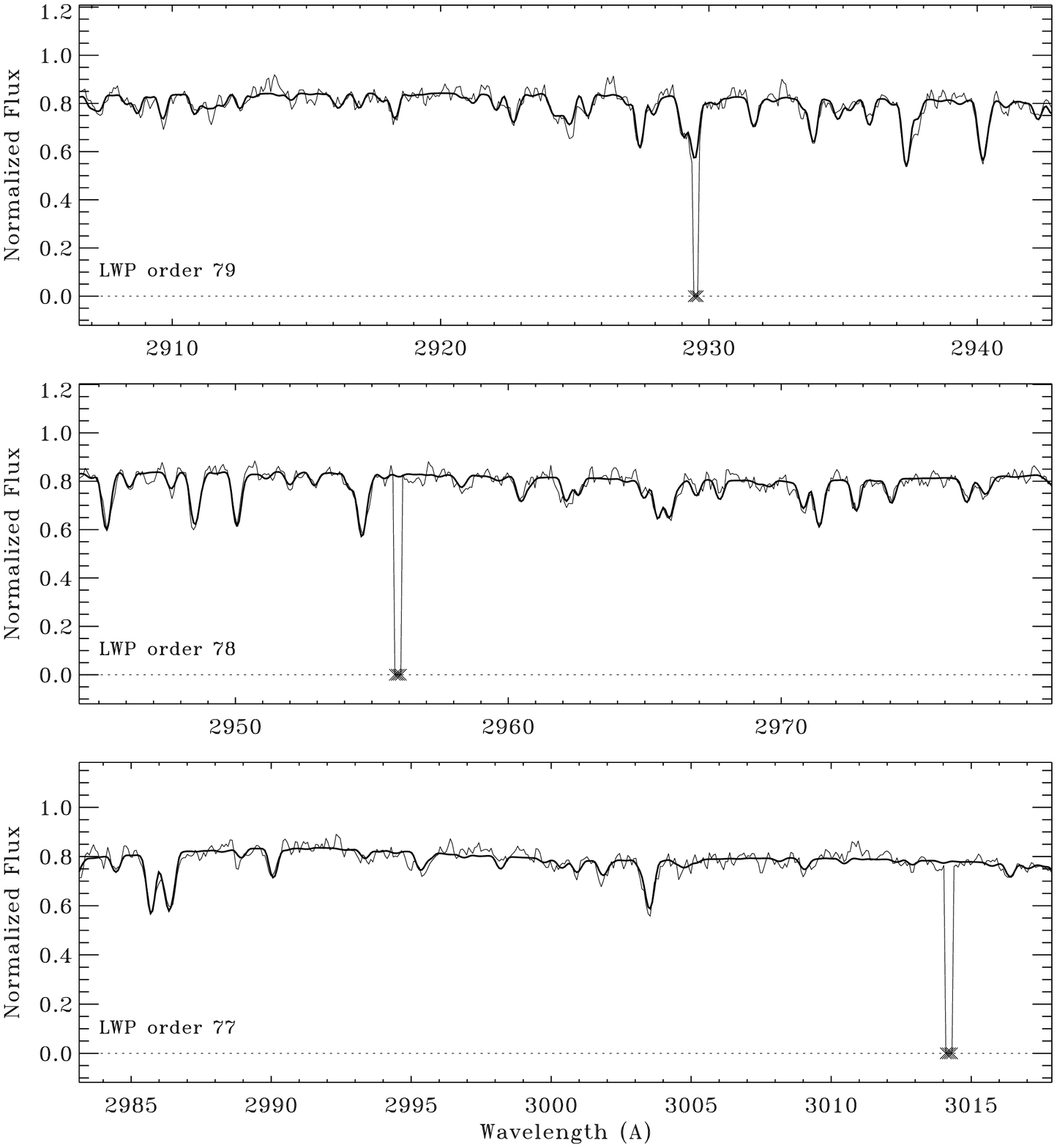}
\caption{Same as Figure \ref{fig_ONLINE1}, but for an additional set of high-dispersion \iue\ echelle orders, identified in the lower left of each panel.}
\end{figure}

\begin{figure}
\figurenum{Online 2.27}
\epsscale{1.00}
\plotone{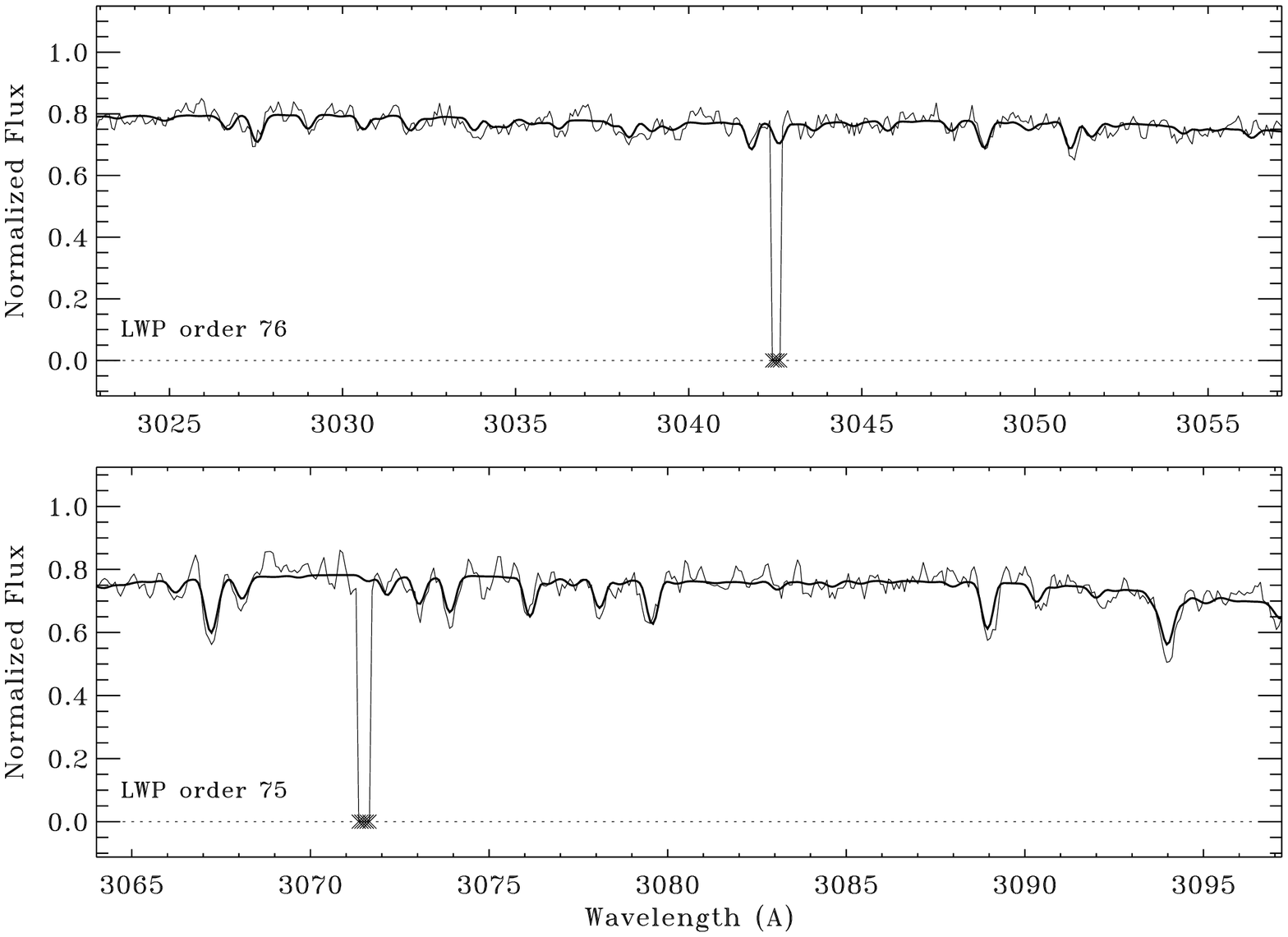}
\caption{Same as Figure \ref{fig_ONLINE1}, but for an additional set of high-dispersion \iue\ echelle orders, identified in the lower left of each panel.}
\end{figure}

\end{document}